\documentclass[authoryear]{elsarticle}

\usepackage[T1]{fontenc}
\usepackage{ae,aecompl}
\usepackage{graphicx}
\usepackage{amssymb}
\usepackage{amsmath}
\usepackage{mathptmx}
\usepackage{epstopdf}
\usepackage{booktabs}
\usepackage{float}
\usepackage{subfig}
\usepackage{epsfig,color,ulem}
\usepackage{tabularx}
\usepackage{enumitem}
\usepackage[utf8]{inputenc}

\newcommand{\msun}{\,{\rm M_{\odot}}}

\newcommand\simlt{\lower.5ex\hbox{$\; \buildrel < \over \sim \;$}}
\newcommand\simgt{\lower.5ex\hbox{$\; \buildrel > \over \sim \;$}}

\def\apjl{ApJL }

\def\apj{ApJ }

\def\apjs{ApJS }

\def\aap{A\&A }
\def\aaps{A\&AS }
\def\nat{Nature }
\def\mnras{MNRAS }
\def\prd{Phys. Rev. D. }

\def\ssr{Space Sci. Rev. }
\def\apss{AP\&SS}
\def\prl{Phys. Rev. Lett. }
\def\physrep{Physics Reports}

\def \x  {x_{l}} 
\def \n  {n_{l}}
\def \taut  {\tilde{\tau}}

\begin{document}

\begin{frontmatter}

\title{Physics of radiation mediated shocks and its applications to GRBs, supernovae, and neutron star mergers}

\author{Amir Levinson \& Ehud Nakar}

\address{School of Physics and Astronomy, Tel Aviv University, Tel Aviv 69978, Israel}

\begin{abstract}
The first electromagnetic signal observed in different types of  cosmic explosions is released upon 
emergence of a shock created  in the explosion from the opaque  envelope enshrouding the central source.
Notable examples are  the early emission from various types of supernovae and low luminosity GRBs, the prompt photospheric 
emission in long GRBs, and the gamma-ray emission that accompanied the gravitational wave signal in neutron star mergers.
In all of these examples, the shock driven by the explosion is mediated by the radiation trapped inside it, and its velocity and structure, that depend on
environmental conditions, dictate the characteristics of the observed electromagnetic emission at early times, and potentially  also their neutrino emission. 
Much efforts have been devoted in recent years to develop a detailed theory of radiation mediated shocks
in an attempt to predict the properties of  the early emission in the aforementioned systems. 
These efforts are timely in view of the  anticipated detection rate of shock breakout candidates by upcoming transient factories, and the potential
detection of a gamma-ray flash from shock breakout in neutron star mergers like GW170817.   This review aims at providing a comprehensive overview of the theory and applications of radiation mediated shocks,  starting from basic principles.  The classification of shock solutions, which are governed by
the conditions prevailing in each class of objects, and the methods used to solve the shock equations in different regimes will be described, with particular emphasis on the observational diagnostics.    
The applications to supernovae, low-luminosity GRBs, long GRBs, neutron star mergers, and neutrino emission will be highlighted.
\end{abstract}


\end{frontmatter}

\tableofcontents

%


\section{Introduction}
Shocks are ubiquitous  in high-energy astrophysics.  They are believed to be the sources of non-thermal photons, 
cosmic-rays and neutrinos observed in a plethora of extreme cosmic phenomena.    Under certain conditions, that prevail
in various astrophysical as well as terrestrial systems, the shock dissipation mechanism  is radiative.
Such shocks, termed {\it radiation mediated shocks} (RMS), 
are responsible for the early emission observed in 
various types of stellar explosions and other violent phenomena.  
The emission released upon the breakout of a RMS  carries a wealth of information regarding the 
properties of the system, e.g., the explosion mechanism and progenitor type in supernovae and low luminosity GRBs, 
the nature of the segregated outflow in neutron star mergers, etc. 
The investigation of RMS, particularly in the relativistic and mildly relativistic regimes, is a newly emerging field which is motivated 
by the recent progress in theory and observations.  
It came into the focus of high-energy astrophysics in the past decade with the detection of shock breakout candidates, such as the recent neutron star merger (GW170817), low-luminosity GRBs and various SNe; the inference of prompt photospheric emission in
long GRBs; and the detection of extra-galactic, diffuse, high-energy neutrino background of an unknown origin. 

RMS form when a  fast shock propagates in a plasma with sufficient optical depth. 
They are mediated by Compton scattering and, under certain conditions, pair creation, and
their properties are vastly different than those of {\it collisionless shocks}, that form in dilute plasmas and
in which dissipation is mediated by collective plasma processes.    
The prompt electromagnetic signal emitted upon the breakout of a RMS and the subsequent emission released when deep layers 
behind the shock approach the photosphere are determined solely by the shock structure.  The latter depends, in turn, on 
the environment in which the shock propagates and on its velocity, that vary significantly between the various systems. 
For instance, shocks that are generated by various types of stellar explosions propagate in an unmagnetized, photon poor medium,
and their velocity prior to breakout ranges from sub-relativistic to ultra-relativistic, depending on the type of the progenitor and 
the explosion energy \citep{nakar2012}. 
Sub-photospheric shocks in GRBs, on the other hand,  propagate in a photon rich plasma, conceivably  with a 
non-negligible magnetization, at mildly relativistic speeds.  Consequently, 
while the menagerie of cosmic phenomena described above 
share a similar underlying physics, predicting their observational characteristics requires detailed analysis of 
the RMS  solution under the specific conditions prevailing in each source.

Early works on RMS date back a half century  \citep{pai1966,zeldovich1967,weaver1976,BP81a,BP81b,riffert1988,lyubarsky1982}.
They were motivated primarily by terrestrial applications, as well as the applications to ordinary supernovae
and accreting neutron stars.    The shocks in all of these systems are highly sub-relativistic, which greatly simplifies the analysis
and reduce the efforts required to solve the shock equations.   In particular, the diffusion approximation has been commonly employed to compute 
the transfer of radiation through the shock.  
Unfortunately, the limited range of shock velocities 
that can be analyzed by employing such techniques renders its applicability to most high-energy transients of little relevance.
In the last decade there has been a growing interest in extending the analysis to the relativistic and mildly relativistic regimes 
\citep{eichler1994,levinson2008,katz2010,budnik2010,levinson2012,nakar2012,keren2014,beloborodov2017a,beloborodov2017b,ito2018a,
granot2018,lundman2018a}.
Various analytical and numerical methods, each suitable for analyzing a specific class of relativistic transients, have been developed and applied to identify observational diagnostics.   Much progress has been made also in the study of shock 
breakout from non-relativistic transients \citep[see, e.g.,][for a recent review]{waxman2017}.    The rest of this 
introductory section presents a more elaborate account of the applications of RMS to specific systems.    In-depth discussions of these systems
is presented in sections \ref{sec:GRBs}-\ref{sec:BNS}.   In section \ref{sec:Theory} we develop the detailed theory of non-relativistic and relativistic RMS.   
Readers who are not interested in the gory details of the analysis can find a schematic overview of the shock physics in section  \ref{sec:basic-principles}, 
and skip the rest of this section.

\subsection{Shock breakout in supernovae and low luminosity GRBs}
\label{sec:Intro_shock_breakoout}
The collapse of a massive star creates a radiation dominated shock wave that propagates in the stellar envelope, breaks out, and ultimately emits the observed supernova light.  In the majority of core-collapse events the breakout occurs at the edge of the stellar envelope, however, in stars that eject a sufficiently intense stellar wind prior to their collapse  the RMS continues to propagate in the wind until reaching a large enough radius at which breakout ensues. 
In general, shock breakout takes place once the optical depth to the observer becomes too small to
prevent substantial leakage of photons through the upstream plasma. 
At this point the radiation trapped inside the shock transition layer is released (roughly over the diffusion time) 
and is seen to a distant observer as a flash.
Observationally, this burst of emission, which is the first electromagnetic signal released by any type of stellar explosion, 
is commonly referred to as "shock breakout", and its 
characteristics (luminosity and spectral evolution) depend solely on the RMS structure during the breakout phase. 
Following this phase the hot gas behind the shock starts expanding gradually, allowing the radiation trapped inside it to 
escape to infinity, at first from the immediate shock downstream and later from inner layers. This emission, known as the "cooling envelope" emission, dominates the luminosity of all non-interacting core-collapse SNe during the first hours to days, and in some cases (such as in type IIp SNe) even for months. The radiation released during the cooling envelope phase was deposited by the RMS prior to its breakout, and the properties of the early cooling envelope emission (first hour to a day) 
reflect the RMS structure (the subsequent emission has enough time to achieve a full thermodynamic equilibrium before escaping the system). 

A common misconception in the literature when referring to actual observations  is to term the different phases of the early emission, which often include only the cooling envelope phase, as the "shock breakout".   From a physical point of view the breakout episode marks a transition from a RMS to a collisionless shock.   From an observational point of view a unique feature of the shock breakout signal in SNe, as opposed to the cooling envelope emission, is a sharp rise of the {\it bolometric} luminosity;
the {\it bolometric} luminosity of the subsequent emission, including from the cooling envelope, declines gradually with time, typically as a power law.
The source of this confusion is improper use of the optical light curve as an indicator of the shock evolution. 
Given that the breakout emission in typical SNe is very hard (peaks at the extreme UV to X-rays), the optical luminosity continues to rise
also during the early cooling phase (hours to days), even though the {\it bolometric} luminosity is already declining, until the radiation cools 
down to a temperature of about $10^4$ k. 
This confusion can lead to wrong consequences regarding the system parameters since the properties of the shock breakout signal and the cooling envelope emission are vastly different.    

Under the conditions prevailing in essentially all SNe types (as opposed to GRBs, see below) the plasma upstream of the shock is photon poor and unmagnetized. This has a profound effect on the shock structure since all the photons are produced inside the shock transition layer and its immediate downstream.  Given these conditions, the RMS structure depends mostly on two parameters; the shock velocity and the density profile at the breakout zone.   The shock velocity in particular dictates the breakout temperature and, hence, the spectrum of the breakout signal.  Three important regimes can be identified: 
\begin{enumerate}[label=(\roman*)]
\item Slow shocks ($\beta_{sh} \lesssim 0.05$), in which the radiation is in thermodynamic equilibrium and the breakout temperature depends rather weakly on the velocity and the density, viz., $T_{bo} \propto \rho^{1/4} v^{1/2}$. For typical SNe parameters in this regime the breakout emission peaks in the extreme UV, $T_{bo} \approx 10-100$ eV. 
\item Fast Newtonian shocks ($0.05 \lesssim \beta_s \lesssim 0.5$), in which the radiation is out of thermodynamic equilibrium and the temperature is determined by the amount of photons produced in the immediate downstream (over one diffusion length roughly). The breakout temperature in this regime depends very sensitively on the shock velocity, ranging from $\sim 0.05$ keV at $\beta=0.05$ to $\sim 50$ keV at $\beta=0.5$, leading 
to a breakout signal that peaks in the X-ray band. 
\item Relativistic shocks  ($\beta_{sh} \gamma_{sh} \gtrsim 0.5$). At these velocities the shock structure and emission are strongly affected by
vigorous pair creation. 
In particular, the freshly created pairs significantly enhance the production of photons inside the shock, thereby 
regulating the downstream temperature. In the rest frame of the downstream plasma it lies in the range $\sim 100 - 200$ keV, practically independent of the shock Lorentz factor. In the observer frame it is boosted by a factor of $\gamma_{sh}$.  Consequently, relativistic breakouts produce $\gamma$-ray flares.
\end{enumerate}
In cases where the explosion is spherical and the breakout occurs at the progenitor's surface, the breakout velocity depends primarily on the progenitor radius 
and the ratio between the explosion energy and the progenitor mass \citep[e.g.,][]{nakar2010,katz2010,nakar2012}.  Before emerging from the star the shock accelerates in the steep density gradient near the edge of the stellar envelope, reaching velocities that can be considerably higher than 
those of the bulk of the ejecta \citep{sakurai1960,matzner1999}.
For a typical SN energy of $10^{51}$ ergs the breakout is always Newtonian. It is slow if the progenitor is extended (e.g., red supergiant [RSG] as in type IIp) and fast if it is compact (e.g., Wolf-Rayet as in type Ib/c). A sufficiently energetic SN from a compact progenitor ($E \gtrsim 10^{52}$ erg and $R_* \lesssim 10^{11}$ cm) can lead to a relativistic breakout.  A relativistic breakout is expected also when a relativistic jet drives a shock into the external parts of the envelope. This is certainly the case if the jet successfully breaks out of the progenitor, such as in long GRBs,  but it is also expected when the jet is choked in the outer layers of the extended progenitor's envelope, as may very well be the case in low-luminosity GRBs \citep{nakar2015}. The duration of the breakout pulse in a spherical breakout from a stellar surface is dominated by the light travel time and is therefore $\approx R_*/c$ for a Newtonian shock. To be more precise, a breakout from a WR progenitor (which is not surrounded by a thick wind) is expected to produce an X-ray pulse with a luminosity of $\sim 10^{44}$ erg/s and duration of $\sim 10$ s, while a breakout from a RSG produces an extreme UV pulse with a luminosity of $\sim 10^{45}$ erg/s and a duration of $\sim 1000$ s.

A different signal is expected when the progenitor is surrounded by a wind. If the wind is thick enough to sustain an RMS then the breakout can take place at a radius much larger than $R_*$. The duration of the breakout signal is significantly longer, $\approx R_{bo}/v_{sh}$, and the energy it releases is considerably larger. Depending on the wind optical depth the breakout duration can range from minutes to weeks and possibly even months. In extreme cases there are events where the entire SN light, over a duration of $\sim 100$ d, is thought to be the shock breakout emission from a very thick wind (e.g., SN 2006gy; \citealt{chevalier2011}). In case of a relativistic shock breakout from a  wind the physics involved in the breakout process is significantly altered.  Most notably, since relativistic shocks build their own opacity trough pair creation, 
photons start leaking from the shock long before complete conversion to a collisionless shock takes place, and 
the emergence of the shock from the wind is very gradual \citep{granot2018}. 

\subsection{Prompt GRB emission}

The nature of the prompt emission in long GRBs is a long standing issue.    Historically, the first fireball
models \citep{paczynski1986,goodman1986}, that asserted adiabatic expansion of 
a pure pair-photon plasma, predicted  that the emerging emission should have 
a  black-body spectrum.  The lack of detection of a black body component in the prompt emission of many GRBs in 
subsequent observations, has led to the hypothesis that the observed emission is produced 
by non-thermal processes in dissipative regions located at relatively large distances from the central 
engine \citep{meszaros1992,rees1992,levinson1993}.     Synchrotron emission by shock accelerated electrons 
has emerged as a leading model.    However, this interpretation has been challenged later on by detailed spectral analysis of
BTSE  sources \citep{preece1998,eichler2000}.   
The main difficulties were (i) the fact that in the majority of the bursts, the portion of the spectrum below the peak 
appears to be much harder than that predicted by the synchrotron shock model \citep{preece1998}, 
and (ii) the apparent clustering of peak energies \citep{frail2001,vanputten2003,ghirlanda2004,eichler2004,yamazaki2004,levinson2005}
that requires fine tuning of 
the model parameters.  Moreover, the anticipated low radiative efficiency of optically thin internal shocks 
has shown to impose stringent constraints on the energetics, that are hard to accommodate in realistic scenarios. 

The difficulties mentioned above have led to re-examination of  photospheric emission models
\citep{eichler2000,ryde2005,ryde2009,ryde2011,peer2011a,peer2011b,lundman2013,levinson2012,beloborodov2013,keren2014,deng2014,ito2018a,ito2018b,parsotan2018,parsotan2018b}.  
It has been proposed that an underlying  thermal component exists essentially in all bursts, and that
its inclusion in the analysis yields a better fit to the overall prompt emission spectrum  \citep{ryde2005} .
The relative strength of this component determines the spectral shape; 
while in the few bursts that exhibit prominent thermal emission it dominates, in
all others it is overwhelmed by the nonthermal emission produced above the photosphere.
How constrained those fits are and whether they can be considered good indicators of underlying thermal emission 
is yet an open issue. 

While the presence of a thermal component strongly implies photospheric emission, the opposite is not true. 
It has been shown that a broad, non-thermal spectrum can be produced
by sub-photospheric dissipation under conditions anticipated to prevail in GRB outflows.
Early work \citep[e.g.,][]{peer2006,giannios2012,beloborodov2013,vurm2013,vurm2016}  attempted
to compute the evolution of the photon density below the photosphere, assuming dissipation by some unspecified mechanism.   
They generally find significant broadening of the seed spectrum if dissipation commences in sufficiently opaque regions
and proceeds through the photosphere.    However, these models commonly invoke soft photon production by nonthermal
electrons, surmised to be accelerated below the photosphere by shocks or some other process, which is questionable \citep{levinson2008,levinson2012}.
A variation of this idea has been considered by \cite{keren2014}, who demonstrated
that breakout of a RMS train can naturally generate a Band-like
spectrum, and may also account for some features observed in a sub-sample of bursts. 
More recent work \citep{ito2015,lazzati2016,parsotan2018,ito2018b}
combines hydrodynamic (HD) and Monte-Carlo codes to compute the emitted spectrum.
In this technique, the output of the HD simulations is used as input for the Monte-Carlo radiative transfer calculations.  
These calculations illustrate that a Plank distribution, injected at a large optical depth, evolves into a 
Band-like spectrum owing to bulk Compton scattering on layers with sharp velocity shears, mainly 
associated with re-confinement shocks.   However, one must be cautious in applying those results, since the emitted
spectrum is sensitive to the width of the boundary shear layers \citep{ito2013}, which is unresolved in those simulations.
Furthermore, the radiative feedback on the shear layer is ignored.  Ultimately, the structure of those radiation mediated 
reconfinement  shocks needs to be resolved to check the validity of the results.

As discussed in depth in section \ref{sec:GRBs}, from a theoretical perspective, formation of sub-photospheric shocks is a likely outcome
in weakly magnetized GRB jets, or in magnetically driven jets that undergoes 
a conversion into kinetic-flux dominated jets well below the photosphere \citep[e.g.,][]{granot2011,levinson2013b,bromberg2016}.    
Hydrodynamic simulations of  jet propagation in collapsars \citep[e.g.,][]{lazzati2009,morsony2010,harrison2018,gottlieb2019}
indicate that a considerable fraction of the bulk energy dissipates in recollimation shocks just
below the photosphere, giving rise to a substantial photospheric component in the prompt emission.
The emerged spectrum should depend on the detailed structure of the shock, which is unknown at present,
but conceivably mediated by the radiation.   An additional dissipation mode is internal sub-and-mildly relativistic RMS, that are 
produced by intermittencies of the central engine.  These are expected to form at modest optical depths
below the photosphere if the Lorentz factor of the outflow is not exceptionally large \citep{eichler1994,morsony2010,bromberg2011}.
Detailed Monte-Carlo simulations \citep{beloborodov2017a,lundman2018a,ito2018a} indicate that under the conditions anticipated in GRBs, both collimation and internal RMS 
should produce a broad, non-thermal spectrum that peaks at a few to a few tens keV in the shock frame, depending on upstream conditions.
Further discussion on the structure observational diagnostics of collimation and internal shocks is deferred to section \ref{sec:GRBs}.

\subsection{Binary neutron star mergers}

The recent association of the gamma-ray burst GRB 170817A with the gravitational wave source GW170817  \citep{abbott2017GW,abbott2017grays,goldstein2017,savchenko2017}, 
and the subsequent detection of macronova/kilonova and afterglow emission, have lent strong support to the long-standing hypothesis that 
binary neutron stars and possibly neutron star black mergers are the progenitors  of short 
gamma-ray bursts \citep{eichler1989}.    However, the unusually low brightness of  GRB 170817A \citep{goldstein2017} 
indicated that in this object the emission source may be different 
than in typical sGRBs.   Of the various explanations offered shortly after the announcement of  GW170817 detection, two 
are consistent with the jet structure (see \citealt{eichler2018} for an alternative view), as inferred from the afterglow; (i) jet emission from regions that are outside of the jet core, where the energy is lower compared to the core but the angle to the observer is  smaller \citep[e.g.][]{ioka2019,kathirgamaraju2019},
and (ii) shock breakout emission \citep{kasliwal2017,gottlieb2018b,pozanenko2018,beloborodov2018}. 
In the latter scenario the shock is driven by an inflating cocoon that forms during the propagation of the relativistic jet in the merger ejecta. 
As in SNe and LGRBs, the shock is mediated by radiation, owing to the large optical depth of the ejecta, and its structure and dynamics during the breakout
phase dictate the properties of the observed gamma-ray flash.  
Shock breakout emission is always anticipated to accompany the emergence of a successful jet
from the merger ejecta, and is likely to dominate the observed signal when the viewing angle from the jet axis is large enough (although it may be overwhelmed by emission from a stratified jet in certain circumstances), but it might also be detected in certain cases even if the jet is choked (see \S \ref{sec:BNS} for a discussion and \citealt{gottlieb2018b}). 

The physics of shock breakout in BNS mergers in similar to that in SNe, with one important difference; while in SNe the unshocked medium (upstream)
is static with respect to the observer, in BNS mergers it is moving at a fraction of the speed of light, perhaps even relativistically if a fast tail
exists (e.g., \citealt{hotokezaka2013}), as discussed in some detail in \S \ref{sec:BNS}.   This can affect the shock dynamics and introduce additional boost of the observed radiation
that needs to be accounted for. 

While the shock breakout emission in BNS mergers can have a range of properties that depend on specific details (e.g., shock velocity, ejecta structure and velocity profile, etc.),  there are several qualitative features that are common to all the shock breakout episodes in BNS mergers (some of which are common also to other systems, e.g., SNe) that we henceforth summarize:

\begin{itemize}
\item \textbf{Low energy:} The energy released in the shock breakout is always a very small fraction of the total energy released in the explosion. The reason is that the breakout emission is generated by energy deposited by the shock into a very small fraction of the total mass. 

\item \textbf{Smooth light curve:} The breakout signal is not highly variable. It may contain a temporal structure, e.g. due to inhomogeneities in the ejecta, but high variability such as seen in the prompt emission of many LGRBs is not expected. 

\item \textbf{Hard to soft evolution} The spectra of the breakout emission and the subsequent cooling emission from the expanding shocked ejecta, show a hard to soft evolution. The spectrum of the breakout emission, which is contributed by the first layers that emerge following shock breakout (see \S \ref{sec:BNS} for details), is harder and does not resemble  a thermal spectrum. The emission from the spherical phase, which follows the breakout emission, is softer (and continues to soften with time) and its spectrum is more similar to a Wien spectrum. 

\item \textbf{Delay between the GW signal and the gamma-rays:} The energy of the breakout emission depends sensitively on the breakout radius. Assuming a mildly relativistic breakout velocity, a detectable signal at a distance of $\sim 100$ Mpc requires a breakout radius of $\gtrsim 10^{11}$ cm (see Eq. \ref{eq:Ebo} in \S \ref{sec:BNS}). This radius implies a delay of about a second or longer between the merger time, as defined by termination of the GW signal, and the gamma-rays emitted by the shock breakout \citep{nakar2019}.

\item \textbf{Relatively wide angle:} The beaming cone of the cocoon breakout emission is much larger than that of the relativistic jet, and it is quite  likely that at relatively large viewing angles from the jet axis it dominates over the jet off-axis emission.
 
\end{itemize}

\subsection{Implications for high-energy neutrino emission}
The recent detection of high-energy neutrinos of extragalactic origin by IceCube \citep{aartsen2013,aartsen2014} appears to confirm old predictions 
\citep{berezinsky1977a,berezinsky1977b,eichler1978a,margolis1978,eichler1978b,eichler1978c}. 
Yet, the nature of the neutrino sources remains elusive.   Potential candidates discussed in the literature include 
galaxy clusters, starburst galaxies \citep[e.g.,][]{waxman2015}, GRBs \citep[e.g.,][]{waxman1997,dermer2003,levinson2003,globus2015}, AGNs \citep[e.g.,][]{halzen1997}, micro-quasars \citep[e.g.,][]{levinson2001,distefano2002}, tidal disruption events and energetic supernovae \citep[e.g.,][]{murase2013,seno2016}.

It has been proposed \citep{eichler1999,meszaros2001} that a burst of TeV neutrinos can be produced in long GRBs during the propagation of the GRB jet 
in the stellar envelope.    In this scenario, protons accelerated at internal shocks that form in the jet, interact with radiation 
emitted from the termination shock at the jet's head.  
This mechanism applies to both, successful and choked jets.   While stacking analysis seems to
rule out bright GRBs as the main neutrino sources \citep{aartsen2017}, it still leaves room for the possibility that low luminosity GRBs and ultra-long GRBs,
which are too faint to be detected by current gamma ray satellites, are viable sources.  
However, in early models the fact that internal and collimation shocks that
are produced below the photosphere are mediated by radiation has been overlooked.    
As shown in \S \ref{sec:basic-principles}, in such shocks particle acceleration is highly suppressed by virtue of the large RMS width, that exceeds
any kinetic scale by several orders of magnitude \citep{levinson2008,katz2010}, which imposes severe restrictions on neutrino 
production in GRBs \citep{murase2013,globus2015}.
This problem can be avoided in 
ultra-long GRBs \citep{murase2013} and in low-luminosity GRBs \citep{nakar2015,seno2016}, if indeed produced by choked GRB
jets, as in the unified picture proposed by \cite{nakar2015}.  In the latter scenario, the progenitor star is ensheathed by an extended 
envelope that prevents jet breakout.  If the jet is choked well above the photosphere, then internal shocks produced inside the jet 
are expected to be collisionless.   The photon density at the shock formation site may still be high enough to contribute the photo-pion 
opacity required for production of a detectable neutrino flux. 

Substantial magnetization of the flow may alter the above picture, because in this case 
formation of a strong collisionless subshock within the RMS occurs \citep{beloborodov2017a}.   While PIC simulations \citep[e.g.,][]{sironi2009}
indicate a strong suppression of particle acceleration in relativistic collisionless shocks having upstream magnetization in excess of $\sim10^{-5}$,
effective particle acceleration may still be possible in sub-and-mildly relativistic shocks with relatively 
high magnetization.   If this is indeed  the case, and given that internal subshocks that form in the GRB jet are expected 
to be sub or mildly relativistic, the problem of neutrino production in GRBs should be reconsidered. 



\section{Physics of radiation mediated shocks}
\label{sec:Theory}
In this chapter we shall outline the theory of RMS.  After introducing the notation, 
we will describe the conditions under which RMS form, derive the basic equations that govern the structure 
and emission of RMS, discuss the different regimes of shock solutions and highlight the main physical processes 
that operate in each regime,
present analytical and numerical solutions that apply to different astrophysical situations, 
and summarize the numerical methods developed recently to study these systems. 

\subsection{Definitions and notation}
In the  case henceforth considered, the fluid inside and downstream of the shock transition layer is a mixture of 
ions, electrons, newly created e$^\pm$ pairs, and radiation.   The different components interact with each other 
through various processes that will be described below.   The local 4-velocity of the plasma with respect to shock rest frame, henceforth measured 
in units of $c$, is denoted by $u^\mu = \gamma (1,{\bf \beta})$.
Throughout this section, we shall use proper thermodynamic parameters
(e.g., density, pressure, etc.) in the shock equations, unless otherwise stated.    The proper baryon, electron, 
pair and radiation densities will be denoted by $n$, $n_e$, $n_\pm$ and $n_\gamma$, respectively, subject to the charge neutrality
condition, $n_e =n$ and $n_-= n_+ $.  Here we assume pure $H$ composition for simplicity.  If heavy 
elements are present then the charge neutrality condition, $n_e=n$, should be modified accordingly.
Other quantities (pressure, temperature, energy, etc.) will be denoted likewise.
In addition, far upstream quantities will be 
designated by a subscript $u$, and downstream quantities by subscript $d$, e.g., $n_u, n_d, n_{\gamma u}, n_{\gamma d}$, etc.
As shown below, the temperature of the downstream fluid may vary over scales much larger than the width of the shock 
transition layer even in an infinite planar shock, by virtue of photon generation through bremsstrahlung emission of the hot 
electrons and positrons.    In our notation $T_d$ will refer to the immediate downstream temperature.   
This is appropriate for most cases discussed in the 
following sections.  When post shock temperature variations will be considered, specific notation will be used where necessary.  

\subsection{Basic principles and assumptions}
\label{sec:basic-principles}
Before delving into the detailed theory of RMS, it is instructive to elucidate the conditions under which such shocks
are expected to form.   In general, radiation dominance prevails when a major fraction of the  bulk kinetic energy of the 
upstream flow is converted into trapped radiation behind the shock.  This occurs in sufficiently fast, optically thick shocks.
Since, as will be presently shown, radiation dominance occurs already in the Newtonian regime, it
is sufficient to assess these conditions for non-relativistic shocks.    

A crude estimate of the pressure $p_d$ behind a Newtonian shock 
can be obtained by balancing the momentum flux across the shock, neglecting the ram pressure of the downstream plasma, and
assuming that the radiation is in thermodynamic equilibrium with the gas (to be justified later):
\begin{equation}
n_u m_p c^2 \beta_u^2 \simeq   p_d = n_d kT_d + a T_d^4/3,
\label{eq:RMS_cond}
\end{equation}
 where $T_d$ is the downstream temperature and $a=7.56\times10^{-15}$ erg cm$^{-3}$ K$^{-4}$ is the radiation constant.  Radiation dominance implies  $n_d kT_d <  a T_d^4/3$.
Combining the latter condition with Eq. (\ref{eq:RMS_cond}) yields
\begin{equation}
\beta_u > \left(\frac{k n_d}{n_u m_pc^2}\right)^{2/3}\left(\frac{3n_u m_pc^2}{a}\right)^{1/6} \simeq  2\times10^{-4} 
\left(\frac{n_u}{10^{15}\, {\rm cm^{-3}}}\right)^{1/6},
\label{eq:beta_u_cond}
\end{equation}
where $n_d/n_u=7$ has been adopted, as appropriate for a high Mach number shock with an 
adiabatic index of $\gamma_{ad}=4/3$ (see Eq.(\ref{eq:jump_NR}) below).  Note that even in non-relativistic RMS the 
downstream pressure is dominated by relativistic constituents (photons), hence $\gamma_{ad}=4/3$.

The tacit assumption made in deriving the above result is that the radiation is trapped inside the shock.   This imposes a constraint on 
the optical depth of the system.  To be precise, since the upstream flow is decelerated by radiation that originates from the immediate 
post shock region and diffuses against the plasma stream, the shock width, $L_s$, can be estimated by equating the photon
diffusion time across the shock, $t_D\sim \sigma_T n_u L_s^2/c$, with the shock crossing time, $t_s= L_s/c\beta_u$.  This readily yields
a shock thickness of
\begin{equation}
\Delta \tau_s = \sigma_T n_u L_s \simeq \beta_u^{-1}.
\label{eq:width_rms}
\end{equation}
The optical depth of the entire system should exceed this value.  
To verify that this indeed gives the deceleration length of the flow, note that the force
exerted on the plasma by the diffusing radiation is $ \sim n \sigma_T \beta e_\gamma$,
where $e_\gamma = 3p_\gamma$ is the local energy density of the radiation and 
$\beta e_\gamma$ is the diffusive flux of photons inside the shock.  Recalling that $ n \beta $ is the
conserved particle flux, the mean force acting on a proton  by the radiation is thus $-m_pc^2 d\beta/dz \sim \sigma_T e_\gamma$,
where the $z$ coordinate is along the shock normal and increases towards the downstream.
Energy conservation yields $e_\gamma = n_u m_p c^2 \beta_u^2$ in the immediate post shock region, with which one obtains 
$-d\beta/dz\sim \beta_u/L_{dec} \sim \sigma_T n_u \beta_u^2$.  Thus, $L_{dec}\simeq L_{s}$, as required.
This scaling is confirmed by detailed calculations \citep{weaver1976,BP81b} that will be presented in section \ref{sec:NR_RMS}.

The salient point of the above arguments is that shocks having a velocity larger than the value defined in 
Eq. (\ref{eq:beta_u_cond}), and which are produced in a medium having a Thomson depth $\tau > \Delta \tau_s \simeq \beta_u^{-1}$, are mediated by radiation. This is particularly true for relativistic shocks
that form in a region where $\tau>1$.   As will be shown in the following sections, in sufficiently relativistic shocks, Klein-Nishina effects and pair creation  alter this simple scaling. 

\subsubsection{Assumptions}

The characteristic width of a RMS, $L_s\simgt 10^9 n_{u15}^{-1}$ cm,  here $n_{u15}=n_{u}/(10^{15} cm^{-3})$, is vastly larger than the scales 
over which electromagnetic interactions are mediated, most notably the skin depth, $l_p=c/\omega_p\sim 1 n_{u15}^{-1/2}$ cm,
and the Larmor radius of thermal protons, $r_L\sim 3 \gamma_u\beta_u(B/10^6 G)^{-1}$ cm.   
Due to this vast separation of scales, it is practically 
infeasible to incorporate microphysical processes associated with collective plasma interactions into the analysis of RMS, even by exploiting 
the most advanced numerical methods and computational platforms.  Hence, some assumptions are needed in order
to determine the energy distribution of  ions, electrons and positrons.  The assumptions commonly made are:

(i) Particles do not accelerate to  nonthermal energies at the shock front,
as in the case of collisionless  shocks (e.g., \cite{levinson2008}).  
The reason  is that over plasma scales the change in the flow velocity is so tiny that in practice 
any energy gain by the converging flow is expected to be completely negligible. 
This does not apply to second order fermi acceleration by plasma turbulence inside the shock. 
However, no potential turbulence source is naturally identified under the anticipated conditions.
Moreover, appreciable subshocks may form during the breakout phase, in which particle acceleration may ensue.  

(ii) Magnetic fields can be neglected. This assumption is most likely justified in case of shock breakout in supernovae
and NS mergers, where the magnetization is anticipated to be small, but not necessarily in long GRBs.   
As argued recently \citep{beloborodov2017a,lundman2018b} moderate magnetization can give rise 
to considerable alteration of the shock solution. 

(iii) The plasma constituents (electrons and  positrons in particular) are in local thermodynamic equilibrium. 
This assumption is justified by the large separation of scales pointed out above.  Specifically, 
since the coupling between the charged particles
is mediated by electromagnetic forces, it is generally anticipated that they 
will equilibrate on timescales much shorter than the radiative and flow timescales.  
In particular, newly created pairs are assumed to join the thermal pool instantaneously. 

(iv) In most analyses a planar geometry is invoked to simplify the calculations.  While this is justified in sufficiently opaque regions, where the shock width is much smaller than the overall scale of the system, it may be questionable during the breakout phase. 
In particular, deviation from planar geometry might be important in breakout from a wind (see section \ref{sec:wind_breakout}).

(v) Steady-state is also commonly assumed when computing the shock structure.  This assumption is 
justified as long as the evolution time of the fluid parameters far upstream, as measured in the shock frame, is longer 
than the shock crossing time.  Incorporation of dynamical effects might be feasible within the diffusion 
approximation (e.g., \cite{sapir2011}), but otherwise introduces a great computational challenge.

We shall adopt these assumptions in what follows, with the exception of Sec. \ref{sec:magnetization}.   
The above assumptions may not apply to extremely dense shocks, such as shock breakout in colliding neutron
stars.  At the anticipated densities, $\rho\simgt10^{10}$ g cm$^{-3}$, the skin depth becomes comparable to, or even larger than, 
the Thomson length. Hence, such shocks involve different physics.

\subsubsection{Schematic structure}
The detailed structure of a RMS depends on its velocity and the upstream conditions.  We distinguish between two types of shocks:
{\it photon rich} RMS in which photon advection by the upstream flow dominates over photon generation inside and just downstream of the
shock, and  {\it photon starved} RMS in which photon generation dominates.   The former type is expected in GRBs whereas the latter type
in most other systems.  The key parameter that determines the type of shock is the photon-to-baryon density ratio in the far upstream flow; in photon starved shocks it is well below the p-e mass ratio, $m_p/m_e$, whereas the opposite holds in photon rich shocks.   An elaborated discussion 
is given in section \ref{sec:Ramifications}.  As mentioned in
section \ref{sec:Intro_shock_breakoout},  there are, in general, three domains of RMS solutions: slow shocks, 
in which the thermalization time is much
shorter than the shock crossing time and the fluid (plasma and radiation) is in a full thermodynamic equilibrium inside the shock; 
fast Newtonian shocks, in which full thermodynamic equilibrium is reached only far downstream; relativistic shocks in which
pair production and Klein-Nishina effects play a dominant role.     The transition from slow to fast shocks occurs at a
shock velocity of $\beta_u\simeq 0.07$ (given by Eq. (\ref{eq:beta_crit}) below with $\Lambda_{ff}\simeq10$), while relativistic effects 
start becoming important at $\beta_u\simeq0.5$.

\begin{figure}
\centering
\vspace*{-25mm}
\includegraphics[width=11cm]{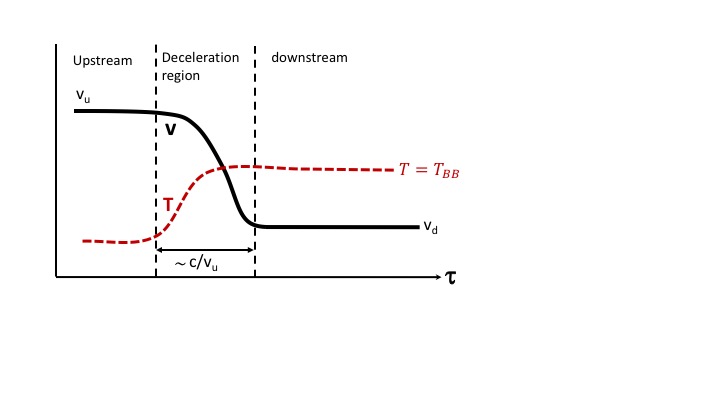} 
\includegraphics[width=11cm]{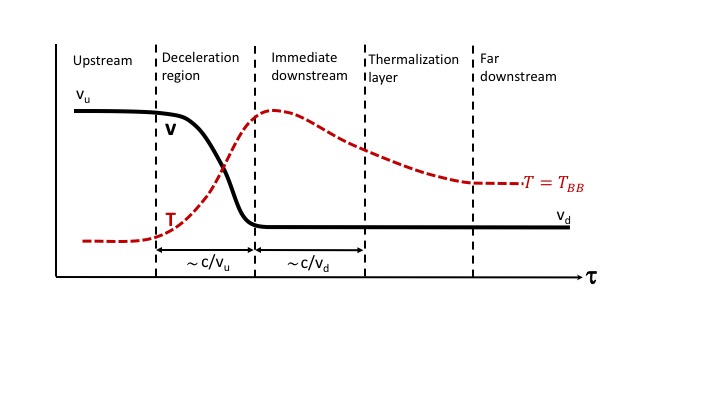} 
\includegraphics[width=11cm]{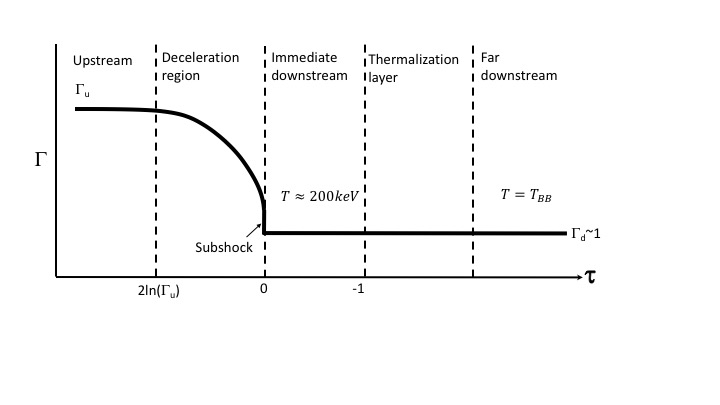} \vspace*{-5mm}
\caption{Schematic illustration of the structure of a slow (upper panel), fast (middle panel) 
and relativistic (lower panel) RMS.  
The five distinct regions (three in a slow shock) are indicated. The solid black  and dashed red lines delineate the 
velocity and temperature profiles, respectively.  In a slow shock the temperature approaches the black body limit inside the shock,
whereas in fast and relativistic shocks this limit is reached far downstream; the temperature in the immediate downstream can be considerably higher, depending on the shock velocity and the photon-to-baryon ratio far upstream.
In a relativistic, photon starved RMS the immediate 
downstream temperature is regulated by copious pair production and ranges from about $100$ keV for $\beta_u\simeq 0.5$
to $200$ keV for $\gamma_u\beta_u>>1$.  The newly created pairs also dominate the shock opacity.  The horizontal 
axis gives the optical depth traversed by a photon moving towards the upstream.}
\label{fig:RMS_struc}
 \end{figure}

In general the basic structure of an infinite RMS consists of five distinct regions, 
as  shown schematically in Fig. \ref{fig:RMS_struc}.
As seen from the shock frame these are:\\
(i) The upstream - unshocked  plasma moving at 4-velocity $\gamma_{u}\beta_u$.
The energy density in this region is dominated by the bulk kinetic energy of baryons, and is given by 
$\gamma_{u}(\gamma_u-1)n_{u}m_{p}c^{2}$ (which reduces to $n_u m_pv_u^2/2$ in non-relativistic shocks).
In photon starved shocks the upstream is cold and devoid of photons, and the radiation is produced inside the shock. In photon rich
shocks the upstream flow advects photons at a rate well in excess of the photon generation rate.\\
(ii)  The deceleration region - the velocity $\gamma \beta$  decreases from its upstream value $\gamma_{u}\beta_u$
to the downstream value, $\beta_d\simeq \beta_u/7$ in non-relativistic shocks, and $\beta_d \simeq 1/3$
in highly relativistic shocks.  In sufficiently sub-relativistic shocks the deceleration is due to the pressure force exerted on the
plasma by the diffusing radiation.  In relativistic and mildly relativistic shocks, where the anisotropy of the radiation inside the shock is
substantial and the diffusion limit is inapplicable, it is due to the interaction of counter streaming photons, 
that originate from the immediate downstream, with the plasma that incident into the shock.
This interaction involves Compton scattering by electrons (and positrons if exist), and in case of sufficiently relativistic shocks 
also substantial pair loading via $\gamma \gamma$ annihilation.   Under certain conditions a weak collisionless  subshock forms 
at the end of the deceleration zone, which has very little effect on the overall RMS structure. This subshock may become important in
the presence of considerable radiative losses.\\
(iii) The immediate downstream - the region just downstream of the shock from which the counter-streaming photons that 
mediate the shock originate. Its optical depth is $\tau \sim \beta_d^{-1}$.  In photon starved shocks
the immediate downstream temperature is set by the rate of photon production, mostly through free-free emission, over the available time (roughly the advection time). 
It approaches $\sim 200$ keV in highly relativistic shocks, and is largely insensitive to the shock Lorentz factor by virtue of 
opacity self-generation (Sec. \ref{sec:RRMS_starved}).  In photon rich shocks the temperature is set by the photon-to-baryon
ratio far upstream (see Eq. (\ref{eq:shock_rich_temp})).  In slow shocks this region formally exists, but doesn't play any decisive role.\\
(iv) The thermalization layer - the region behind the immediate downstream over which photons are continuously being generated
and the radiation gradually approaches thermodynamic equilibrium.  The  photons from this region cannot stream back to the deceleration region 
and do not affect the shock.  In slow shocks the radiation thermalizes well inside the shock and this
region is essentially absent.\\
(v)  The far downstream - the zone where the gas and radiation are in full thermodynamic equilibrium,
and the radiation energy density satisfies $e_\gamma=a_{BB}T_{BB}^{4}$, with $T_{BB}$ bieng the black body temperature, which for 
relativistic shocks (as well as fast Newtonian shocks) is vastly smaller than the immediate downstream temperature.  In many circumstances the downstream region may not be thick enough for a full thermodynamic equilibrium to be established and the temperature will exceed $T_{BB}$ everywhere.

\subsection{Detailed analysis}
\subsubsection{Governing equations}
In this section  we derive the general equations that govern the structure and spectrum of an unmagnetized RMS.   Inclusion 
of magnetic fields will be considered separately in section \ref{sec:magnetization}.   
As explained above, some assumptions about the energy distribution of charged particles are needed for
the calculation of the various radiation processes.
A customary prescription is to approximate the distribution function of  electrons, $f_e$, and pairs, $f_\pm$, by 
a Maxwell-J$\ddot{\rm u}$ttner (i.e., relativistic Maxwell-Boltzmann) distribution:
\begin{eqnarray}
 f_{e(\pm)}({\bf p_{e(\pm)}}, T) = \frac{1}{4\pi (m_e c)^3 \Theta  K_2(1/\Theta)}  {\rm exp}\left(- \frac{\epsilon_{e(\pm)} }{kT} \right) ,
 \label{eq:MB_rel}
\end{eqnarray}
where $T(t,{\bf x})=m_ec^2\Theta(t,{\bf x})/k$ is the local  temperature of the eletcron-positron plasma, 
$K_2$ is the  2nd order modified Bessel function of the second kind, ${\bf p}_{e(\pm)}$ is the particle momentum, 
and $\epsilon_{e(\pm)} = \sqrt{(p_{e(\pm)} c)^2+m_e^2c^4}$ the corresponding energy.
The local temperature is dictated by the interaction of the radiation with the charged leptons  at any given time and location. As for the protons,
their temperature is determined by energy exchange with electrons.   It is unclear at present what is the characteristic 
timescale for proton equilibration by this process.   If it is longer than the characteristic flow time then the protons may be 
considered cold, and their pressure might be neglected (e.g., \citealt{budnik2010}).   If an infinitely strong coupling is 
assumed, then the proton temperature should 
be taken equal to the pairs temperature.   We shall adopt the latter prescription.  
At any rate, in most circumstances the thermal energy of the protons has only little effect on
the shock structure and emission.   Another practical issue concerns the equation of state of the pairs.   In relativistic shocks the 
thermal pairs may become relativistic inside the shock transition layer, while sub-or-mildly relativistic in other regions.  This raises
the need for an equation of state that describes the relation between specific energy, $e_\pm$, and pressure, $p_\pm$, in 
the intermediate regime, between the non-relativistic limit, $e_\pm =3p_\pm/2$, and the relativistic limit, $e_\pm=3 p_\pm$.
The exact relationship can be derived using the Maxwell-Juttner distribution, but, unfortunately, no simple analytic 
expression can be obtained.   A useful fitting function that interpolates between the non-relativistic and relativistic 
regimes with an accuracy better than a fraction of a percent at all temperatures 
has been provided in \cite{budnik2010}:
\begin{equation}
g(T)=\frac{1}{2} \tanh\left(\frac{\ln\Theta + 0.3}{1.93}\right) + \frac{3}{2}.
\end{equation}
In terms of this function the specific energy of electrons and positrons can be expressed as $e_\pm = 3 g(T) p_\pm /2$ and 
likewise $e_e = 3 g(T) p_e /2$.

The fluid equations are most conveniently expressed in terms of the energy-momentum tensors of the neutral ion-electron plasma, $T_b^{\mu\nu}$,
the pair fluid, $T_\pm^{\mu\nu}$, and the radiation $T_\gamma^{\mu\nu}$, explicitly given by:
\begin{align}
T_b^{\mu\nu} & = \left[n(m_p+m_e)c^2+\frac{5}{2}p_i +\left( 1 +\frac{3}{2}g(T)\right)p_e \right]u^\mu u^\nu + g^{\mu\nu}(p_i+p_e),\label{eq:Tmunu-plasma}\\
T_\pm^{\mu\nu} & = \left[n_\pm m_e c^2 + \left( 1 +\frac{3}{2}g(T)\right)p_\pm \right]u^\mu u^\nu + g^{\mu\nu}p_\pm,\label{eq:Tmunu-pairs}\\
T_\gamma^{\mu\nu} & =  \int k^\mu k^\nu  f_\gamma(k,x) \frac{d^3 k}{k^0},\label{eq:Tmunu-rad}
\end{align}
where $k^\mu = \frac{\nu}{c}(1,\hat{\Omega})$ denotes the photon 4-momentum, $f_\gamma(k,x)$ the phase space distribution function, 
$p_i =nkT$, $p_e=nkT$ and $p_\pm=n_\pm kT$ the ion, electron and pair pressure, respectively,
and $g^{\mu\nu}=$ diag$(-1,1,1,1)$ the Minkowski metric.   
Conservation of baryon number, energy and momentum is governed by the equations
\begin{eqnarray}
& &\frac{\partial}{\partial x^\alpha}\left(n u^\alpha\right)=0,\label{eq:cont}\\
& &\frac{\partial}{\partial x^\alpha}\left(T^{\mu\alpha}_b+T^{\mu\alpha}_\pm+T^{\mu\alpha}_\gamma \right)=0.
\label{eq:eng-mom}
\end{eqnarray}
These conservation laws must be augmented by additional equations that account for the interactions between the different components. 

The evolution of the photon distribution function $f_\gamma$ is described by a transfer equation that includes scattering,  
changes associated with pair creation and annihilation, and photon emission and absorption by electrons and positrons.  
Neglecting stimulated scattering it reads (for details see, e.g., \citealt{melrose2008,vanputten2012})
\begin{eqnarray}
k^\mu\frac{\partial f_\gamma(k)}{\partial x^\mu}=\int d^3k_1\int d^3p \, k^0 w_c(p,p_1,k,k_1)
\{f_l({ p}_1)f_\gamma({ k}_1)\cr -f_l({p})f_\gamma({ k})\}+C_{pp}[f_\gamma,f_\pm,k]+S_k,
\label{eq:transfer}
\end{eqnarray}  
with the final state of the scattering electron, ${\bf p}_1$, fully determined by the kinematic conditions: $p_1^\mu= p^\mu + k^\mu - k_1^\mu$ .
Here $f_{l}=f_e + f_++f_-$ is the distribution function of scatterers (electrons and positrons), which under the strong coupling assumption is
given by Eq. (\ref{eq:MB_rel}),
the operator $C_{pp}[f_\gamma,f_\pm,k]$ accounts for the change in $f_\gamma$ due to e$^\pm$ pair creation and annihilation, and $S_k$ is a source term associated with all other processes that create or destroy photons (e.g., free-free emission and absorption).   It is given explicitly below for processes relevant to RMS.  
The quantity $w_c(p,p_1,k,k_1)$ denotes the probability per unit time for scattering of a photon in a state ${\bf k}$ to the state ${\bf k}_1$ by 
an electron in a state ${\bf p}$.  In the rest frame of the electron, here denoted by prime, it is given by
\begin{equation}
w_c=\frac{3\sigma_Tm_e}{32 \pi p_1^{\prime0}k^{\prime0}k_1^{\prime0}}
\left[\frac{k^{\prime0}}{k_1^{\prime0}}+\frac{k_1^{\prime0}}{k^{\prime0}}-\sin^2\theta^\prime\right]     
\delta(p^{\prime0}_1-p^{\prime0}+k^{\prime0}_1-k^{\prime0}),\label{R_c}
\end{equation}
where $\theta^\prime$ is the angle between ${\bf k}^\prime$ and ${\bf k}_1^\prime$, and
the kinematic conditions implies 
$p_1^{\prime0}=m_e c+(k^{\prime0}k_1^{\prime0}/m_ec)(1-\cos\theta^\prime)$.
The first term on the RHS of Eq. (\ref{eq:transfer}) accounts for scattering into state ${\bf k}$ of photons in state ${\bf k}_1$, whereas the 
second term accounts for scattering out of state ${\bf k}$ into state ${\bf k}_1$.
The operator $C_{pp}$  has two contributions.  The first one accounts for photon attenuation:
\begin{equation}
\left(k^\mu\frac{\partial f_\gamma({k})}{\partial x^\mu}\right)_{\gamma\gamma} = \int (1-\cos\theta_{\gamma\gamma})\sigma_{\gamma\gamma}({ k},{k}_1) f_\gamma({k}_1) f_\gamma({ k}) d^3k_1,
\end{equation}
where $\theta_{\gamma\gamma}$ denotes the angle between the propagation directions of the incident and target photons, and 
the cross section is given in terms of the pair velocity with respect to the center of momentum frame, $\beta_{\rm cm}= \sqrt{1 - 2 m_e^2 c^2/ [(1-{\rm cos}\theta_{\gamma \gamma})k^0 k_1^0]} $, as
\begin{eqnarray}
\sigma_{\gamma \gamma}({ k},{ k}_1) =
\frac{3}{16}
  \sigma_T (1- \beta_{\rm cm}^2) 
   \left[   (3 - \beta_{\rm cm}^4)
  {\rm ln}\left( \frac{1 + \beta_{\rm cm}}{1 - \beta_{\rm cm}} \right) 
    - 2 \beta_{\rm cm} (2- \beta_{\rm cm}^2) \right],
\end{eqnarray}
subject to the threshold condition $\sigma_{\gamma\gamma}=0$ at
$k_1^0 <  2 m_e c/ [k^0 (1 - {\rm cos}\theta_{\gamma \gamma}) ]$.
The integral over $k$ gives the net pair production rate per unit volume, viz., 
\begin{equation}
(\dot{n}_\pm)_{pp}= c \int d^3k\int d^3k_1(1-\cos\theta_{\gamma\gamma}) \sigma_{\gamma\gamma}({ k},{k}_1) f_\gamma({ k}_1) f_\gamma({ k}).
\end{equation}
The second contribution to $C_{pp}$ comes from pair annihilation.  The total annihilation rate per unit volume
 is evaluated as a function of the pair number density and temperature:
\begin{eqnarray}
\left( \dot{n}_{\pm} \right)_{\rm ann}
 = - (n_e + n_{-}) (n_{+}) c \sigma_{\pm} (\Theta).
\label{Qann}
\end{eqnarray}
 For a thermal pair distribution, Eq. (\ref{eq:MB_rel}), the pair annihilation cross section $\sigma_\pm$ can be approximated 
by an analytical function introduced in \cite{budnik2010} based on the formula derived in \cite{svensson1982}:
\begin{eqnarray}
\sigma_{\pm} = \frac{3 \sigma_T}{4} \left[1 + \frac{2 \Theta^2}{{\rm ln}(2 \eta_E \Theta + 1.3)} \right]^{-1},
\end{eqnarray}
with $\eta_E = e^{-\gamma_E} \approx 0.56146$,  where $\gamma_E \approx 0.5772$ is the Euler's constant.
It is noted that  the above quantity  is Lorentz invariant.  
The energy spectrum of the  photons produced via pair annihilation can be computed by employing the
fitting formula derived in \cite{svensson1996}, which approximates the exact emissivity over a wide range of temperatures (see 
also \citealt{ito2018a} for details).
The evolution of the pair density can be expressed in terms of the pair creation and annihilation rates as,
\begin {eqnarray} 
\frac{\partial}{\partial x_\alpha}\left(n_\pm u^\alpha\right)=-\int{C_{pp}[f_r,f_\pm,k]\frac{d^3k}{k^0}}=(\dot{n}_\pm)_{pp} + (\dot{n}_\pm)_{ann}
\label{eq:n_pm}.
\end{eqnarray}

The above set of equations augmented by appropriate boundary conditions upstream provides a complete description of the shock transition layer.  

\subsubsection{Jump conditions}
\label{sec:jump_cond}

\begin{figure}[ht]
\centering
\includegraphics[width=10cm]{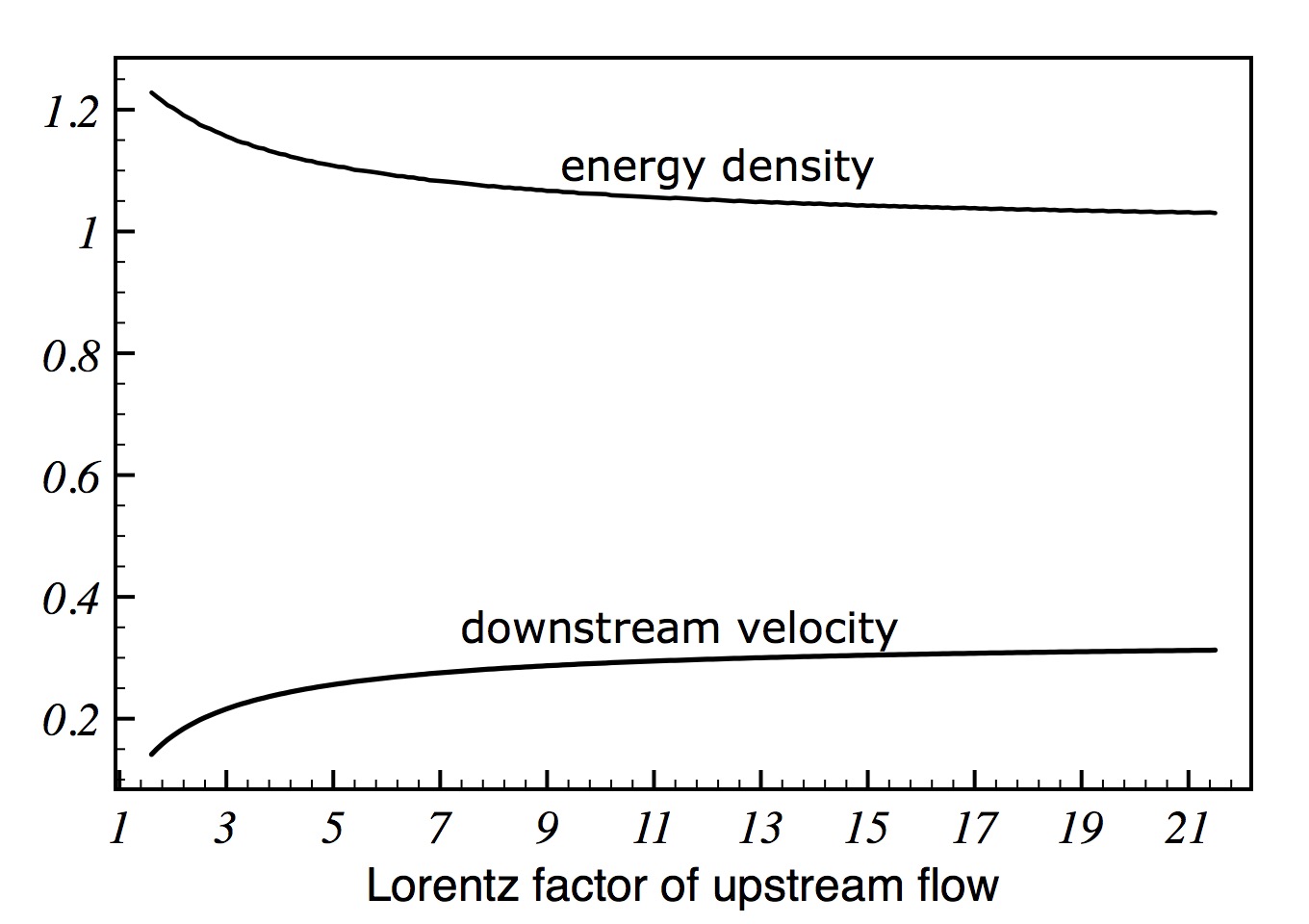}
\caption{\label{fig:f1} Downstream velocity, $\beta_d$, and normalized radiation energy density, $e_{d}/(2m_pc^2n_u\gamma_u^2\beta^2_u)$, 
as functions of the upstream Lorentz factor $\gamma_u$.}
\label{fig:jump}
 \end{figure}

Integration of Eqs. (\ref{eq:cont})-(\ref{eq:eng-mom})  across the shock transition layer yields the shock jump 
conditions, that determine the values  of the fluid parameters downstream, given a set of upstream conditions.  
Useful and insightful relations can be obtained by considering a steady, planar shock. 
Quite generally, the radiation in the upstream and downstream regions becomes isotropic in the 
fluid rest frame (and, hence, fully advected with the flow) over a few Thomson lengths. 
Thus, the energy momentum tensor of the radiation downstream of the shock can be approximated as
$T_{\gamma d}^{\mu\nu} = 4 p_{\gamma d} u_d^\mu u_d^\nu +g^{\mu\nu} p_{\gamma d}$, and likewise in the upstream region. 
For clarity of presentation we omit the contribution of the electrons to the rest mass density in Eq. (\ref{eq:Tmunu-plasma}), neglect 
the ion pressure, and invoke a relativistic equation of state for the pairs.  The jump conditions, expressed in the shock frame, then read:

\begin{align}
 n_u\gamma_u\beta_u  &= n_d\gamma_d \beta_d,\label{eq:jump-1}\\
 (n_u m_pc^2 +4p_{\gamma u})\gamma_u^2\beta_u^2 + p_{\gamma u}  & =  (n_d m_p c^2 + n_{\pm d} m_ec^2  +4p_d)\gamma_d^2\beta_d^2 + p_{d},\\
  (n_u m_pc^2 +4p_{\gamma u})\gamma_u^2\beta_u  & =  (n_d m_p c^2 + n_\pm m_ec^2 +4p_d)\gamma_d^2\beta_d,\label{eq:jump-3}
\end{align}
with $p_d=p_{\gamma d}+ p_{e d} +p_{\pm d}$ is the total pressure downstream.  In most practical situations, including photon rich RMS in long GRBs, the specific radiation enthalpy upstream, $4p_{\gamma u}$, is much smaller
than the rest mass energy density and can be neglected (but see \citealt{beloborodov2017a} and \citealt{ito2018a} for an account of excluded cases).    
Equations (\ref{eq:jump-1})-(\ref{eq:jump-3}) can then 
be solved analytically in the non-relativistic and ultra-relativistic limits.  In the latter case, with
$e_d  = 3p_d $, one has
\begin{align}
\begin{split}
e_d &= 2m_pc^2 n_u\gamma_u^2\beta_u^2,\\
\beta_d &=1/3.
\label{eq:jump_R}
\end{split}
\end{align}
This solution provides a reasonable approximation even at modest Lorentz factors, as indicated by figure \ref{fig:jump}, where
a plot of the exact solution of  Eqs. (\ref{eq:jump-1})-(\ref{eq:jump-3}) with $p_{\gamma u}=m_e=0$ is displayed. 

In the non-relativistic limit ($\beta_u<<1$) the pair content vanishes and the pressure downstream is dominated by the radiation.   A shock can form provided
the upstream flow is supersonic, that is, $c^2\beta_u^2 > c_s^2= 4p_{\gamma u}/3n_um_p$. In terms of the upstream Mach number, $M = c\beta_u/c_s$, 
the solution to Eqs. (\ref{eq:jump-1})-(\ref{eq:jump-3}) reads
\begin{align}
\begin{split}
e_{\gamma d} &= \frac{18}{7}m_pc^2 n_u \beta_u^2(1- 1/8M^2), \\
\beta_d &=\frac{\beta_u}{7}(1+6 /M^{2}). \label{eq:jump_NR}
\end{split}
\end{align}
In high Mach number shocks the compression ratio, $\beta_u/\beta_d$, approaches $7$.  This is merely a consequence of the 
relativistic equation of state, $e_d=3p_d$, of the downstream plasma. 

It should be emphasized that the jump conditions, while yielding the downstream radiation pressure, do not tell us anything about the temperature.
The latter depends on the photon generation rate inside the shock, that involves additional physics.   We will return to this point later on. 

\subsubsection{\label{sec:Phot_generation} Photon generation and thermalization length}
Under most circumstances, photon generation in unmagnetized RMS  is dominated by bremsstrahlung emission.  Double Compton emission
might be important in sufficiently photon rich shocks at high enough temperatures \citep{bromberg2011,levinson2012}. 
Substantial magnetization can lead to formation of a sub-shock and consequent emission of synchrotron photons \citep{lundman2018b}. 
The latter process is mostly relevant to sub-photospheric shocks in long GRBs, and will be discussed in \S \ref{sec:magnetization}.

Thermal bremsstrahlung in relativistic plasmas includes contributions from $e^\pm p$, $e^\pm e^\pm$, and $e^+ e^-$ encounters
(e.g., \citealt{svensson1983,dermer1984,skibo1995}).
With our notation, the photon generation rate (number per unit volume per unit time per frequency per solid angle) in a pure 
hydrogen plasma can be expressed as 
\begin{equation}
\dot{n}_{ff}(\hat{\Omega},\nu)=\frac{1}{\pi^2}\sqrt{\frac{2}{\pi}}\alpha_e\sigma_Tcn^2\frac{e^{-h\nu/kT}}{\nu\Theta} \lambda_{ff}
\label{eq:bremss_rel}
 \end{equation}
 with 
 \begin{equation}
\lambda_{ff}(x_+,\Theta)=(1+2x_+)\lambda_{ep}+[x_+^2+(1+x_+)^2]\lambda_{ee}+x_+(1+x_+)\lambda_{+-},
\label{eq:bremss_lambda}
 \end{equation}
where $\alpha_e$ is the fine structure constant, $x_+ = n_+/n$ denotes the positron-to-proton density ratio, and the coefficients  $\lambda_{ij}$ are functions of temperature and frequency $\nu$.  Fitting formulae for $\lambda_{ff}$ in different regimes are derived in \cite{skibo1995}.
In case of a mixture of fully ionized ions with abundance $X_i$ charge  $Z_i$ and mass number $A_i$ for ion species $i$ with a 
number density $n_i$, the total number density of 
ions is $n=\Sigma_i n_i$, the number density of electrons is  $n_e=\Sigma_iZ_in_i=n\Sigma_iZ_iX_i$, and 
the total mass density (neglecting the contribution of electrons) is $\rho=m_p n \Sigma_iA_i X_i$.   Denoting $\langle A\rangle = \Sigma_iX_iA_i$
and likewise for $\langle Z\rangle$, the factor $n^2$ in Eq. (\ref{eq:bremss_rel}) should be replaced by $(\rho/m_p)^2\langle Z^2\rangle \langle Z\rangle/ \langle A\rangle^2$, where the factor $\langle Z^2\rangle$ comes from the Larmor formula for the sum of ions.  To keep the notation
simple we shall assume a pure hydrogen composition in what follows, with the exception of \S \ref{sec:BNS}, 
where a detailed treatment of the effect of r-process elements on the shock temperature in BNS  merger ejecta is given.

The net photon generation rate, $\dot{n}_{ff}$, is obtained upon
integrating Eq. (\ref{eq:bremss_rel})  over frequency and solid angle.   
When computing the net photon generation rate a special care must be taken in dealing with the infrared divergence of the 
emission spectrum.    A fully self-consistent treatment requires inclusion of free-free absorption, stimulated emission and Compton scattering. 
Such practice is commonly used in numerical computations of radiation dominated flows (e.g., \citealt{budnik2010,vurm2013}).
However, in many circumstances approximate analytic expressions for $\dot{n}_{ff}$ are desired in order to simplify the analysis.
 Some scheme is then needed to decide which fraction of the emission spectrum should be included 
 in the integral of Eq. (\ref{eq:bremss_rel}).
A common approach \citep{katz2010,bromberg2011,levinson2012} is to introduce a lower cutoff frequency, $\nu_c$, in the spectrum of bremsstrahlung 
emission, $\dot{n}_{ph}(\Omega,\nu)$,
below which newly generated soft photons will be re-absorbed before being boosted to the thermal peak by inverse Compton scattering. 
Newly created photons at frequencies above the cutoff will quickly thermalize. 
This cutoff frequency is determined from the condition $\alpha_\nu^{ff} \lambda_T (m_ec^2/4kT) < 1$, where $\lambda_T =(\sigma_T n)^{-1}$ is
the Thomson length, $\alpha_\nu^{ff}$ is the free-free absorption coefficients and $(m_ec^2/4kT)$ is the average number of scatterings  
over which a photon of energy $h\nu << kT$ doubles its energy.  The latter criterion implicitly assumes that 
 the Compton y parameter is large enough, specifically, $y > \ln(kT/h\nu_c)$ - a condition that must be verified when applying this scheme.  
In non-relativistic RMS, as well as mildly relativistic photon rich shocks, where pairs are absent, this yields 
\begin{equation}
\dot{n}_{ff} \simeq \int^\infty_{\nu_c}d\nu\int d\Omega \, \dot{n}_{ff}(\Omega,\nu) \simeq \alpha_e\sigma_T c n^2 \Theta^{-1/2} \Lambda_{ff}
\label{eq:ff_rate}
\end{equation}
in terms of the coefficient $\Lambda_{ff}=E_1(h\nu_c/kT) g_{ff}$, where $E_1(x)$  is the exponential integral 
of $x$, which satisfies  $E_1(x) \simeq -\ln x $ at $x<<1$, and $g_{ff}$ is the usual Gaunt factor. 

In fast enough shocks, the density of photons produced inside and just behind the shock is well below the black body limit,
$n_\gamma < n_{BB}=aT^3/2.7k$.   As a consequence, the temperature in the immediate post shock region is well above the black-body value.
Full thermodynamic equilibrium will ultimately be reached further downstream, since photons continue to be generated as
the flow moves away from the shock.
The size of the thermalization layer (i.e., the distance from the shock at which a black body spectrum is established) is 
given by  $L_{ff}= c\beta_d t_{ff}$, where $t_{ff} = n_{BB}/\dot{n}_{ff}$ is the thermalization time.    The latter depends on
the black body temperature that corresponds to the specific upstream conditions. 
The energy density of the radiation behind the shock can be computed using the jump conditions.  It 
can be expressed as $e_{\gamma d}=\eta_s m_pc^2 n_u (\gamma_u\beta_u)^2$, where $\eta_s =2$ for relativistic shocks, $\gamma_u\beta_u>>1$, and $\eta_s = 18/7$ for non-relativistic shocks, $\gamma_u\beta_u <<1$ (e.g., \citealt{katz2010,budnik2010,levinson2012}).
The corresponding black-body temperature is then determined from the relationship $T_{BB}=(e_{\gamma d}/a)^{1/4}$. 
Upon substituting into Eq. (\ref{eq:ff_rate}) and using $n_{BB}=aT_{BB}^3/2.7k$, $L_{ff}$ is obtained.
 It is convenient to express the thermalization length in units of the Thomson length downstream.  For non-relativistic RMS 
 with $\gamma_u=1$, $\beta_u=7\beta_d$ and $\eta_s=18/7$, one finds,
\begin{equation}
\tau_{ff} \equiv \sigma_T n_d L_{ff}\approx 10^5 \left(\frac{n_u}{10^{15}\, {\rm cm^{-3}}}\right)^{-1/8}  \Lambda_{ff}^{-1}  \beta_u^{11/4}.
\label{eq:tau_ff}
\end{equation}
A similar expression is obtained for relativistic RMS, with $\beta_u^{11/4}$ replaced by $7.5 \gamma_u^{3/4}$ \citep{levinson2012}.    Equation (\ref{eq:tau_ff}) indicates that in fast RMS photon generation is very slow compared with the shock 
crossing time.   The velocity above which substantial deviations from a full thermodynamic equilibrium are expected can be estimated by equating
the shock width and the thermalization length, viz., $\tau_{ff} = \Delta \tau_s \simeq \beta^{-1}_u$. This yields
\begin{equation}
\beta_{u} > 0.04 \left(\frac{n_u}{10^{15}\, {\rm cm^{-3}}}\right)^{1/30}  \Lambda_{ff}^{4/15}.
\label{eq:beta_crit}
\end{equation}
In RMS that satisfy this criterion the  temperature behind the shock exceeds the black-body temperature.   As will be shown later on,
this has a profound effect on the spectrum emitted during shock breakout. 

Double Compton (DC) emission might be important in certain situations and in certain regions behind the shock \citep{bromberg2011,levinson2012}.
The rate per unit volume can be approximated as
\begin{equation}
\dot{n}_{DC}=\frac{16}{\pi}\alpha_f\sigma_Tcn_{l}n_{\gamma }\Theta^{2}\Lambda_{DC},\label{eq:rate-DC}
\end{equation}
with $\Lambda_{DC}$  given in \cite{bromberg2011}.  As the ratio of the DC and bremsstrahlung rates satisfies $\dot{n}_{DC}/\dot{n}_{ff}\sim (n_\gamma/n_l)\Theta^{5/2}$,
it is readily seen that DC emission is only important in regions where the photon density largely exceeds the total lepton density,
$n_\gamma \gtrsim n_l\,\Theta^{-5/2}$.    Such conditions prevail in the near downstream of sufficiently photon rich shocks \citep{levinson2012}.   
In non-relativistic shocks, the thermalization length by DC alone is given by \citep{levinson2012}
\begin{equation}
\tau_{DC} \simeq  5\times10^6 \left(\frac{n_u}{10^{15}\, {\rm cm^{-3}}}\right)^{-1/2}  \Lambda_{DC}^{-1} .
\label{eq:tau_DC}
\end{equation}
A similar expression is obtained for relativistic RMS.   Comparing (\ref{eq:tau_ff}) and (\ref{eq:tau_DC}) it is seen that thermalization by 
DC may become important only at extremely high densities. 
In photon starved shocks, where $n_\gamma\simeq n_l$ and $\Theta\simeq 0.2$, DC emission is subdominant and can be neglected.

\subsection{Regimes of shock solutions}
\label{sec:Ramifications}

The structure and emission of RMS are dictated by the shock velocity and the conditions in the flow far upstream. 
The radiation far upstream is commonly characterized by two important parameters (e.g., \citealt{ito2018a}): the photon-to-baryon density ratio,
\begin{equation}
\tilde{n}=n_{\gamma u}/n_u,
\end{equation}
and the ratio of radiation energy density, $e_{\gamma u}=3p_{\gamma u}$, and bulk kinetic energy density,
\begin{equation}
\xi_u=\frac{\gamma_u\,e_{\gamma u}}{(\gamma_u-1)m_pc^2n_u}.
\end{equation}
It is noteworthy that in the non-relativistic limit the latter quantity is related to the upstream Mach number through $\xi_u = 9/2M^2$.
Three different regimes can be identified in which different processes dominate the behaviour of the shock solutions.
In the first one, termed photon starved shocks, the photon density 
in the immediate downstream  is dominated by photon production inside the shock, mainly through 
bremsstrahlung emission.     In practice, this means setting $\tilde{n}=\xi_u=0$.  This regime 
is most relevant to shock breakouts in stellar explosions (supernovae, hypernovae, and low luminosity GRBs in choked jet scenarios), 
as well as in BNS mergers, where the upstream flow is expected to be cold. 
As will be shown in section \ref{sec:RRMS_starved}, 
 in relativistic, photon starved  shocks the downstream temperature is regulated via exponential pair creation
 at  $ kT_d\sim 200$ keV \citep{katz2010,budnik2010,granot2018},
photon scattering is in the deep KN regime, and the shock opacity is dominated by the pairs created in the shock
transition layer. 

\begin{figure}[ht]
\centering
\includegraphics[width=10cm]{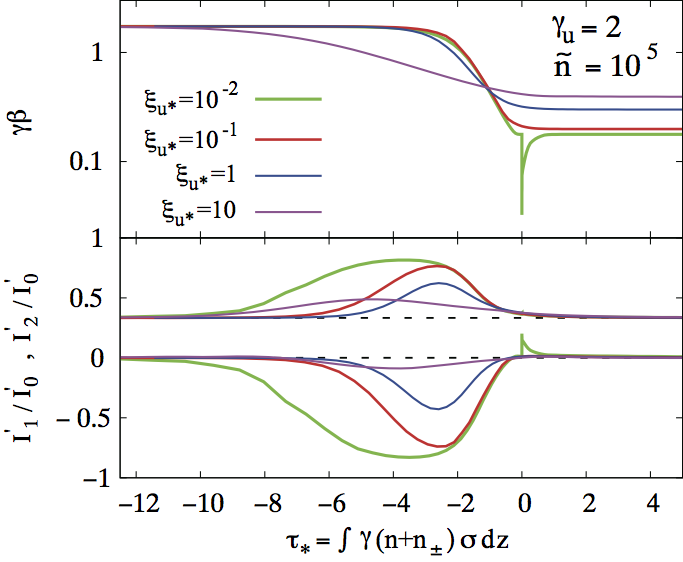}
\caption{\label{fig:fg_isot} Dependence of the 4-velocity profile ({\it top}) and 
 the normalized comoving  1st and 2nd moments of the radiation intensity, $I_{1}^{'}/I_{0}^{'}$ and   $I_{2}^{'}/I_{0}^{'}$ ({\it bottom}),
 on the far upstream photo-to-baryon inertia ratio $\xi_{u }$, for  $\gamma_u = 2$ and $\tilde{n}=10^5$.  The horizontal axis gives 
 the pair loaded Thomson optical depth.
 These results were obtain using Monte-Carlo methods developed to compute the structure and emission of RMS \citep{ito2018a}.
 For a given pair of lines in each model in the bottom panel, the upper one corresponds to the second moment $I_{2}^{'}/I_{0}^{'}$, and the lower one to
 the first moment $I_{1}^{'}/I_{0}^{'}$.  The two dashed lines in the bottom panel mark the values of the radiation moments of an isotropic 
 radiation field ($I_{1}^{'}/I_{0}^{'}=0$ and $I_{2}^{'}/I_{0}^{'}=1/3$).  The sharp feature at $\tau_\star=0$ in the case $\xi_u=10^{-2}$ corresponds
 to a weak subshock.  From \cite{ito2018a}.}
 \end{figure}

The other two regimes correspond to the case where the photon 
density in the immediate downstream is dominated by photon advection rather than photon production 
(photon rich shocks), as expected e.g., in sub-photospheric shocks in GRBs \citep{bromberg2011,levinson2012,beloborodov2017a}.  
In the Newtonian limit, $\beta_u<<1$, RMS can form (i.e., the upstream flow is supersonic) if $\xi_u < 4.5$.
In relativistic photon rich RMS one should distinguish between two cases; one in which the energy density 
of the upstream flow is dominated by the radiation, $\xi_u > 1$,   and
the second one in which the radiation energy density is sub-dominant, $\xi_u << 1$.   In the former case strong anisotropy 
cannot develop within the shock, since a small departure from isotropy is sufficient to give significant impact on the bulk 
flow of the plasma.  
The shock transition is therefore gradual, occurring over a relatively large optical depth, 
and the diffusion limit applies \citep{beloborodov2017a,ito2018a}.   In the second case, the upstream radiation does not have sufficient
energy to affect the bulk flow, and the extraction of the shock energy is accomplished by back-streaming  
photons that propagate from the immediate downstream to the upstream.   Consequently, the width of the shock transition
layer is determined by scattering of the back-streaming photons, and is of the order of one Thomson length roughly.
The radiation inside the shock is highly anisotropic in this case, as seen in the lower panel of Fig. \ref{fig:fg_isot},
that exhibits the first and second intensity moments for different values of $\xi_u$, where the nth moment is defined as
\begin{eqnarray}
I^\prime_{n} = 2 \pi \int \int I^\prime_{\nu} {\rm cos}^{ n}\theta^\prime ~d\nu^\prime d\Omega^\prime,
\end{eqnarray}  
and the prime indicates that it is measured in the local fluid rest frame.   The values $I^\prime_1=0$, $I^\prime_2/I^\prime_0 =1/3$ correspond 
to complete isotropy, whereas $I^\prime_2/I^\prime_0 =1$ and $I^\prime_1/I^\prime_0 = - 1$ to a perfect beaming. 

As explained above, the downstream region of a relativistic RMS is inherently non-uniform, because the thermalization length 
over which the plasma reaches full thermodynamic equilibrium is larger than the width
of the shock transition layer. 
However, Eq. (\ref{eq:tau_ff}) implies that for typical astrophysical conditions, the thermalization length exceeds the 
shock width by several orders of magnitude, so that for any practical
purpose photon generation in the downstream plasma can be ignored.  
This readily implies that to a good approximation the photon number is conserved across the 
shock transition layer:
\begin{equation}
n_{\gamma d}\gamma_d\beta_d=n_{\gamma u}\gamma_u\beta_u.
\label{eq:phot_cont}
\end{equation}
Combined with baryon number conservation, Eq (\ref{eq:jump-1}), one finds $\tilde{n}=n_{\gamma u}/n_{u}=n_{\gamma d}/n_{d}$.
The downstream temperature can now be computed using Eqs. (\ref{eq:jump_R}) and (\ref{eq:phot_cont}):
\begin{equation}
\Theta_d =\frac{e_{\gamma d}}{3n_{\gamma d}\,m_ec^2}=\frac{2m_p}{3\,m_e}\frac{(\gamma_u\beta_u)(\gamma_d\beta_d)}{\tilde{n}}
\simeq 430\,\frac{\gamma_u\beta_u}{\tilde{n}},
\label{eq:shock_rich_temp}
\end{equation}
where $\beta_d=1/3$ was adopted to obtain the numerical factor in the rightmost term. 
Thus, $\Theta_d \ll 1$ as long as $\tilde{n} \gg 430\gamma_u\beta_u$.    This result is a consequence of the 
fact that the upstream energy of a baryon, $m_pc^2\gamma_u$, is shared among $\tilde{n}$ photons behind the shock,
each having an energy of $\sim m_ec^2 \Theta_d$ on average. 

\begin{figure}[ht]
\centering
\includegraphics[width=8cm]{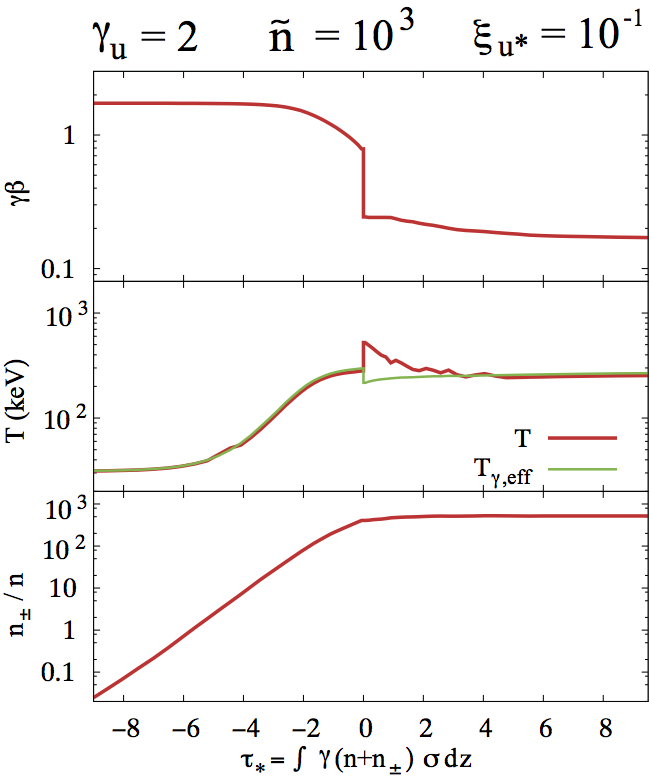}
\caption{\label{fig:fg_sub} Velocity (upper panel), temperature (middle panel) and pair density (bottom panel) as a function of 
the pair-loaded optical depth, in a photon rich shock with a photon-to-baryon ratio $\tilde{n} = 10^3$ and no photon generation, 
computed using Monte-Carlo simulations.   The formation of a strong collisionless subshock and a pair loaded precursor are evident. 
Downstream of the subshock a pair equilibrium state at a temperature of $kT\sim m_ec^2/3$ is reached.  From Ito et al. 2018.}
 \end{figure}

Further insight into the transition from photon rich to photon starved shocks can be obtained by considering 
the minimum value of $\tilde{n}$ required in order that counterstreaming photons will be able to
decelerate the upstream flow.   Let $\eta$ denote the fraction of downstream photons that 
propagate towards the upstream.   The energy each  photon can extract in a single collision 
is at most $\gamma_um_ec^2$.   Thus, the number of downstream photons required to decelerate the upstream flow satisfies
$\gamma_d\,n_{\gamma d}>\eta^{-1} (m_p/m_e)\gamma_u\,n_{u}$ (assuming $\xi_\gamma \ll 1$).  By employing Eq. (\ref{eq:phot_cont}) we find that the shock can be mediated by the advected 
photons provided 
\begin{equation}
\tilde{n}>  \frac{m_p}{m_e}\frac{\beta_d}{\eta \beta_u}\simeq \frac{m_p}{m_e},
\label{ncrit}
\end{equation}
adopting $\beta_d/\eta =1$. 
Equation (\ref{eq:shock_rich_temp}) implies that at the critical number density, $\tilde{n}\sim m_p/m_e$, the average photon energy, 
$3 k T_d \simeq 2\eta\,m_ec^2 \gamma_u\beta_u$, 
is in excess of the electron mass.  Under this condition a vigorous pair production is expected to ensue inside 
and just downstream of the shock, that will significantly enhance photon generation, thereby
reducing the downstream temperature.    This trend is seen in Fig. \ref{fig:fg_sub}, that exhibits results of 
Monte-Carlo simulations reported in \cite{ito2018a}, of a photon rich shock with $\tilde{n}=10^3$ and no photon generation.
As seen, a strong collisionless subshock forms, indicating that bulk Comptonization alone cannot mediate the shock. 
Downstream of the subshock pair equilibrium is established, with $n_\pm/n_\gamma\simeq 1$ ($n_\pm/n \simeq \tilde{n}/2$).   
In reality, these newly created pairs will generate sufficient photons (via bremsstrahlung emission) to decelerate  the flow
and eliminate the subshock, as indeed found in \cite{budnik2010}.   This case roughly marks the transition 
between photon rich and photon starved shocks. 

\subsection{Newtonian RMS}
\label{sec:NR_RMS} 
In non-relativistic shocks the radiation is nearly isotropic.   The moments of the radiation intensity, as measured 
in the shock frame,  can be expanded in powers of the local flow velocity $\beta$.   The transfer equation (\ref{eq:transfer})
can then be solved to a desired accuracy by invoking some closure condition of the moment equations.  
To compute the structure of the shock it is sufficient to solve the transfer equation to second order in $\beta$ in the diffusion limit. 
A detailed derivation of the diffusion equation is outlined in \cite{BP81a}.  The net photon flux (spectral flux 
integrated over frequency) obtained in this approximation can be expressed as
\begin{equation}
{\bf j}_\gamma = {\bf \beta} c n_\gamma - \frac{c}{3n_e \sigma_T} \nabla n_\gamma.
\end{equation}
The first term on the right hand side accounts for advection by the flow (advection flux) and the second term for diffusion (diffusion flux). 
In a steady state, this flux changes according to:
\begin{equation}
\nabla j_\gamma =    \dot{n}_\gamma,
\label{eq:transfer_eq}
\end{equation}
where $\dot{n}_\gamma$ is a photon source that accounts for all emission and absorptions processes.   
To the same order  the energy and momentum fluxes, Eq. (\ref{eq:Tmunu-rad}), reduce to 
\begin{align}
\begin{split}
T_\gamma^{0i} &=  4\beta_i  p_\gamma - \frac{1}{n_e \sigma_T}\frac{\partial p_\gamma}{\partial x_i} ,\\
T_\gamma^{ij} &= \delta_{ij}\,  p_\gamma,
\label{eq:T_gam_diff}
\end{split}
\end{align}
with the usual closure condition, $3p_\gamma=e_\gamma$. 
For simplicity,  we restrict the analysis to a planar geometry, wherein the flow moves in the positive $z$ direction, ${\pmb \beta} = \beta \hat{z}$.
Taking $n_\pm=0$, neglecting the electron rest mass density and the plasma pressure in $T_b^{\mu\nu}$ in Eq. (\ref{eq:Tmunu-plasma}),
and using Eq. (\ref{eq:T_gam_diff}), the shock equations (\ref{eq:cont}) and (\ref{eq:eng-mom}) reduce to:
\begin{eqnarray}
\begin{split}
n \beta= n_u \beta_u ,\\
\frac{d}{dz}(n m_pc^2 \beta^2 + p_\gamma)= 0 ,\\
\frac{d}{dz}\left(m_pc^2 n \beta^3/2 + 4 p_\gamma \beta -\frac{1}{n \sigma_T} \frac{d p_\gamma}{dz}\right)=0.
\end{split}
\end{eqnarray}
These equations admit the analytic solution 
\begin{align}
\begin{split}
\beta/\beta_u &= \frac{1}{7}(4+ 3 M^{-2}) -  \frac{3}{7}(1- M^{-2}) \tanh[(1-M^{-2})3\tilde{\tau}/2], \label{eq:RMS_NR_v}\\
p_{\gamma } &= m_pc^2 n_u\beta_u(\beta_u-\beta) +p_{\gamma u},
\end{split}
\end{align}
originally obtained in \cite{BP81b}, here expressed in terms of  the upstream Mach number 
$M = (3m_pc^2n_u \beta_u^2/4p_u)^{1/2}$, and the fiducial optical depth $\tilde{\tau} = \beta_u\tau = \beta_u \int \sigma_T n dz$.
The jump conditions (\ref{eq:jump_NR}) are recovered in the limit $\tilde{\tau} >> 1$.
Equation (\ref{eq:RMS_NR_v}) confirms that the width of the shock transition layer is indeed $\Delta \tau \simeq \beta_u^{-1}$, as qualitatively 
derived above (see  Eq. (\ref{eq:width_rms})) using heuristic arguments.

\cite{BP81b} have shown that when the advected photon density is sufficiently large, $n_{\gamma u}/n_u > m_p/m_e$,
photon generation can be ignored (photon rich shock).   They then 
computed the transmitted spectrum for fast shocks in which  bulk Comptonization dominates over
thermal Comptonization, and found that it tends to a power law with a spectral index $\alpha$ that depends on the Mach 
number $M$ as: $\alpha=(M^2-1/2)(M^2+6)/(M^2 -1)^2$.    This process is reminiscent of Fermi acceleration of cosmic rays in converging 
flows \citep{blandford1987}.   The maximum cutoff energy of the power law spectrum is determined by equating the average energy gain per collision with 
the average energy loss due to Compton recoil.   This yields $h\nu_{max}\simeq 0.2 m_ec^2\beta_u^2$.

\begin{figure}[h]
\centering
\includegraphics[width=10cm]{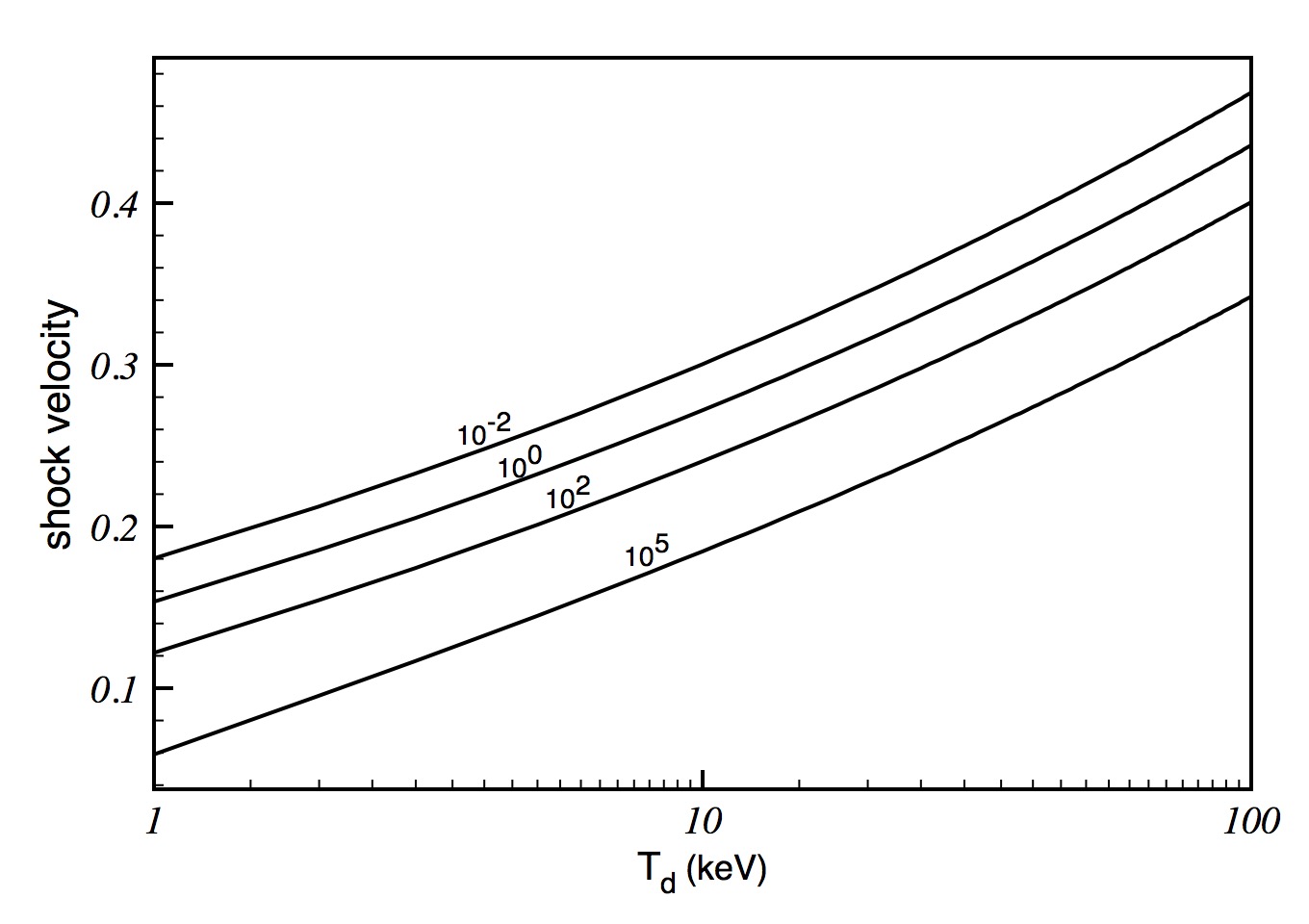}
\caption{\label{fig:fg_Tb} A plot of shock velocity $\beta_u$ versus  downstream temperature $T_d$, Eq. (\ref{eq:Theta_d}),
obtained using $\Lambda_{ff} = \ln(kT_d/h\nu_c) g_{ff}$, where $g_{ff}=\frac{\sqrt{3}}{\pi}\ln(kT_d/h\nu_c)$ is the Gaunt factor,
and the cutoff frequency $\nu_c$ is given by Eq. (11) in Katz et al. (2010).  The numbers that label the curves indicate values 
of $n_{u15}$,  the upstream density in units of $10^{15}$ cm$^{-3}$.}
 \end{figure}

While such conditions may prevail in some specific situations, 
e.g., subrelativistic shocks in GRBs, in many other sources (e.g., supernovae, BNS merger) the upstream flow is 
expected to be cold and devoid of photons.  The upstream conditions then simplify to $n_{\gamma u}=0$ and $M\rightarrow \infty$
in the above equations.   In \S \ref{sec:Phot_generation} it was argued that when the shock velocity is smaller than the value given by Eq. (\ref{eq:beta_crit})
photon production is fast enough to establish a full thermodynamic equilibrium inside the shock.  The radiation can then be 
treated as a black body \citep{pai1966,zeldovich1967,weaver1976}.
At higher velocities a Bose-Einstein distribution
will be established locally with a chemical potential that depends on the photon production rate \citep{weaver1976,thorne1981,katz2010}.
In order to compute the temperature profile inside and downstream of the shock in such cases one must first solve the photon diffusion equation 
(\ref{eq:transfer_eq}) to obtain the density profile  $n_\gamma (z)$.   The temperature is then given by $k T(z) = p_\gamma(z)/n_\gamma(z)$.
A complete treatment requires incorporation of absorption and stimulated emission in the source 
term $\dot{n}_\gamma$ in addition to bremsstrahlung and double
Compton emissions, which considerably complicates the analysis.  
A simple treatment is to modify the photon generation rate to include a suppression factor that accounts for 
absorption \citep{weaver1976,katz2010}, specifically, $\dot{n}_{\gamma} = \dot{n}_{ff}(1-3kn_\gamma/aT^3)$, where $\dot{n}_{ff}$ 
is given by Eq. (\ref{eq:ff_rate}), and then integrating Eq. (\ref{eq:transfer_eq}) using the analytic shock 
profile, Eq. (\ref{eq:RMS_NR_v}), and appropriate boundary conditions.
The reader is referred to \cite{katz2010} for details.   Double Compton emission has been neglected 
as it is sub-dominant in these shocks (see Eq. (\ref{eq:tau_DC})). 
A crude estimate of the temperature just downstream of the shock can be obtained upon 
assuming that the dominant contribution to photon production comes from a layer of width  $L_{ph}\sim (3\beta_d \sigma_T n_d)^{-1}$ near the immediate post shock region, within which $\beta\simeq\beta_d$ and $T\sim T_d$  \citep{katz2010}.
Integration of Eq.  (\ref{eq:transfer_eq}) then yields $n_{\gamma d}\sim \dot{n}_{ff} L_{ph}/c\beta_d$,
with $\dot{n}_{ff}\simeq \alpha_e\sigma_T c n_d^2(kT_d/m_ec^2)^{-1/2}\Lambda_{ff}$ from Eq. (\ref{eq:ff_rate}).
For a high Mach number shock the jump conditions (\ref{eq:jump_NR}) are reduced to 
$n_{\gamma d}kT_d=e_{\gamma d}/3= 6 m_pc^2 n_u\beta_u^2/7$ and $\beta_d = \beta_u/7$.  Combining with the above results this yields
\begin{equation}
\Theta_d \simeq \left(\frac{18 m_p}{m_e\alpha_e \Lambda_{ff}}\right)^2 \beta_d^8 \simeq 4\times10^6 \Lambda_{ff}^{-2}\beta_u^8.
\label{eq:Theta_d}
\end{equation}
Note that this relationship is formally implicit since $\Lambda_{ff}$  depends on $\Theta_d$ and $n_d$.  Although this dependence is
logarithmic it has a no-negligible effect on the scaling of $\Theta_d$.   
A plot of $kT_d = m_ec^2\Theta_d$ as a function of $\beta_d$ is displayed
in Fig. \ref{fig:fg_Tb}.  It is worth emphsizing that Eq. (\ref{eq:Theta_d}) holds only at low temperatures, $\Theta_d <<1$, where pair production is negligible.  

\subsection{Relativistic RMS}
\label{sec:RRMS}

There are vast differences between relativistic and non-relativistic RMS that 
render the methods commonly employed to solved the shock equations in the Newtonian regime inadequate for relativistic shocks. 
In recent years new techniques have been developed to compute the structure and emission of relativistic RMS under different conditions,
both analytically and numerically, as will be described below in more detail.    In this section we present a concise review of these methods.  
But before delving into the theory of RRMS, it is  instructive to highlight some notable 
differences between relativistic and Newtonian RMS.  The main differences can be summarized as follows:

\begin{enumerate} 
\item  While in non-relativistic shocks the photon distribution function inside the shock is nearly isotropic, in relativistic shocks 
it is anticipated to be highly anisotropic, owing to the fact that  the shock thickness $\Delta \tau\sim1$, and that 
 the average change in photon energy in a single scattering is large, $\Delta \nu/\nu >1$.
As a result, the diffusion approximation commonly used to compute the structure of
Newtonian RMS (see section \ref{sec:NR_RMS}), is rendered inapplicable when the shock velocity $\beta_u$ approaches unity.
Obtaining a closure of the hydrodynamic shock equations then becomes an involved issue (see \citealt{levinson2008} for details).   
 Additional complication arises from the anisotropy of the optical depth itself. 
The optical depth of a fluid slab  having a Lorentz factor $\gamma>1$ depends on the angle $\theta$ between the photon 
direction and the flow velocity as $d\tau\propto\gamma(1-\beta\cos\theta)dx$.  This means that while 
the  shock transition layer is opaque to backstreaming photons, it is  transparent to photons moving in the flow direction, an
effect that needs to be treated properly.

\item In relativistic RMS photon scattering is in the deep Klein-Nishina (KN) regime.  
This means that a full account of KN effects is required when solving the RMS equations \citep{budnik2010,nakar2012,granot2018}

\item Pair creation may become important if the photon energy exceeds the 
pair creation threshold.   In photon rich shocks this applies mainly to bulk Comptonized photons, 
as the temperature behind the shock is well below the electron mass.   Under such conditions pair creation 
becomes significant only when the upstream Lorentz factor is large enough, $\gamma_u>2$ \citep{ito2018a}.
In photon starved shocks  the downstream temperature is higher, and copious pair creation ensues already at mildly
relativistic speeds, $\gamma_u\beta_u\sim 1$ \citep{katz2010,budnik2010,nakar2012,granot2018}.
As will be shown below, in these shocks pair production plays a key role in regulating
the downstream temperature and governing the shock opacity.
\end{enumerate}

\subsubsection{\label{sec:RRMS_starved} Photon starved RMS}

Equation (\ref{eq:Theta_d}) indicates that the downstream temperature approaches the electron mass as the shock velocity $\beta_u\simgt 0.3$,
implying that accelerated pair creation should be anticipated.    A pair equilibrium will be established in the immediate post shock 
region, whereby the  pair-to-photon ratio is given by $n_\pm/n_\gamma \simeq  K_2(\Theta_d^{-1})/\Theta_d^2 $, where $K_2$ is 
the modified Bessel function of the second kind that asymptotes to $K_2(\Theta_d^{-1})\simeq \sqrt{\pi/2} \Theta_d^{1/2} \exp(-\Theta_d^{-1})$
at $\Theta_d \ll1$.   Now, the newly created pairs will emit additional photons that will tend to reduce the temperature, 
giving rise to an exponentially feedback on the number of pairs.  Thus, this exponential pair creation acts as a thermostat that 
regulates the downstream temperature.  Formally, the downstream temperature can be evaluated by solving the set of equations
\begin{align}
\begin{split}
n_{\gamma d} &=p_{\gamma d}/kT_d,\\
n_{\gamma d} &\simeq \dot{n}_{ff} (3\beta_d^2 c\sigma_Tn_d)^{-1},\\
n_{\pm d}/n_{\gamma d}&\simeq  K_2(\Theta_d^{-1})/\Theta_d^2 ,
\end{split}
\end{align}
for the three unknowns, $\Theta_d$, $n_{\gamma d}$ and $n_{\pm d}$, in conjunction with the shock jump conditions that determines 
$p_{\gamma d}, n_d, \beta_d$,  and the integral  of Eq. (\ref{eq:bremss_rel}) over $\nu$ and $\hat{\Omega}$ that gives the net
photon generation rate, $\dot{n}_{ff}$, which includes the contribution of all leptons (i.e., electrons and newly created pairs).
The solution yields a downstream temperature of $\Theta_d\sim 1/3$ for relativistic shocks, 
which is largely insensitive to the shock Lorentz factor  \citep{katz2010,budnik2010}.
This regulation mechanism ceases to operate once the temperature exceeds the value above which the dependence of the pair production rate on temperature becomes linear
rather than exponential.   The analysis of \cite{budnik2010} indicates that exponential pair creation is expected at least 
up to $\gamma_u=30$.    For such shocks one can safely assume that photons  just behind the shock have 
a mean energy of $\sim m_ec^2$.   This readily implies that scattering inside the shock, 
where $\gamma >>1$, is in the deep Klein-Nishina regime.  Furthermore, 
within the shock transition layer, where the flow is sufficiently relativistic with respect to the shock frame ($\gamma >2$), the
radiation is anticipated to be strongly beamed.   It is then possible to compute analytically the structure of a planar shock by applying 
the two stream approximation \citep{nakar2012,granot2018}, that greatly simplifies the transfer equation (\ref{eq:transfer}).  In this approach, one stream (the primary beam) consists of the plasma constituents (protons, electrons and pairs) and the back-scattered photons, all of which move towards the downstream, while the counterstream contains photons, each having an energy of $\sim m_ec^2$ in the shock frame, that were generated in the immediate downstream and move towards the upstream.  As we shall now show, these two beams interact in a manner that fixes the shock profile. 

Consider a planar shock moving in the positive $z$ direction, such that in the shock frame ${\pmb \beta}= - \beta\hat{z}$.
Following \cite{granot2018} we denote the proper density of photons streaming with the flow (i.e., moving from 
the upstream to the downstream) by $n_{\gamma\rightarrow d}$ and the proper density of counterstreaming photons by $n_{\gamma\rightarrow u}$.
The counterstreaming photons  are inverse Compton scattered by the inflowing electrons and positrons, and are converted into e$^\pm$ pairs  via interactions
with scattered photons that are moving with the bulk flow.    The equations are solved in the shock frame, and to shorten the notation we
designate, in the present account, by  a subscript "prime" the local densities in that frame; that is, 
$n^\prime_e=\gamma n_e,  n^\prime_\pm=\gamma n_\pm$, etc., 
The change in the number density  of counterstreaming
photons is then governed by the equation
\begin{equation}
\frac{dn^\prime_{\gamma\rightarrow u}}{dz}=-(1+\beta)[\sigma_{KN} (n^\prime_\pm+n^\prime_e)+\sigma_{\gamma\gamma}n^\prime_{\gamma\rightarrow d}]n^\prime_{\gamma\rightarrow u},
\label{eq:starved_rate1}
\end{equation}
where $\sigma_{KN}$, $\sigma_{\gamma\gamma}$ are the full cross-sections for Compton scattering and pair-production, respectively.
The change in the density of downstream moving photons and newly created pairs are likewise given by
\begin{equation}
\frac{dn^\prime_{\gamma\rightarrow d}}{dz}=- (1+\beta)[\sigma_{KN} (n^\prime_\pm+n^\prime_e)-\sigma_{\gamma\gamma}n^\prime_{\gamma\rightarrow d}]n^\prime_{\gamma\rightarrow u},
\label{eq:starved_rate3}
\end{equation}
and
\begin{equation}
\frac{dn^\prime_\pm}{dz}= -2(1+\beta)\sigma_{\gamma\gamma}n^\prime_{\gamma\rightarrow d}n^\prime_{\gamma\rightarrow u}.
\label{eq:starved_rate2}
\end{equation}
For clarity, pair annihilation has been neglected as it is insignificant inside the shock, and in any case does not change the final 
result.  It can be easily included in the analysis if one desires a more formal derivation.   The sum of the last two equations gives the change in the 
net density of quanta, $n_l=n_\pm+n_{\gamma\rightarrow d}$, produced inside the shock via conversion of counterstreaming photons:
\begin{equation}
\frac{dn^\prime_{l}}{dz}=-(1+\beta)[\sigma_{KN} (n^\prime_\pm+n^\prime_e)+\sigma_{\gamma\gamma}n^\prime_{\gamma\rightarrow d}]n^\prime_{\gamma\rightarrow u}.
\label{eq:starved_ratel}
\end{equation}
In an infinite shock counterstreaming photons cannot escape to infinity, hence their density vanishes far upstream.   
The appropriate boundary condition in this case is:
$n^\prime_{\gamma\rightarrow u}(z\rightarrow\infty) = n^\prime_{l}(z\rightarrow\infty) =0$.   
Subtracting Eq. (\ref{eq:starved_ratel}) from Eq.  (\ref{eq:starved_rate1}), and using the latter boundary condition, yields
a conservation law for the total number of quanta: $n^\prime_{\gamma\rightarrow u} - n^\prime_{l} = 0$.  The physical interpretation 
of this conservation law is straightforward;
every counterstreaming  photon is ultimately converted into either a photon, an electron or a positron that move towards the downstream.
In terms of the net optical depth for conversion of counterstreaming photons,
\begin{equation}
d\tau=(1+\beta)[\sigma_{KN}(n^\prime_\pm+n^\prime_e)+\sigma_{\gamma\gamma} n^\prime_{\gamma\rightarrow d} ]dz,
\label{app:tpt-opacity}
\end{equation}
and the fraction $\x=n^\prime_l/n^\prime = n_l/n$,  the above rate equations reduce to the single equation
\begin{eqnarray}
\frac{d \x}{d\tau}= -\x.
\label{eq:dx_l/dz}
\end{eqnarray}

To proceed, we must employ the energy equation \footnote{The assumption invoked in the analytic model, that the 
photon distribution can be approximated as two perfect beams, renders the momentum equation redundant.}.   
Neglecting the proton pressure and the electron rest mass 
energy in Eq. (\ref{eq:Tmunu-plasma}) and (\ref{eq:Tmunu-pairs}), and denoting $\mu=m_e/m_p$, yields 
the net energy flux:
\begin{equation}
T^{0z} = T^{0z}_b+T^{0z}_l+T^{0z}_{\gamma\rightarrow u} = m_pc^2 n \gamma^2\beta[1+(\x+1)\mu \Theta].
\end{equation}
Equation (\ref{eq:eng-mom}) ascertain that this flux is conserved.  By applying the boundary conditions $x_l(z\rightarrow \infty)=\Theta(z\rightarrow\infty)=0$, $\gamma(z\rightarrow\infty)=\gamma_u$, and using the baryon conservation law, Eq. (\ref{eq:cont}), one arrives at:
\begin{equation}
\gamma [1+(\x+1)\mu \Theta]=\gamma_u.
\label{eq:Gamma-cons}
\end{equation}

To close the set of shock equations the temperature $\Theta$ must be determined.    Granot et al. (2018) proposed the form
\begin{equation}\label{eq:T}
	\Theta=\frac{\eta \gamma n_{\gamma\rightarrow u}}{\n + n_e+n} = \eta\frac{\gamma \x}{\x+2},
\end{equation}
where $\eta$ is an order unity factor that depends on the exact energy and angular distributions of pairs and photons inside the shock, 
as well as other details ignored in the analytic model.    The reasoning behind that choice is that every collision of a counterstreaming photon with 
the primary beam adds, on the average, additional quanta of proper energy $\eta \gamma m_e c^2$ to the primary beam\footnote{This is
because the interaction of counterstreaming photons with the primary beam is in the deep Klei-Nishina regime.},
which is shared among its entire constituents.  The numerical results of \cite{budnik2010} indicate that $\eta$ lies in the range 
$0.45$ to $0.55$ for the range of shock Lorentz factors they analyzed. 

\begin{figure}[ht]
\centering
\includegraphics[width=6cm]{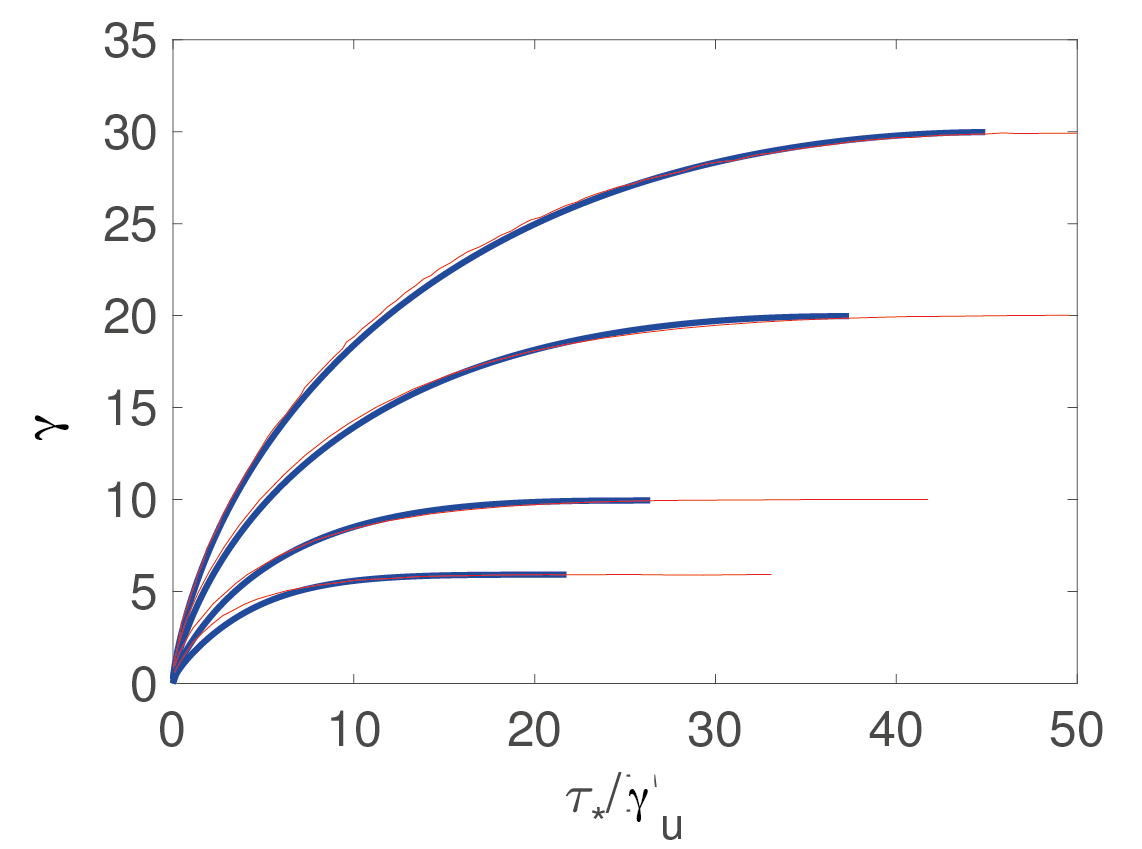} \includegraphics[width=6cm]{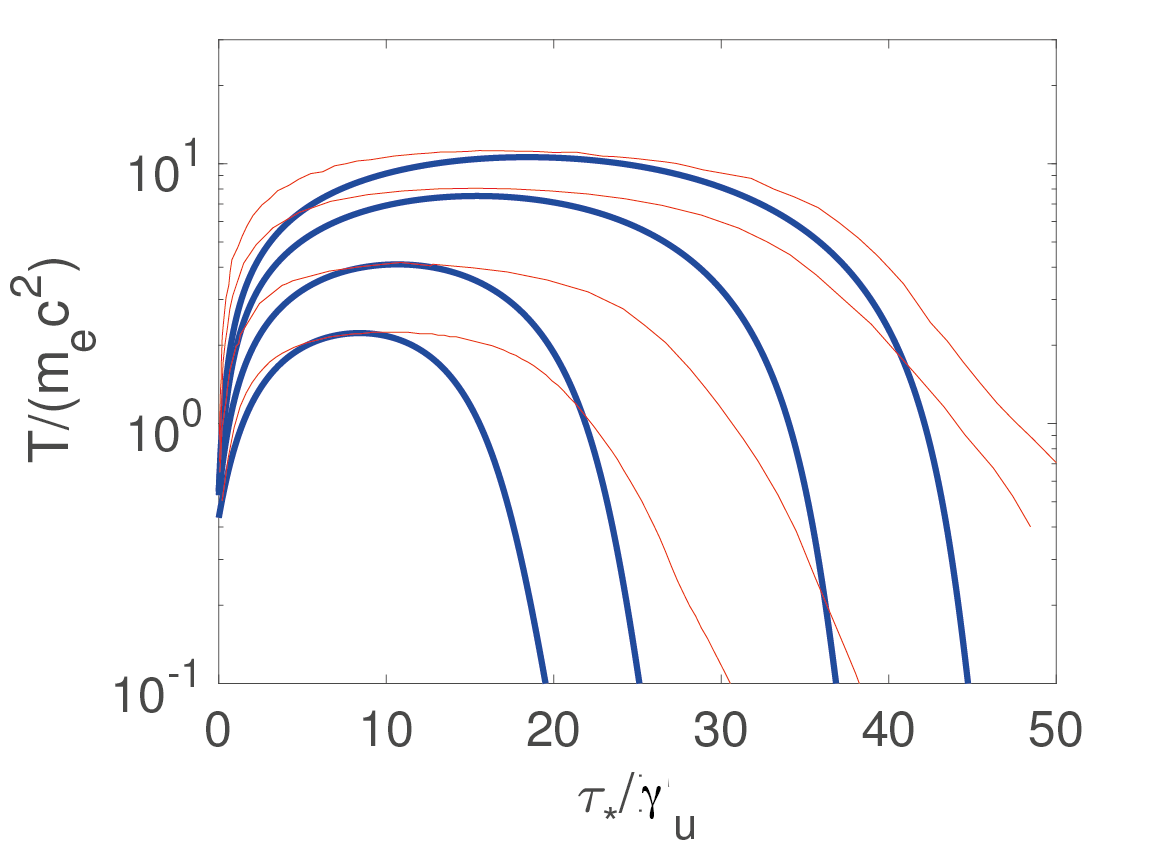}
\caption{\label{fig:Gamma_starved} Lorentz factor (left) and temperature (right) profiles plotted as
functions of the pair loaded Thomson depth $\tau_\star$, for 
upstream Lorentz factors $\gamma_u=6, 10, 20, 30$.  The blue solid lines depict the analytic solution and 
the thin red lines the numerical solution obtained by Budnik et al. (2010). From Granot et al. (2018).}
 \end{figure}

Equations (\ref{eq:Gamma-cons}) and (\ref{eq:T}) readily yield the relation
\begin{equation}\label{eq:G_x}
	\gamma(\x) = \frac{\sqrt{1+16\mu\gamma_{u}\eta\frac{\x(\x+1)}{\x+2}}-1}{8\mu\eta\frac{\x(\x+1)}{\x+2}},
\end{equation}
that formally holds in the region where $\gamma$ is large enough.   If extended to the immediate post shock location $\tau_0$ 
where $\gamma(\tau_0)=1$, it implies $x_0\equiv x_l(\tau_0)\simeq \gamma_u/4\mu\eta$.   This probably underestimates the
actual value of $x_0$, as it ignores the contribution of counterstreaming photons there, which may not be negligible. 
However, it is not expected to alter this result by more than a factor of 2.   Choosing for convenience $\tau_0=0$, 
one obtains from Eq. (\ref{eq:dx_l/dz})
\begin{equation}\label{eq:x_tau}
	\x=\frac{\gamma_u}{4\mu\eta} e^{-\tau}.
\end{equation}
It is now seen that the flow undergoes exponential deceleration in the shock transition layer, specifically, $\gamma(\tau)\simeq e^{\tau/2}$ at
$1\le\gamma \le \gamma_u$.   Hence, the width of the shock measured in terms of $\tau$ is $\Delta\tau\simeq 2\ln\gamma_u$.
A comparison of the analytic solution derived above, Eqs.  (\ref{eq:T}) - (\ref{eq:x_tau}), 
and the numerical solution obtained by \cite{budnik2010} is
shown in Fig. \ref{fig:Gamma_starved}, where for the sake of comparison 
the Lorentz factor and temperature profiles are plotted in terms of the pair loaded
Thomson optical depth, $d\tau_\star=(\sigma_T/\sigma_{KN})d\tau$, using Eq. (8) from \cite{granot2018} for $\sigma_{KN}$.

It is reminded that $\tau$ is the sum of scattering and pair creation opacities that include KN effects.  The physical scale of the shock 
can be inferred when expressing the solution in terms of the pair unloaded optical depth, approximately given by
\begin{equation}
d\tilde{\tau} =\frac{\sigma_T}{\sigma_{KN}(x_l+1)}d\tau,
\label{eq:unload_tau}
\end{equation}
upon invoking $\sigma_{KN} =\sigma_{\gamma\gamma}$, which at high energies is accurate to better than a factor of two.
Upon combining the chain rule $d\gamma/d\tau = (d\gamma/dx_l)(dx_l/d\tau)$ with Eqs.  (\ref{eq:dx_l/dz}), (\ref{eq:G_x}) and  (\ref{eq:unload_tau}),
$\gamma(\tilde{\tau})$ can be obtained \citep{nakar2012,granot2018}.    It can be readily shown then that the shock width scales as
\begin{equation}
\Delta \tilde{\tau}_s \simeq 10\eta \mu \gamma_u^3\simeq \frac{\gamma_u^3}{400}.
\label{eq:RRMS_width_inft}
\end{equation}
A factor $\gamma_u^2$ comes from KN effects\footnote{Inside the shock the temperature is approximately $m_ec^2 \gamma$, hence the collision
energy, as measured in the shock frame, is $\sim m_ec^2 \gamma^2$.},  and another power from the scaling of the pair loading profile, $x_l(\gamma)$,
in the deceleration zone. 

\begin{figure}[ht]
\centering
\includegraphics[width=10cm]{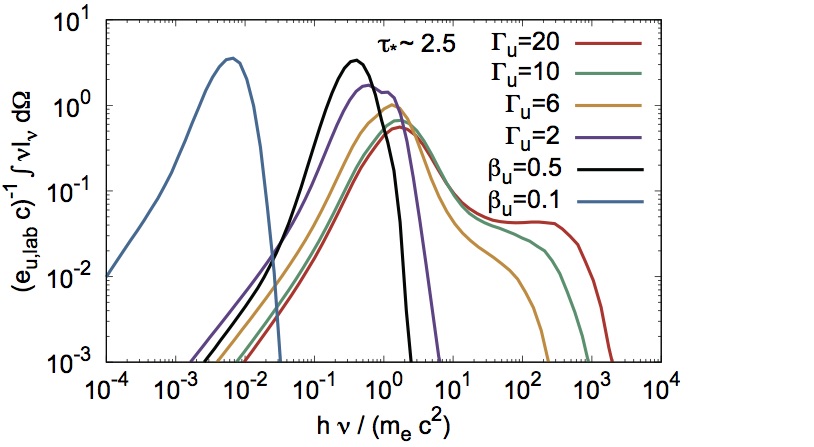}
\caption{\label{fig:spect_starved} Angle averaged spectra in the immediate downstream of a relativistic, photon starved RMS, obtained from Monte-Carlo simulations, for different Lorentz factors of the upstream flow $\gamma_u$.  The spectra are exhibited in the shock frame. From \cite{ito2020}.}
 \end{figure}

Computing the spectrum is a far more involved problem, that requires numerical techniques.  The first attempt to compute the 
spectrum of a relativistic RMS was undertaken by \cite{budnik2010}, who solved the kinetic equations across the shock transition 
layer using iteration methods.  Their analysis elucidated the main spectral features, but was limited to sufficiently high Lorentz factors ($\gamma \ge6$).  
\cite{beloborodov2017a} and \cite{lundman2018a}  employed direct time-dependent hydro simulations coupled to Monte-Carlo 
radiative transfer and pair creation; they followed the process of shock formation and obtained the steady-state shock structure.
Their results are limited to mildly relativistic, highly rich RMS.
A different method that can treat also sub-and-mildly relativistic shocks has been developed subsequently for photon rich RMS 
by \cite{ito2018a} and generalized recently to photon starved shocks \citep{ito2020}.  In this method
the shock structure and spectrum are computed in a self-consistent manner using a Monte-Carlo code that incorporates 
an energy-momentum solver routine that allows adjustments of the shock profile in each iterative step.  
An example is shown in Fig. \ref{fig:spect_starved}. It confirms the expectation that the immediate downstream temperature 
should be regulated by pair creation at sufficiently high Lorentz factors.  It also indicates formation of a power law tail above the peak,
in agreement with the results of \cite{budnik2010}.   Note, however, that the spectrum inside the shock is highehly anisotropic, and
that the power law tail is only present in the spectrum of photons moving with the plasma flow (i.e., from the upstream to the downstream; \citealt{budnik2010}).
At Lorentz factors below $5$  or so the peak energy becomes smaller and the 
power law tail is small or absent.   The mean photon energy is about $200$ keV (or $kT_d\approx 75$ keV) 
at shock velocity $\beta_u=0.5$ and about $4$ keV at $\beta_u=0.1$.

\subsubsection{Photon rich RMS}
In photon rich shocks with a large photon-to-baryon ratio, $\tilde{n} \gg m_p/m_e$, the downstream temperature is well below the 
electron mass (see  Eq. (\ref{eq:shock_rich_temp})).  Consequently, pair production by thermal photons is negligibly small.
Pairs may nonetheless be produced via annihilation of nonthermal (bulk Comptonized) photons \citep{beloborodov2017a,lundman2018a,ito2018a,lundman2018b}, however, the density 
of pairs thereby produced is typically much smaller than 
the density of the radiation, and while under certain conditions they can dominate the opacity inside the shock and affect 
its structure, they contribute very little to the total  energy budget of the shock.    This fact can be used to simplify analytical approach
to rich RMS calculations.

\begin{figure}[ht]
\centering
\includegraphics[width=10cm]{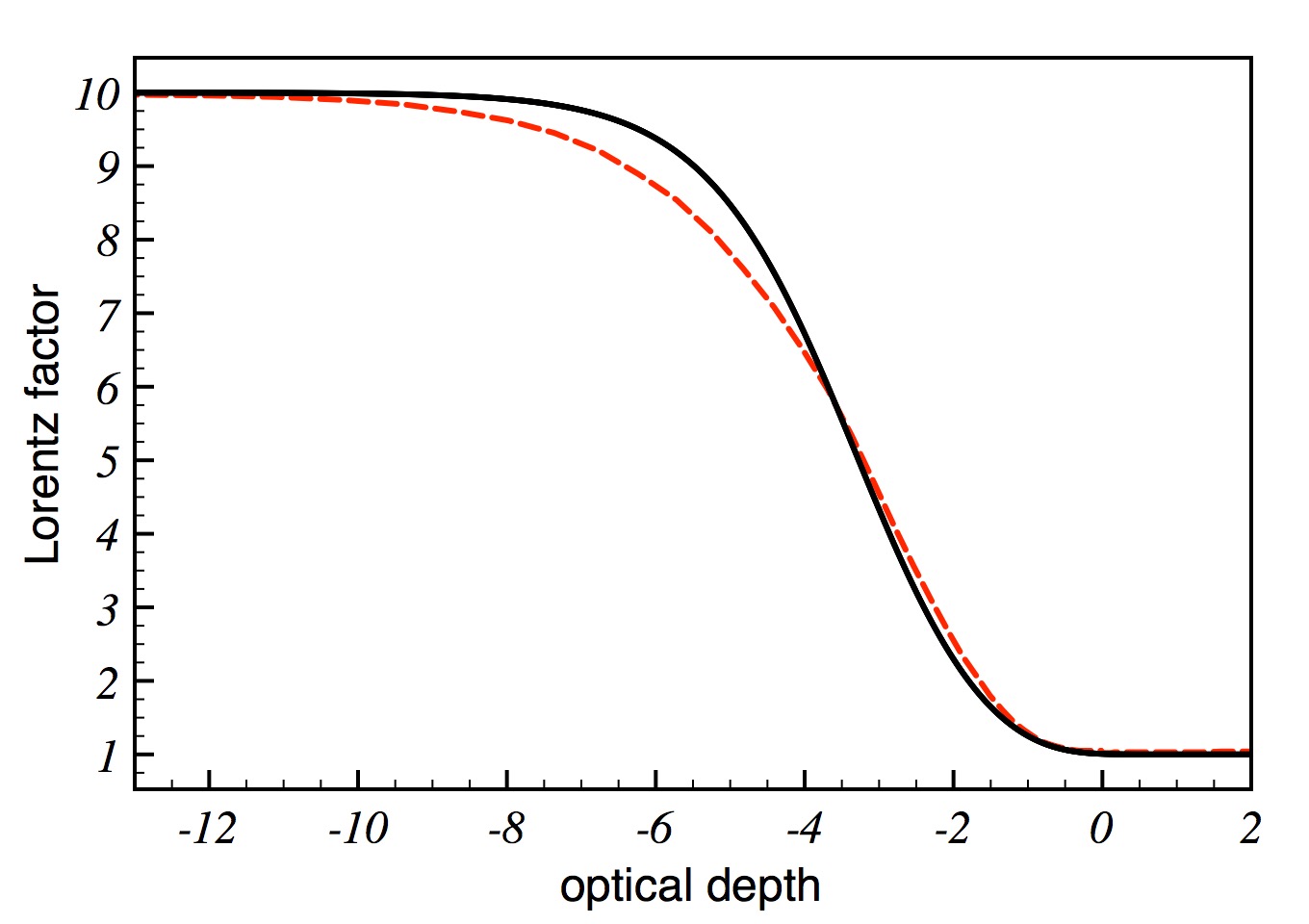}
\caption{\label{fig:f2} The solid black line delineates the solution given by Equation (\ref{analyt-prof-gam}) with $\gamma_u=10$ and
$\kappa_\gamma\epsilon=0.2$.  The dashed red line is the shock profile obtained from a Monte-Carlo simulation \citep{ito2018a}.}
 \end{figure}

An approximate analytic solution of the shock structure can be 
obtained in a manner similar to that used in the previous section \citep{ito2018a}.    Neglecting the nonthermal tail,
the change in the photon density is given by
\begin{equation}
\frac{dn^\prime_{\gamma\rightarrow u}}{dz}=-(1+\beta)\sigma_{KN}(n^\prime_e+n^\prime_\pm) n^\prime_{\gamma\rightarrow u}.
\end{equation}
For sufficiently photon-rich shocks the scattering of bulk photons is in the Thomson regime, thus  $\sigma_{KN}\simeq \sigma_T$. 
In terms of the optical depth $d\tau = (1+\beta)\sigma_{T}\,(n^\prime_e+n^\prime_\pm) dz$, and the energy density of 
the counterstreaming photons, $u^\prime_{\gamma\rightarrow u}=<\epsilon_\gamma>n^\prime_{\gamma\rightarrow u}$, one then has
\begin{equation}
\frac{du^\prime_{\gamma\rightarrow u}}{d\tau}=-u^\prime_{\gamma\rightarrow u}.\label{dugamdtau}
\end{equation}
The total inverse Compton power emitted by a single electron (positron) inside the shock is approximately 
\begin{equation}
P_{Comp}= \kappa_\gamma c\sigma_T (\gamma\beta)^2\,u^\prime_{\gamma\rightarrow u},\label{eq:Pcomp}
\end{equation}
where the pre-factor $\kappa_\gamma$ ranges from $4/3$ for isotropic radiation to $4$ for completely beamed radiation.
For illustrative purposes, it can be assumed constant throughout the shock.  
Neglecting the internal energy of the plasma inside the shock, the energy flux of the fluid can be expressed as 
\begin{equation}
T_b^{0z}=m_pc^3n \gamma^2 \beta=Jc^2 \gamma,
\end{equation}
in terms of the conserved mass flux $J=m_pc\,n \gamma \beta$.
Energy conservation implies $d\,T_b^{0x}/dz=\gamma(n_e+n_\pm) P_{comp} $,
or, using Equation (\ref{eq:Pcomp}),
\begin{equation}
Jc^2 \frac{d\,\gamma}{d\tau}=\kappa_\gamma (\gamma^2-1) u_{\gamma\rightarrow u}.\label{dgdtau}
\end{equation}
The boundary condition reads: $\gamma(\tau\rightarrow\infty)=\gamma_u$.
Denoting $\alpha=\kappa_\gamma u_{\gamma\rightarrow u}(\tau=0)/Jc^2$, and 
\begin{equation}
\zeta(\tau)=\ln\left(\frac{\gamma_u+1}{\gamma_u-1}\right)+2\alpha\,e^{-\tau},
\label{analyt-shock-prof}
\end{equation}
the solution of Eqs (\ref{dugamdtau}) and (\ref{dgdtau}) can be expressed as
\begin{equation}
\gamma(\tau)=\frac{e^\zeta+1}{e^\zeta-1}.
\label{analyt-prof-gam}
\end{equation}
From the jump conditions we have $\alpha=2\kappa_\gamma\epsilon \gamma_u$, where 
$\epsilon=u_{\gamma\rightarrow u}/u_{\gamma d}$ is roughly the 
fraction of downstream photons that propagate backwards.  
The black solid line in  Fig. \ref{fig:f2} shows the analytic shock profile obtained for $\kappa_\gamma \epsilon=0.2$.
The red line is the result of a MC simulation performed by \cite{ito2018a}.
A comparison of numerical solutions obtained by \cite{beloborodov2017a}, \cite{lundman2018a} and \cite{ito2018a} using different methods 
also shows good agreement.

\begin{figure}[ht]
\centering
\includegraphics[width=6cm]{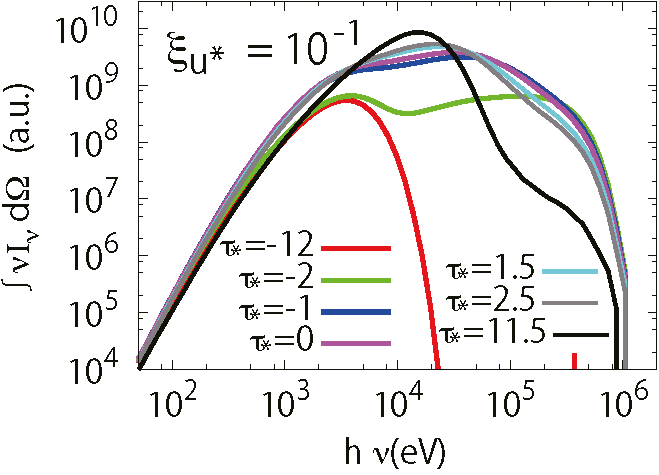} \includegraphics[width=6cm]{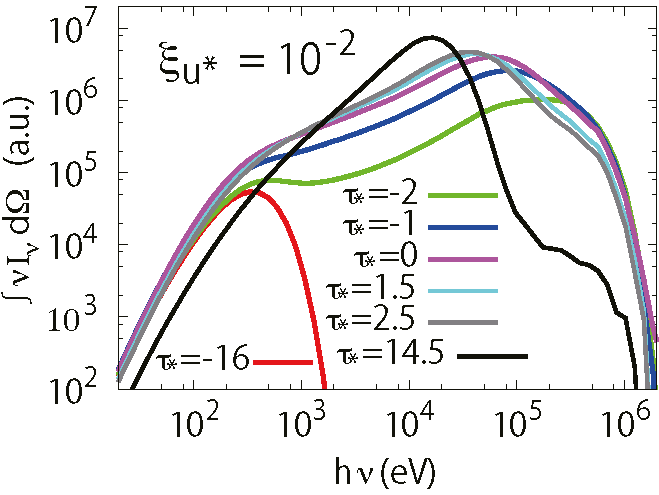}
\caption{\label{fig:spect_rich} Local, angle integrated SEDs of a photon rich shock, for  $\gamma_u=2$, $\tilde{n}=10^5$, and two
values of $\xi_u$, as indicated. The red and black lines show, respectively, the spectra near the upstream and downstream 
boundaries of the simulation domain in each case.    The green, blue, magenta, cyan and gray lines
display spectra which were computed at locations $\tau=-2$, -1, 0, 1.5, 2.5 around the shock transition layer.   The downstream
region is located at  $\tau \ge 0$.
The scale on the vertical axis is given in arbitrary units.  The absolute value can be specified once the number density of either 
baryons or photons at far upstream is given.  From \cite{ito2018a}.}
 \end{figure}

In shocks with $\xi_u < 1$ a significant fraction of the upstream bulk energy  is converted, via 
bulk Comptonization of counter streaming photons, to high-energy radiation.
The resultant photon spectra exhibit a broad, non-thermal component 
that extends up to an energy of $\sim (\gamma_u -1)m_e c^2$, as seen in the example depicted in Fig. \ref{fig:spect_rich}.
Sufficiently far downstream the radiation thermalizes and the local spectrum approaches the Wien  spectrum. Nonetheless, the 
spectrum integrated over the entire shocked slab, even if it has a relatively large optical depth,  still appears nonthermal. 
 Quite generally, the spectrum inside the shock becomes harder for lower values of  $\xi_{u}$,
 leading to enhanced pair creation by virtue of the increased number of 
 photons with energies in excess of the pair production threshold.   
 As shown in \cite{ito2018a}, the large pair enrichment in models with high $\gamma_u$ and low $\xi_{u}$ 
 gives rise to a signature of the 511 keV annihilation line in the spectrum.

\subsection{Finite shocks with photon escape}
\label{sec:RMS-with-escape}

The analyses outlined in the preceding sections assume complete trapping of the radiation inside the shock and, hence, 
are suitable for shocks propagating well below the breakout radius, where the optical depth is much larger than the shock thickness. 
During the breakout phase an increasing fraction of the radiation produced inside the shock escape the system,
and this should affect the shock solution.   If the breakout
occurs gradually, as in the case of sub-photospheric shocks in long GRBs, or shock breakout from a stellar 
wind in supernovae, then the shock has time to adjust to local changes and the 
evolution of its structure may be approximated as quasi-steady.    Steady shock solutions that incorporate photon losses may then be sought.

Such a treatment has been applied recently to shock breakout from a stellar wind, both in the Newtonian regime, where the
diffusion approximation applies \citep{ioka2018}, and the relativistic regime \citep{granot2018} where the tow-stream approximation can be invoked  (section \ref{sec:RRMS}).  In case of a non-relativistic shock the analysis outlined in section \ref{sec:NR_RMS} can be generalized
to include radiative losses from an upstream boundary, whereby the analytic solution (\ref{eq:RMS_NR_v}) is modified.  The temperature profile is then computed by solving a transfer equation in the diffusion limit, using the analytic velocity and pressure profiles \citep{ioka2018}. 
The resultant solution indicates a significant decline in the observed temperature with increasing radiative losses.   However, the quasi-steady approximation 
of the shock evolution may be questionable in this regime and needs to be verified by dynamical calculations which are extremely challenging. 

In relativistic RMS a complete breakout occurs at a radius at which the total optical depth 
ahead of the shock is $\tau_w\sim (m_e/m_p)\gamma_{u}$, rather than $\tau_w\sim 1$ as might be naively expected, 
provided that the shock remains relativistic at this location \citep{granot2018}.  The reason why the shock is maintained radiation mediated
even at radii where $\tau_w<<1$, is that it self-generates its own opacity via accelerated pair creation.   The fact that the breakout
radius is altered by opacity self-generation has important observational consequences that will be discussed in \S \ref{sec:wind_breakout_R}.

\begin{figure}[ht]
\centering
\includegraphics[width=6cm,height=3.7cm]{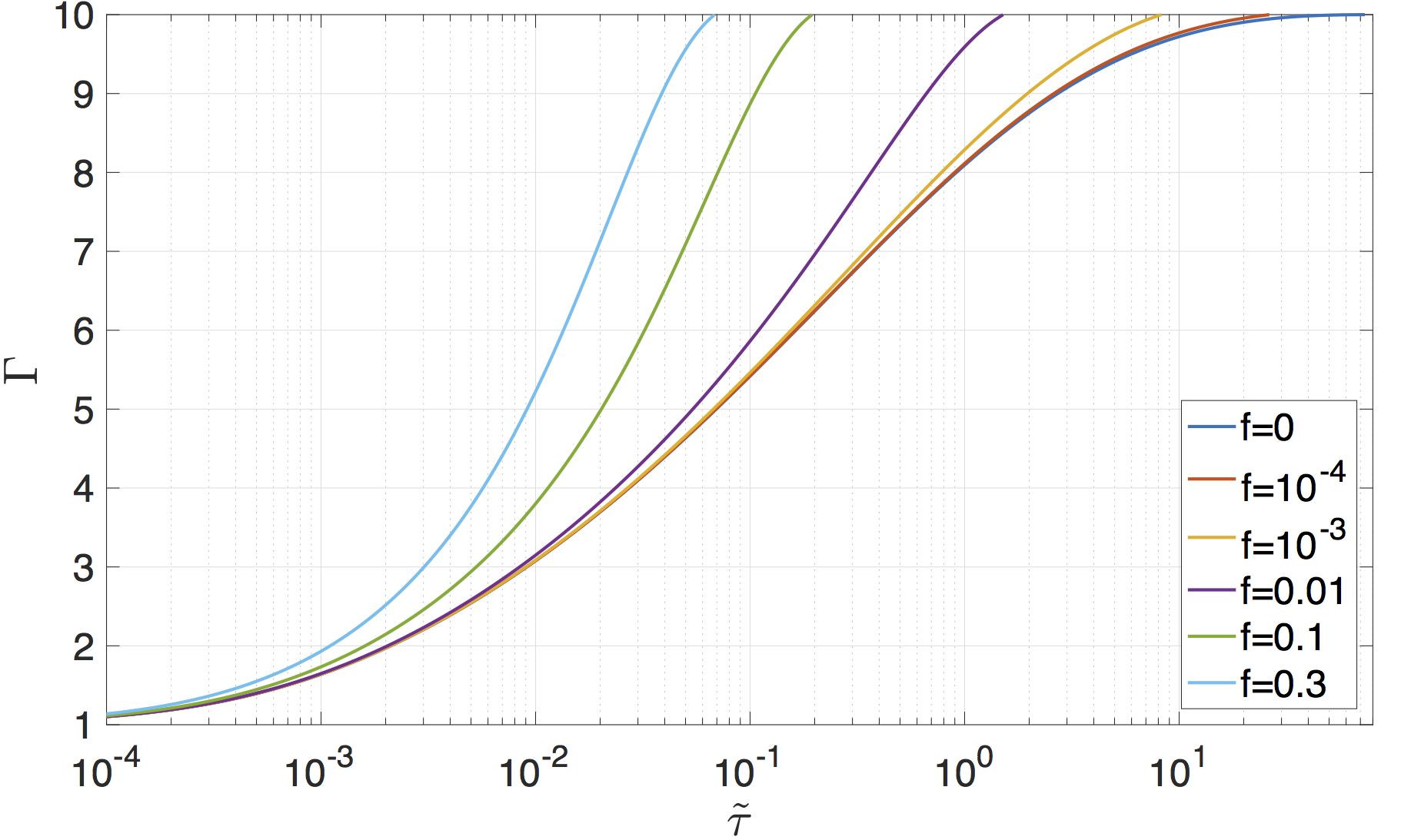} \includegraphics[width=6cm,height=3.8cm,]{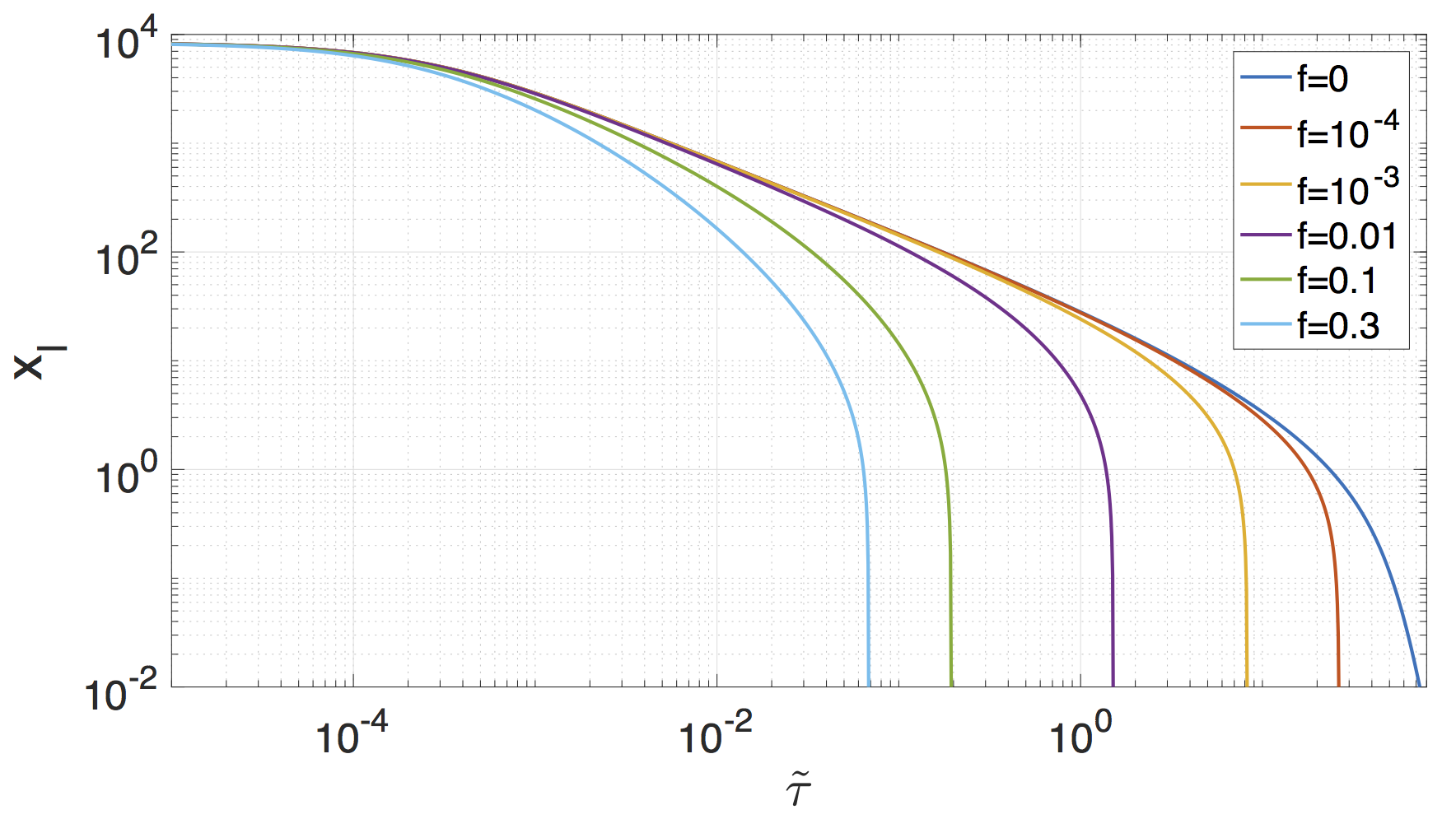}
\caption{\label{fig:RMS_escape} Lorentz factor (left) and pair loading (right) profiles, plotted as functions of the 
pair-unloaded Thomson depth, for different values of the 
escape parameter $f$.  From \cite{granot2018}.}
 \end{figure}

The analysis of relativistic, quasi-steady finite shocks is similar to that of infinite shocks (see \S \ref{sec:RRMS_starved}), with the exception that the boundary  condition $n^\prime_{\gamma\rightarrow u}=0$
must be replaced by $n^\prime_{\gamma\rightarrow\ u}=n^\prime_{esc}$, where $n^\prime_{esc}$ designates the number density of counterstreaming
photons (as measured in the shock frame) that escape from the shock and never return.   With this modification, Eq. (\ref{eq:dx_l/dz}) generalizes to
\begin{eqnarray}
\frac{d \x}{d\tau}= -(\x + x_{esc}),
\label{eq:dx_l/dz_esc}
\end{eqnarray}
where $x_{esc}=n_{esc}/n$.  Combined with baryon number conservation, Eq. (\ref{eq:cont}), and energy conservation, Eq. (\ref{eq:Gamma-cons}), one obtains the modified shock solution in terms of the escape parameter $f=n_{esc}/n_{\gamma d}$. .   The resulting Lorentz factor and pair loading profiles are exhibited in Fig. \ref{fig:RMS_escape} for  $\gamma_u=10$ and different values of the escape parameter $f$.  Substantial modification
of the shock structure (compared with the infinite shock solution)  is expected once $f > \gamma_u^{-2}$.  In particular, the shock width,
measured in terms of the pair unloaded Thomson depth, satisfies:
\begin{equation}\label{eq:Dtaut_f}
\Delta\taut_s=\left\{
\begin{array}{lr}
10\eta\mu\gamma_{u}^3\hspace{1em}&  f\ll\frac{1}{\gamma_{u}^{2}}, \\
&\\
\frac{\mu \gamma_{u}}{f}\hspace{1em}& f\gg\frac{1}{\gamma_{u}^{2}},
\end{array} \right.
\end{equation} 
where the limit $f\ll\gamma_u^{-2}$ coincides with the infinite shock solution, Eq. (\ref{eq:RRMS_width_inft}). 

The accelerated pair creation seen in Fig. \ref{fig:RMS_escape}, and the fact that all profiles, even for large values of the escape parameter $f$, 
converge to that of an infinite shock, suggests
that for relativistic RMS the quasi-steady approximation is good provided the density profile of the unshocked medium is not too steep. .

\subsection{Effects of finite magnetization}
\label{sec:magnetization}
When the fluid is sufficiently magnetized the interaction of the electric charges and the electromagnetic field modifies the dynamics of the flow.
To be consistent with our previous notation, we denote the electric and magnetic fields in the shock frame by ${\bf E}^\prime$
and ${\bf B}^\prime$, and in the fluid rest frame by ${\bf E}$ and ${\bf B}$.  They are related by appropriate Lorentz transformation.   
In the ideal MHD limit (which assumes that the fluid is a perfect conductor), the comoving electric field vanishes, ${\bf E}=0$, and
a Lorentz transformation yields the well know result ${\bf E}^\prime=-{\pmb \beta}\times{\bf B}^\prime$.  
If, in addition, the magnetic field  ${\bf B}^\prime$ is perpendicular to the flow velocity ${\pmb \beta}$,
then ${\bf B}^\prime=\gamma {\bf B}$, where, as before, $\gamma=(1-\beta^2)^{-1}$ denotes the bulk Lorentz factor.
The energy density  and Poynting flux of the electromagnetic field can be written in terms of ${\pmb B}$ as:,
\begin{align}
u_{EM}^\prime &=\frac{E^{\prime 2}+B^{\prime 2}}{8\pi} =\frac {\gamma^2B^2}{4\pi}-\frac{B^2}{8\pi},\\
{\bf S}^\prime &=c\frac{{\bf E}^\prime\times{\bf B}^\prime }{4\pi}=c\frac{B^2 }{4\pi}\gamma^2 {\pmb\beta}.
\end{align}
The force acting on the flow is given by ${\bf F}^\prime _{EM}=\rho^\prime_e {\bf E}^\prime+{\bf j}^\prime\times{\bf B}^\prime$, and the associated power by ${\bf j}^\prime\cdot{\bf E}^\prime$,
where $\rho_e^\prime$ and ${\bf j}^\prime$ are the electric charge and current densities, respectively. 
Let $T^{\mu\nu}_M=T^{\mu\nu}_b+T^{\mu\nu}_\pm+T^{\mu\nu}_\gamma$ denotes the  combined energy-momentum tensor of
the mixed plasma and radiation, where terms on the right hand side are defined explicitly in Eqs. (\ref{eq:Tmunu-plasma}) - (\ref{eq:Tmunu-rad}).
Then, the temporal and special components of Eq (\ref{eq:eng-mom}) must be modified according to 
\begin{eqnarray}
\begin{split}
\partial_t T^{00}_M + \partial _i T_M^{i0}= {\bf j}^\prime\cdot{\bf E}^\prime,\\
\partial_t T^{0j}_M + \partial _i T_M^{ij}= F^{\prime j }_{EM}.
\end{split}
\end{eqnarray}
Using Maxwell's equations the source terms can be expressed as~ 
${\bf j}^\prime\cdot{\bf E}^\prime= - \partial_t u_{EM}^\prime + \nabla\cdot{\bf S}^\prime$
and  ${\bf F}_{EM}^\prime = \partial_i[(E_i^\prime{\bf E}^\prime + B_i^\prime{\bf B}^\prime)/4\pi] - \partial_t {\bf S}^\prime - \nabla u_{EM}^\prime$. Introducing the energy-momentum tensor of the electromagnetic field,
$T_{EM}^{00}= u^\prime_{EM}$, $T^{0j}_{EM}= S^{\prime j}$, $T^{ij}_{EM}= u^\prime_{EM}g^{ij} -(E^{\prime i}E^{\prime j} + B^{\prime i}B^{\prime j})/4\pi$,
here $g^{ij}=\delta_{ij}$ are the spatiial components of the metric tensor, the MHD equations can be recast in the form $\partial_\mu T^{\mu\nu}=0$,
where $T^{\mu\nu}=T^{\mu\nu}_M + T^{\mu\nu}_{EM}$.

\begin{figure}[ht]
\centering
\includegraphics[width=6cm]{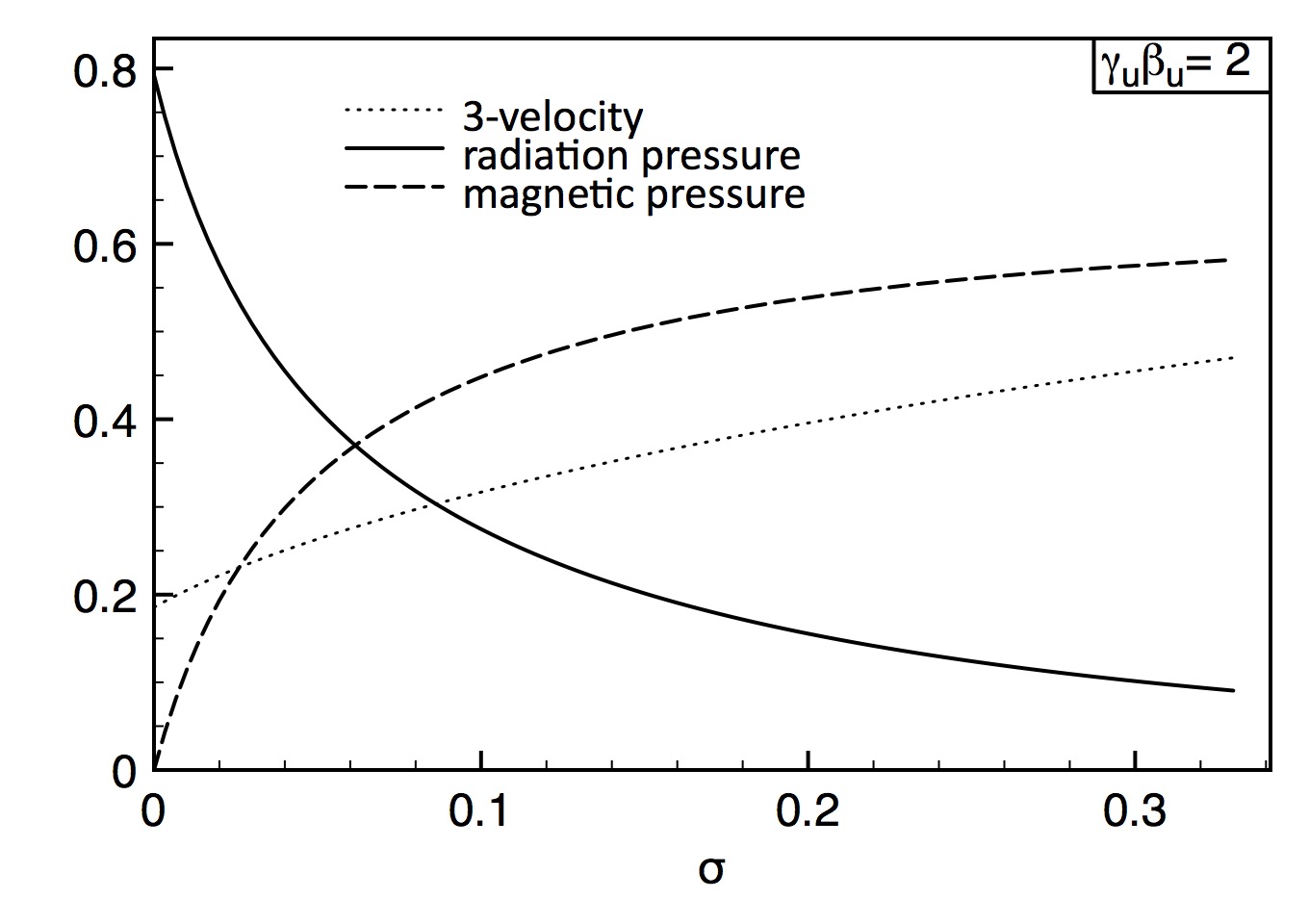} \includegraphics[width=6cm]{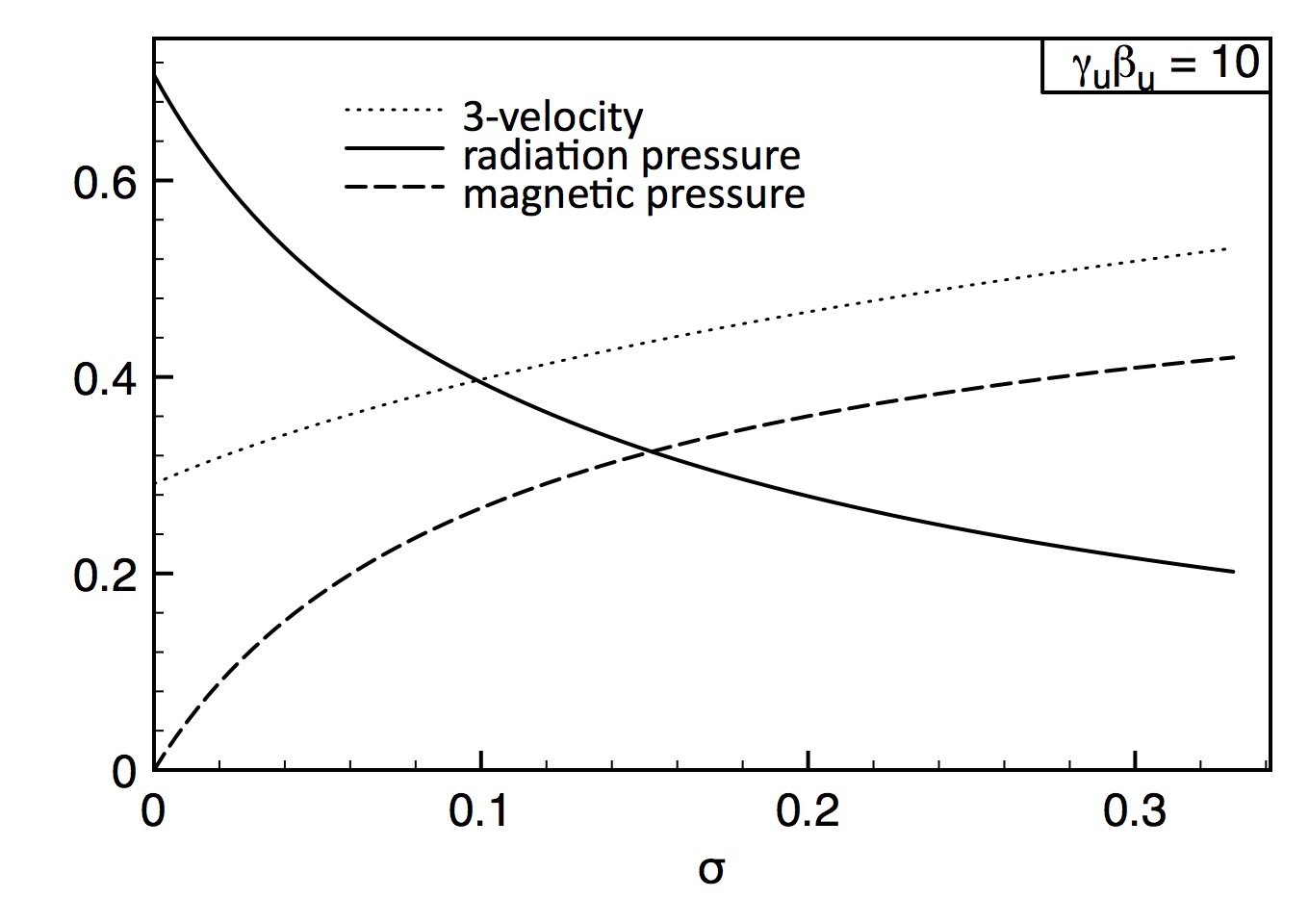}
\caption{\label{fig:RMS_B} Dependence of the downstream 3-velocity (dotted line) , radiation pressure (solid line) and magnetic pressure (dashed line) 
 on the magnetization parameter $\sigma_u$, for a shock Lorentz factor $\gamma=2$ (left panel) and $\gamma_u=10$ (right panel), 
with upstream pressure $p_u=0$.
The radiation and magnetic pressures 
are normalized by the ram pressure of the plasma in the upstream flow, $n_um_pc^2\gamma_u^2\beta^2_u$}.
 \end{figure}

Consider now a steady, planar shock with an upstream velocity ${\pmb \beta}=\beta \hat{x}$, as measured in the shock frame,
and magnetic field ${\pmb B}$ perpendicular to ${\pmb \beta}$.   The energy and momentum fluxes can be expressed in terms of the 
total pressure, $p=p_\pm +p_e +p_\gamma$, and the total dimensionless enthalpy per baryon, $h= 1+\mu x_\pm + 4 p/nm_pc^2$, where  the
 rest mass energy of the pairs, $\mu x_\pm= m_e n_\pm/m_pn$, is included in the definition of the enthalpy, as:
\begin{align}
\begin{split}
T^{0x} &= nm_pc^2(h +\sigma)\gamma^2\beta,\\
T^{xx} &=nm_pc^2(h +\sigma)\gamma^2\beta^2 + (p+B^2/8\pi),
\end{split}
\end{align}
here
\begin{equation}
\sigma=\frac{B^2}{4\pi m_pc^2 n}
\end{equation}
is the magnetization parameter.   Note that with this definition the Alfven 4-velocity is given by $u_A=\sqrt{\sigma/h}$.  Faraday-Maxwell equation, $\nabla\times{\pmb E^\prime}=-\nabla\times(\gamma{\pmb \beta}\times{\pmb B})=0$,
combined with the continuity equation, $\nabla\cdot(n\gamma {\pmb \beta})=0$, can be employed to show that $B/n$ is conserved along streamlines.
Denoting for short $\tilde{h}=h+\sigma$, the jump conditions read:

\begin{align}
\begin{split}
\sigma_u \gamma_u\beta_u &=\sigma_d \gamma_d\beta_d,\\
\tilde{h}_u \gamma_u &= \tilde{h}_d \gamma_d,\\
\tilde{h}_u\gamma_u\beta_u +\frac{\tilde{h}_u+\sigma_u-1 - \mu x_{\pm u}}{4\gamma_u\beta_u} &
= \tilde{h}_d\gamma_d\beta_d + \frac{\tilde{h}_d+\sigma_d-1 -\mu x_{\pm d}}{4\gamma_d\beta_d} .
\label{eq:RMS_B_jump}
\end{split}
\end{align}
This set of equations must augmented by an equation that determines the pair multiplicity $x_\pm$.  
For typical GRB parameters $x_{\pm d}\ll m_p\gamma_u/m_e$, thus the rest mass energy of the pairs can be neglected. 
With $\mu x_{\pm u}=\mu x_{\pm d}=0$, Eqs. (\ref{eq:RMS_B_jump}) can solved to yield $\beta_d$, $\sigma_d$ and $\tilde{h}_d$.
An example is shown in Fig \ref{fig:RMS_B}, where 
the dependence of the downstream 3-velocity $\beta_d$, radiation pressure $p_{\gamma d}$ and magnetic pressure $B_d^2/8\pi$ 
are plotted against $\sigma$, for a shock with a cold upstream plasma ($p_u=0$).  The radiation and magnetic pressures 
are normalized by the ram pressure of the plasma far upstream, $n_um_pc^2\gamma_u^2\beta^2_u$.

As Fig \ref{fig:RMS_B} indicates, the fraction of upstream bulk energy which is converted into radiation downstream, $e_{\gamma d}/n_um_pc^2\gamma_u^2\beta^2_u$, decreases with increasing $\sigma_u$, and is  considerably reduced when $\sigma_u$ 
approaches $\sim 0.1$.   The remainder is used up to compress the magnetic field.  As a consequence, the net force per baryon
acting on the upstream flow, $\sigma_T e_{\gamma d}$, is reduced by the same factor.  This suggests that at high enough magnetization 
the radiation alone will not be able to 
decelerate the upstream flow and a collisionless subshock must form.  \cite{beloborodov2017a} has shown that this happens when the 
magnetization parameter exceeds a few percents. 

The presence of a subshock can lead to copious production of soft photons via synchrotron emission of thermal and non-thermal 
pairs.   While particle acceleration is prohibited at such a high magnetization in relativistic shocks \citep{sironi2009}, it is unclear at present
whether this is true also for the mildly and sub relativistic subshocks expected to form in GRBs.   \cite{lundman2018b} contended that
a considerable fraction, $f_{sub}\sim0.3-0.5$, of the dissipated subshock energy is tapped to produce a quasi-Mawellian 
distribution of pairs, that cool rapidly via synchrotron and inverse Compton emission.  
To estimate the characteristic scale of the cooling layer, the net cooling rate  of an electron (ignoring KN effects which are negligible),
$t^{-1}_c\simeq 4\gamma_{th}\sigma_T(e_{\gamma d}+e_{Bd})/3m_ec$, where 
$\gamma_{th}$ is the thermal Lorentz factor of pairs just behind the subshock and $e_{Bd}=B_d^2/8\pi$,
can be compared  with the mean scattering rate,
$t_{sc}^{-1}= (n_{ed}+n_{\pm d}) \sigma_T c=(1+x_\pm)n_d\sigma_T n_dc$.
This yields:  $t_c/t_{sc} < m_ec^2 (1+x_\pm)n_d/\gamma_{th}e_{\gamma d} \sim m_e(1+x_\pm)/(m_p\gamma_u\gamma_{th})<<1$,
 implying that the width of the cooling layer behind the subshock is vastly smaller than the RMS scale \citep{lundman2018b}.
In deriving the above result the approximations $e_{\gamma d}\sim m_pc^2n_u\gamma_u^2\beta_u=m_pc^2n_d\gamma_u$ 
was adopted (see Fig \ref{fig:RMS_B}), and the fact that in such shocks $x_\pm <<m_p\gamma_u /m_e$  
\citep{ito2018a} was used. 

The thin cooling layer behind the subshock is the source of the soft synchrotron photons.  The synchrotron spectrum
depends on the Lorentz factor $\gamma_{th}$ of thermal pairs, which, in turn, depends on the pair load $x_\pm$, roughly 
as $\gamma_{th}x_\pm \simeq (m_p/m_e)f_{sub} (\gamma_u\beta_u)^2$ \citep{lundman2018b}.   On the other hand, $x_\pm$ 
depends on the overall RMS structure, hence, the subshock emission is nonlinear in nature.    The relation $\gamma_{th}\propto x_\pm^{-1}$ 
stems from the fact that the subshock energy is equally shared among all particles.  \cite{lundman2018b} estimate
that in mildly relativistic RMS ($\gamma_u\beta_u\simlt1$) the average pair energy in the cooling layer spans the range $\gamma_{th}\simeq 20-40$.

The soft synchrotron photons produced in the cooling layer will propagate away from the subshock and will experience energy gain 
through thermal and bulk Comptonization, as well as energy losses through self-absorption, free-free absorption 
and induced downscattering.    The relative importance of the different processes depends on the RMS parameters.   It can be 
shown that synchrotron emission is strongly suppressed in relativistic RMS by virtue of the large pair loading contributed by 
bulk Comptonized photons, which leads to diminution of $\gamma_{th}$ and the associated synchrotron frequency $\gamma_{th}^2 \nu_B$.


\section{ Sub-photospheric emission in long GRBs}
\label{sec:GRBs}

\subsection{Formation and dissipation of GRB outflows}
The formation and dissipation of GRB outflows have been the subject of extensive research since the discovery of GRBs.  
The high Lorentz factors inferred from energy considerations, compactness arguments and afterglow models, $\Gamma\sim 10^2-10^3$, 
require extremely low baryon load at the outflow injection point, which pose a tremendous challenge for outflow formation models.
The conventional wisdom has been that those outflows are powered by magnetic 
extraction of the rotational energy of a neutron star or an accreting black hole, and that the energy thereby extracted is transported 
outward in the form of Poynting flux, which on large enough scales is converted into kinetic energy flux.   An alternative scenario 
asserts that these outflows are driven by the pressure of a relativistically hot electron-positron plasma, which is injected in the polar region via annihilation of neutrinos emitted from the hyper-accretion flow surrounding the black hole 
\citep[e.g.,][]{levinson1993,mochkovitch1993,popham1999,birkl2007,zalamea2011,levinson2013}.
However, it is generally accepted that in long GRBs this model is disfavoured on energetic grounds \citep[e.g.,][]{zalamea2011,kawanaka2013,globus2014}.

In the context of magnetic jets,
an important question concerning the prompt emission mechanism is whether the conversion 
of magnetic-to-kinetic energy occurs above or well below 
the photosphere (e.g., McKinney \& Uzdenski 2012; Levinson \& Begelman 2013;  Bromberg et al. 2014).   
If the claimed evidence for photospheric emission is true, it means that magnetic field conversion should occur well
below the photosphere. 
The mechanism by which magnetic energy is converted to kinetic energy has not been identified yet, but it is generally believed to involve gradual acceleration of the flow \citep[e.g.,][]{heyvaerts1989,chiueh1991,bogovalov1995,lyubarsky2009}, impulsive acceleration 
\citep{granot2011,levinson2010,komissarov2012,granot2012}, and/or non-ideal MHD effects, specifically magnetic reconnection 
\citep[e.g.,][]{drenkhahn2002,lyutikov2003,lyubarsky2010,mckinney2012}.  Note that in the former case (acceleration 
of a steady, ideal MHD flow) the magnetization practically saturates at $\sigma \sim1$  \citep{lyubarsky2009}, 
hence magnetic dissipation is still required in order to allow formation of strong shocks.

Magnetic reconnection requires the formation of small-scale magnetic domains with oppositely oriented magnetic field lines. Such structures may inherently form during outflow injection, e.g., owing to advection of asymmetric magnetic field into the black hole, as postulated in the striped wind model \citep{drenkhahn2002,levinson2016}, or result from current-driven instabilities induced during the propagation of the jet 
\citep{mignone2010,mizuno2012,oneil2012,guan2014,singh2016,bromberg2016}.   While in the former case the extracted power is considerably smaller than the power obtained in a magnetically arrested disk (MAD), it seems sufficient to account for the 
observed luminosities in most objects \citep{perfrey2015}.   Whether magnetic field dissipation occurs above or below the 
photosphere in the striped wind model depends primarily on the 
asymptotic Lorentz factor (or, equivalently, baryon loading) of the flow \citep{drenkhahn2002}.  
In case of an initially stable (ordered) magnetic field configuration, effective magnetic dissipation may ensue via a rapid growth of the current-driven kink instability.  
Recent state-of-the-art numerical simulations \citep{bromberg2016,singh2016} demonstrate that such a rapid growth
is expected to occur in the dense focusing nozzle that forms inside the high-pressure cocoon surrounding the GRB jet.
It is not entirely clear at present what is the final magnetization in the dissipation zone, but if below unity then the GRB outflow is expected to be weakly magnetized when approaching the photosphere. 

If the Poynting flux jet indeed transforms into a weakly magnetized flow below the photosphere, either via 
magnetic reconnection or impulsive acceleration,
then further dissipation, that produces the observed prompt emission,
most likely involves formation of hydrodynamic shocks in the weakly magnetized flow.   
Hydrodynamic simulations of  jet propagation in collapsars \citep[e.g.,][]{lazzati2009,morsony2010,Lopez-Camara2013,ito2015,ito2018b,harrison2018,gottlieb2019},
as well as in the ejecta of neutron star mergers \citep{gottlieb2017,lazzati2017,gottlieb2019}, demonstrate that  a considerable fraction of the bulk energy dissipates in recollimation shocks
below the photosphere, giving rise to a substantial photospheric component in the prompt emission of both, long and short GRBs.
Below we show, using heuristic arguments, that high radiative efficiency in photospheric emission is quite generally expected in collimation shocks, and 
discuss  recent numerical studies that assess the robustness of this conclusion by systematically probing a wide range of conditions.   
A second dissipation mode discussed below is internal RMS, that are 
produced by intermittencies of the central engine.  These are expected to form at modest optical depths
below the photosphere if the Lorentz factor of the outflow is not exceptionally large \citep{eichler1994,morsony2010,bromberg2011}.

\subsection{Conditions at the photosphere}
As explained in section \ref{sec:Theory}, the characteristics of sub-photospheric shock emission depend on the upstream conditions, and in particular 
on the photon-to-baryon density ratio $\tilde{n}$.  The latter can be evaluated if the dynamics of the GRB outflow is known.  An
illustrative example is a conical adiabatic outflow (Levinson 2012).  
As shown below, two important parameters determine the formation radius of internal shocks and the value of $\tilde{n}$; the isotropic equivalent
outflow power, $L_{jiso}$, and the outflow injection radius $R_0$.
The observed isotropic equivalent luminosities of long GRBs span the range $10^{50} < L_{\gamma iso} <10^{54}$ erg s$^{-1}$ \citep[e.g.,][for a recent account]{deng2016,paul2018}; the 
corresponding jet power, $L_{jiso}$, is most likely a few times larger.  The injection radius of the outflow may be associated with the 
outer light cylinder in Poynting flux jets, or the  sonic point in hydrodynamic (e.g., neutrino driven) jets.  Typically, it is located at  a few Schwarzschild 
radii \citep{globus2014}, which for a 10 $M_\odot$ black hole is $R_0\sim10^7$ cm.
Suppose now that a conical outflow having an isotropic equivalent
power $L_{jiso}=10^{53}L_{53}$ ergs s$^{-1}$
is ejected with an initial Lorentz factor $\Gamma_0\sim1$  from a radius
$R_0=10^7R_7$ cm, and that it carries baryons with an isotropic mass loss rate $\dot M_{iso}$.   
The location of the photosphere depends on the ratio $\eta/\eta_c$, where $\eta=L_{jiso}/(\dot M_{iso} c^2)$ and \citep{grimsrud1998}
\begin{equation}
\eta_c=\left(\frac{\sigma_T L_{jiso}\Gamma_0}{4\pi
R_0m_bc^3}\right)^{1/4}=1.8\times 10^3L_{53}^{1/4}R_7^{-1/4}\Gamma_0^{1/4}.\label{eq:eta_c-def1}
\end{equation}
When $\eta>\eta_c$  the fireball will become transparent already during the acceleration phase, before reaching the coasting
radius.   The Lorentz factor in that case  may be close to $\eta_c$ (Nakar et al. 2005), and the emerging emission should
have a roughly black body spectrum, as in the original fireball models  \citep{paczynski1986,goodman1986}.    
On the other hand,
when $\eta<\eta_c$ the outflow is sufficiently opaque, such that the radiation is trapped during the entire acceleration phase.
The major fraction of the explosion energy is
then converted into bulk kinetic energy of the baryons, and the outflow reaches a terminal Lorentz factor $\Gamma_\infty\simeq\eta$ at some
 radius $r_{coast}\simeq\eta R_0/\Gamma_0$, beyond which it continues to coast.
The photosphere is located in the coasting region, at a radius $r_{ph}=  (\eta_c/\eta)^4r_{coast} >r_{coast}$, ignoring spreading for simplicity.
At the coasting radius the optical depth is $\tau(r_{coast})=(\eta_c/\eta)^4$. Thus, the optical depth above the coasting radius, 
where sub-photospheric shocks are likely to form, satisfies $1< \tau < (\eta_c/\eta)^4$.

An approximate estimate of the photon-to-baryon density ratio near the photosphere can be obtained upon assuming that the GRB outflow is adiabatic from its injection point at $r=R_0$ up to the sub-photospheric region where shocks form (the effect of a collimation shock on $\tilde{n}$ is discussed below).   For a purely hydrodynamic flow, the temperature 
in the vicinity of the injection point typically exceeds a few MeV,  hence the radiation is in thermodynamic 
equilibrium with the $e^\pm$ pairs.  As the flow expands the comoving temperature drops and the pairs are gradually converted into photons.
Since by the adiabatic assumption no new photons are being generated as the flow expands, the total number of quanta (that is, electrons, positrons
and photons) is conserved.   This means that the ratio $n_Q/n$, where $n_Q=n_\gamma + n_\pm$, is conserved along streamlines, and is equal to 
$\tilde{n}$ near the photosphere where $n_\pm=0$.
To find $n_Q/n$ we recall that for a conical, adiabatic flow, baryon number conservation and energy conservation yield
\begin{eqnarray}
m_pc~n~\Gamma~\beta 4\pi~r^2=\dot{M}_{iso},\\
n~h~m_pc^3~\Gamma^2\beta~4\pi~r^2=h~\Gamma~\dot{M}_{iso}c^2 = L_{jiso},\label{eq:FB_energ}
\end{eqnarray}
where $h=1+4p/nm_pc^2$ is the dimensionless enthalpy per baryon, and $p$ is the total pressure contributed by pairs and photons.  
The last equation implies 
that the product $h\Gamma$ is conserved and its value is $h\Gamma=\eta$.  At the injection point, $\Gamma=\Gamma_0\sim1$ and
$h=h_0\simeq 4p_0/n_0m_pc^2 \gg1$.   Since, as mentioned above, the pairs and radiation are in thermodynamic equilibrium
at the base of the flow, where the temperature is $T_0 \sim$ a few MeV, we have $p_0=11~a~T_0^4/12=(0.9 n_{\gamma 0}+1.05n_{\pm0})kT_0$.
For simplicity we shall adopt the approximate equation of state $p_0=(n_{\gamma0} + n_{\pm0}) kT_0$, which is accurate enough for our purposes. 
We then have $\tilde{n}=(\eta/4\Gamma_0)(m_pc^2/kT_0)$.  The temperature can be found from Eq. (\ref{eq:FB_energ}) upon 
substituting $p_0=11~a~T_0^4/12$ in $h_0$. Expressing $L_{jiso}$ in terms of $\eta_c$, Eq. (\ref{eq:eta_c-def1}), one finally obtains:
\begin{equation}
\tilde{n}\simeq 3\times10^5(\eta/\eta_c)R_7^{1/4}\Gamma_0^{-1/4}.
\label{eq:ntil_GRB}
\end{equation}
We emphasize that the dimensionless entropy given in Eq. (\ref{eq:ntil_GRB}) depends only on the total power and baryon load of the fireball, and not its structure.  It therefore holds for any outflow geometry.   Moreover, it is worth noting that for a given power $L_{jiso}$, 
Eqs. (\ref{eq:eta_c-def1}) and (\ref{eq:ntil_GRB}) imply that $\tilde{n}\propto R_7^{1/2}$.  This 
means that if dissipation takes place well below the photosphere, e.g., by collimation shocks, such that the acceleration of the 
outflow is significantly delayed, it can be translated to a larger injection radius $R_{0}$ in the above derivation, and, hence, larger $\tilde{n}$,
provided the optical depth exceeds the value required for thermodynamic equilibrium (see discussion in \S \ref{sec:Phot_generation}).  

\subsection{Properties of sub-photospheric shocks}
\subsubsection{Internal shocks}
\begin{figure}[ht]
\centering
\includegraphics[width=10cm]{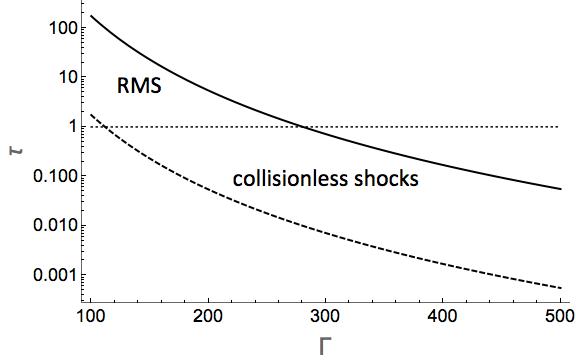}
\caption{\label{fig:f1} Optical depth at the radius of shock formation versus shell Lorentz factor in a conical outflow, for $L_{53}=1$, $\delta t =1 {\rm ms}$ (solid line) and $\delta t=10^2 {\rm ms}$ (dashed line).  The regimes in which radiation mediated shocks (RMS) and collisionless shocks form are indicated. }
\label{fig:tau_shck}
 \end{figure}

Sporadic outflow activity produces waves that steepen into shocks at some distance from the central engine.  This can be caused by
intermittencies of the central engine, or via mixing of jet and cocoon material in the vicinity of the collimation shock, as will be discussed
further in the next subsection.
A simple estimate of the optical depth at the shock formation radius can be made by considering the consecutive ejection of two shells, one
 ejected at time $t_0$ with Lorentz factor $\Gamma_1$, and the other one at time $t_0+\delta t$ with
 Lorentz factor  $\Gamma_2 >\Gamma_1$.  The two shells will collide at a radius 
 $r_d = c\delta t/(\beta_2-\beta_1)\simeq 2\Gamma_1^2c\delta t$.   If the collision occurs in the coasting zone then $\Gamma_1\simeq \eta$,
yielding an optical depth of $\tau(r_d) \simeq \eta_c^4 R_0/\eta^3\Gamma_0 r_d  \simeq (\eta_c^4/2\Gamma_0 \Gamma_1^5)(c\delta t/R_0)^{-1}$ at the shock formation radius \citep{bromberg2011}.
Consequently, the collision will occur below the  photosphere, that is $\tau(r_d)>1$, provided 
\begin{equation}
\Gamma_1< 290 ~ L_{53}^{1/5}(\delta t/1~{\rm ms})^{-1/5}.
\end{equation}
A plot of this relation is shown in Fig. \ref{fig:tau_shck}.  It indicates that if the terminal Lorentz factors of GRB outflows are moderate,
$\Gamma < 300 L_{53}^{1/5}$, then sufficiently rapid intermittencies ($ \delta t < 0.1 s$) should steepen into shocks
below the photosphere, at a moderate optical depth.   Note that this condition is more easily satisfied in brighter sources. 
The above analysis can be readily generalized to collimating flows \citep{levinson2012}, to show that the effect of collimation is 
not very significant. 

The collision of two shells having Lorentz factors $\Gamma_1\gg1$ and $\Gamma_2>\Gamma_1$ 
(as measured in the star frame) creates a shock that propagates at a Lorentz factor $\gamma_u\simeq \sqrt{\Gamma_2/4\Gamma_1}$
with respect to the rest frame of the unshocked shell\footnote{This is true for internally symmetric shells. For a more
general expression see, e.g., \cite{vanputten2012}}.  This implies that internal shocks are likely to have modest Lorentz factors,
$\gamma_u\beta_u\simgt1$.   
Now, if the shock forms below the photosphere, it is mediated by radiation and its width, as
measured in the shock frame, is $\Delta^\prime_s\simeq(\sigma_Tn_{u})^{-1}$, where $n_{u}$ is the 
proper density of the unshocked gas (see section \ref{sec:basic-principles}).  The latter estimate assumes negligible opacity 
by newly created pairs,
which may be justified in case of mildly relativistic shocks \citep{ito2018a}.
In the star frame the shock width is given by $\Delta_s =\Delta^\prime_s/\Gamma_1\simeq r_{ph}/(\Gamma_1\tau)^2$,
where $r_{ph}$ is the photospheric radius, and $\tau=\sigma_Tn_{u}r/\Gamma_1 =(r_{ph}/r)$ is the optical depth at radius $r<r_{ph}$ \citep{levinson2012}.   It is seen that the shock broadens as it approaches the photosphere.  
For shells having a width larger than $r/\Gamma^2$, the net optical depth of the postshock  layer (i.e., the downstream region) is 
$\Delta \tau_d\sim \sigma_T n_d\Delta_d^\prime \sim 
\sqrt{2}\sigma_T n_u r/\Gamma_1\simeq \sqrt{2} \tau$, where $\Delta_d^\prime \simeq r/(2\Gamma_1\gamma_u)$ denotes 
the comoving width of postshock layer\footnote{Note that the shock Lorentz factor in the Lab frame is $\Gamma_{sh}=2\Gamma_1\gamma_u$}, 
and for illustration we assumed a strong shock, $n_d=\sqrt{8}\gamma_u n_u$.  
The fraction of dissipated energy contained inside the shock (i.e., within the shock transition layer) at a radius $r$  is roughly $\Delta\tau_s/\Delta\tau_d\sim (\sqrt{2}\tau)^{-1}$, which can be significant near the photosphere.   This has important consequences for the observed spectrum.

As the shock propagates from its formation site to the photosphere it suffers adiabatic losses.  
For a shell of width $\Delta_1 \simgt \delta t$ and Lorentz factor $\Gamma_1$, shock breakout will occur at a radius $r_b\simeq r+\Gamma_1^2 \Delta_1 >2r$,
where $r$ is he shock formation radius.  The optical depth at breakout is $\tau_b = r_{ph}/r_b < \tau/2$.  Significant adiabatic losses are 
expected if $r_b<<r_{ph}$ ($\tau_b>>1$). Such shells will not contribute to the observed emission, unless experiencing additional collisions 
at larger radii.  
However, if $\tau_b$ is modest  these
losses are expected to be minor, owing to the fact that the total swept-up mass 
increases with radius \citep{levinson2012}.  For instance, in case of a conical outflow the shock velocity 
is approximately constant, resulting in a constant dissipation per unit mass, $dE/dm \simeq $ const.  
At the same time, the mass enclosed below the photosphere scales as $m(\tau)\propto 1-\tau^{-1}$, implying that most of the 
emitted energy is accumulated just below the photosphere.   In fact, if $\Delta_1 >>\delta t$ most of the dissipation may occur 
after the shock becomes collisionless, otherwise, if $\Delta_1 \simgt \delta t$, shocks that form below the photosphere emit when 
still mediated by radiation.  A more involved shock dynamics may alter these estimates,
but the salient lesson is that RMS which form not too deep beneath the photosphere should be radiatively efficient.

The immediate downstream temperature of a sub-photospheric shock can be found now from Eqs. (\ref{eq:shock_rich_temp}) 
and (\ref{eq:ntil_GRB}):  $(kT_d/m_ec^2) \simeq 1.4\times10^{-3}(\eta_c/\eta)(R_7/\Gamma_0)^{-1/4}$.  For a coasting shell the observed 
temperature is boosted by the factor $\Gamma_\infty \simeq \eta$, 
\begin{equation}
kT_{d,ob}=\eta kT_d\simeq 1 \gamma_u \beta_u  L_{53}^{1/4} \Gamma_0^{1/2}R_7^{-1/2}\quad {\rm MeV},\label{eq:kTsR}
\end{equation}
and it is seen that it is independent of the bulk Lorentz factor of the unshocked shell.   By employing Eq. (\ref{eq:shock_rich_temp}) it is tacitly assumed
that the shock is highly relativistic. For mildly relativistic shocks ($\gamma_u\beta_u\sim1$) this overestimates the actual temperature
by a factor of about 2.   With $\gamma_u\beta_u\sim1$ Eq. (\ref{eq:kTsR}) predicts observed temperatures of $kT_{obs}\simlt1$ MeV.
However, if the flow is dissipative $\tilde{n}$ may be  larger than the value given in Eq. (\ref{eq:ntil_GRB}) and the 
temperature lower.    
 Moreover, mild magnetization of the outflow may lead to formation of subshocks \citep{beloborodov2017a,beloborodov2017b}
 and the consequent emission of soft synchrotron photons that may also enhance $\tilde{n}$ \citep{lundman2018b}.

\subsubsection{Collimation shocks}
\begin{figure}[ht]
\centering
\includegraphics[width=12cm]{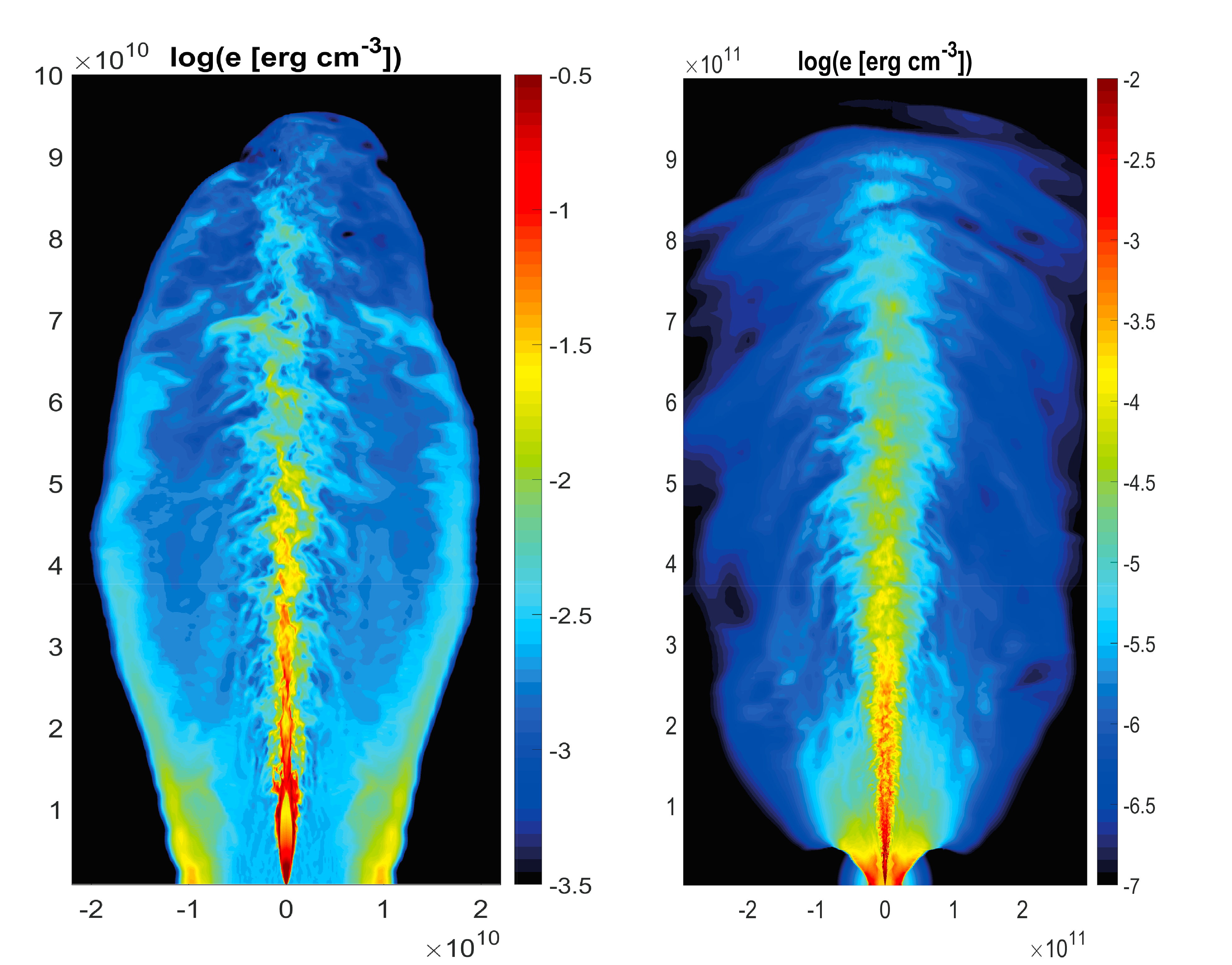}
\caption{Snapshots from a 3D hydrodynamical simulation of jet propagation inside a collapsed
star of radius $R_\star=10^{11}$ cm.  The jet power and opening angle in this run are $L_j=10^{50}$ 
erg s$^{-1}$ and $\theta_0=0.14$, respectively.  The jet is injected from a radius of $r_{inj}=10^{-2} R_\star$
with initial Lorentz factor $\Gamma_0=5$ and specific enthalpy $h_0=10^2$.  The left panel exhibits 
energy map at breakout time, about 20 seconds after the beginning of the simulation, and the right panel
at time $t=65~ s$, when the head of the jet has reached a radius of $10 R_\star$.  A strong collimation shock
located at $r_s=0.1 R_\star$ is clearly visible in the left snapshot.   The shock moved to a radius $r_s\simeq 0.3 R_\star$
by the time the jet head reached $10R_\star$ (right panel).  Wobbling of the jet, caused by mixing of jet and cocoon material
in the collimation zone, is also seen in these snapshots.}
\label{fig:collim}
 \end{figure}

Collimation shocks are generic features in GRB jets \citep[e.g.,][]{lazzati2009,morsony2010,bromberg2011b,Lopez-Camara2013,ito2015,harrison2018,gottlieb2019}.   They result from supersonic deflection of streamlines by 
the overpressured cocoon that forms as the jet propagates through the dense medium enshrouding the central engine - the
stellar envelope in long GRBs and the merger subrelativistic ejecta in binary neutron star mergers.    
As indicated by recent RHD simulations \citep{gottlieb2019}, the evolution of the collimation shock depends primarily 
on the density profile of the confining medium:
In typical long GRBs the collimation shock propagates outwards slowly as the outflow expands, reaching a radius of about one tenth stellar radii 
by the time the outflow breaks out of the star (Fig \ref{fig:collim}).  Subsequently, it continues to expand as the pressure 
in the cocoon gradually declines, however, in most cases it
 remains inside the progenitor's envelope for the entire duration of the burst.  In case of particularly bright bursts, with 
 isotropic equivalent energy in excess of  $10^{54}$ ergs,  the collimation shock may ultimately break out
 of the star and propagate at a mildly relativistic speed to the vicinity of the photosphere.   In short GRBs the shock breaks out quickly
 and reaches the photosphere by the time of emission.   As explained in \cite{gottlieb2019}, the injection radius of the outflow in RHD simulations 
 needs to be sufficiently small ($<0.01 R_\star$ in collapsar simulations)  in order to reach convergence; improper choice of
 the injection radius may result in an artificially different structure at late times.

As mentioned above, the formation of a collimation shock changes the relative location of the coasting radius and the photosphere, and can 
significantly enhance the efficiency of photospheric emission.    In addition, substantial mixing of jet and ambient 
matter, as indicated by recent 3D simulations \citep{gottlieb2019},  can also alter the photospheric conditions.  In particular, it leads to
stratification of the flow and, consequently, to a strong angular dependence of the radiative efficiency at the photosphere (for
details see \citealt{gottlieb2019}).   The simple considerations below elucidate some of these effects.

The Lorentz factor of a flow of  initial opening angle $\theta_0=0.1\theta_{-1}$ that enters a collimation shock at some 
radius $r_s=10^{11}r_{s11}$ cm,  drops to $\Gamma_s\approx \theta_0^{-1}$ behind the shock \citep{bromberg2011b}.  
After passing the shock the outflow re-accelerates.    
For illustration, suppose that it is conical with the same opening angle as the initial one, $\theta_0$.
If no additional baryon loading occurs during the collimation process, then $\eta_s=\eta_0$, where 
$\eta_0=h_0\Gamma_0$ denotes the initial load set up at the outflow injection point and $\eta_s=h_s\Gamma_s$
defines the local load downstream of the collimation shock.  
However,  mixing at the collimation throat can increase the load, so that $\eta_s$ can vary among 
different fluid elements, but must satisfy $\eta_s \le \eta_0$.   The new coasting radius  of a re-accelerating fluid element above the 
collimation shock will be located at $r_{s,coast}\approx \eta_s r_s/\Gamma_s \approx \eta_s r_s \theta_0$. 
The location of the photosphere of the re-accelerating flow can be grossly estimated by employing 
Eq. (\ref{eq:eta_c-def1}) with $R_0$ and $\Gamma_0$ 
replaced by the new injection radius $r_s$ and Lorentz factor $\Gamma_s$.    One then finds that the
photospheric radius $r_{ph}$ will coincide with the new coasting radius $r_{s,coast}$ when $\eta_s = \eta_{s,c}$, where
\begin{equation}
\eta_{s,c}=\left(\frac{\sigma_T L_{jiso}\Gamma_s}{4\pi m_pc^3 r_s}\right)^{1/4} =320 ~L_{53}^{1/4}(r_{s11}\theta_{-1})^{-1/4}.
\label{eq:eta_sc}
\end{equation}
When $\eta_s>\eta_{s,c}$ the radiation is released in the acceleration zone ($r_{ph}<r_{s,coast}$) 
with high efficiency, roughly $1-r_{ph}/r_{s,coast}$ at $r_{ph}<<r_{coast}$.
 When $\eta_s<\eta_{s,c}$ the photosphere is located in the coasting zone, at $r_{ph}\approx (\eta_{s,c}/\eta_s)^4r_{s,coast}$,
 and the efficiency is suppressed; it is  roughly given by $(\eta_s/\eta_{s,c})^{8/3}$ for $(\eta_s/\eta_{s,c})^{8/3}<<1$ before radial
 spreading of fluid shells commences  \citep{levinson2012}.  
\begin{figure}[ht]
\centering
\includegraphics[width=12cm]{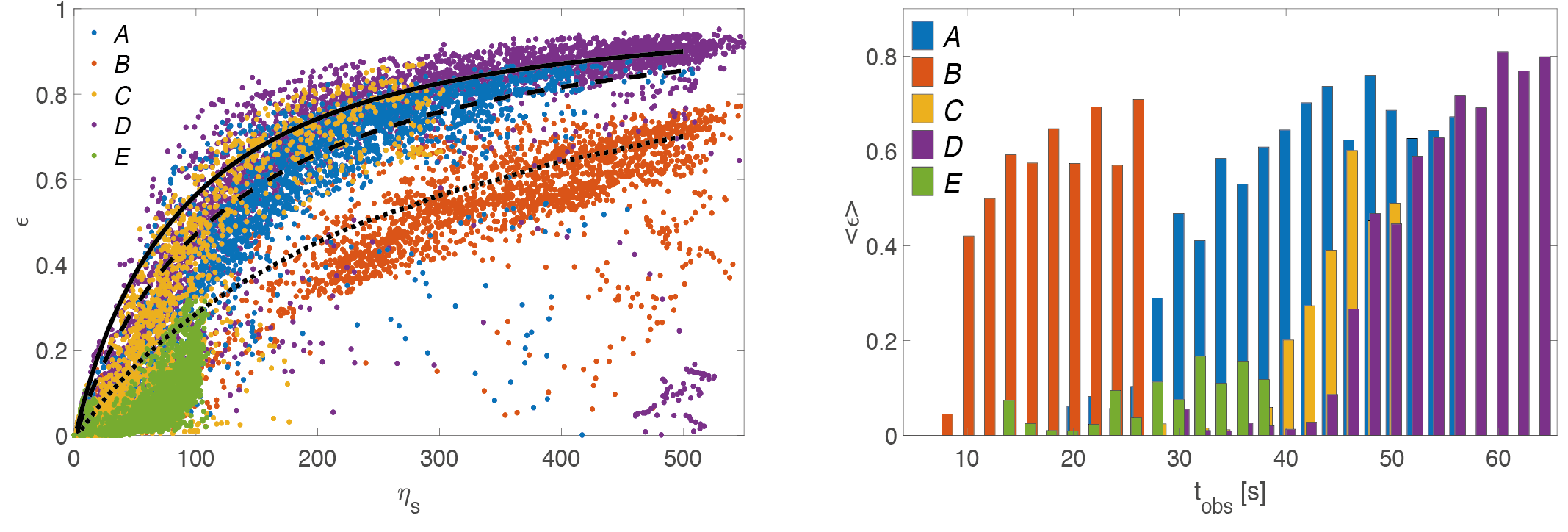} 
\caption{Left: The dependence of the radiative efficiency on the load parameter $\eta_s$ for fluid elements along the jet axis, 
obtained from 3D RHD simulations of jet propagation in a collapsed star or radius $R_\star=10^{11}$ cm.
The different models correspond to the following parameters:  A ($L_j=10^{50}$ erg/s, $\eta_0=500$, $\theta_0=0.14$),
B ($L_j=5\times10^{50}$ erg/s, $\eta_0=500$, $\theta_0=0.14$), C ($L_j=10^{50}$ erg/s, $\eta_0=300$, $\theta_0=0.24$), D ($L_j=10^{50}$ erg/s, $\eta_0=500$, $\theta_0=0.24$), E ($L_j=10^{50}$ erg/s, $\eta_0=100$, $\theta_0=0.14$).
  The variation in $\eta_s$ between different fluid elements in each model is caused by mixing at the collimation throat.  The black lines delineate the analytic result obtained from the integration of the adiabatic fireball equations, with $\eta_{s,c}$ adopted from Eq. (\ref{eq:eta_sc}) for the parameters
of the different models (the solid line corresponds to models $ C$ and $ D $, the dashed line to models $ A $ and $ E $, and the dotted line to model $ B $; see \cite{gottlieb2019} for details).
Right: The temporal evolution of the efficiency in the observer frame, presented in bins of two seconds for clarity. The observer time is measured with respect to the jet launch to show the full delay in the  onset of emission.  Reproduced with permission from \cite{gottlieb2019}.
}
\label{fig:eps-eta}
 \end{figure}
Figure \ref{fig:eps-eta}, taken from \cite{gottlieb2019},  shows the dependence of the radiative efficiency $\epsilon$ on the load parameter $\eta_s$ for fluid along the axis (left), and its temporal evolution in the observer frame (right),
obtained from 3D RHD simulations of different collapsar jet models, as indicated in the caption.   It confirms the expectation for high efficiency 
of photospheric emission based on the simple analytic criterion derived in Eq. (\ref{eq:eta_sc}). 
The efficiency is found to be smaller along streamlines with larger inclination angles, 
but is substantial up to an angle of about one half the opening angle at injection \citep{gottlieb2019}.  These results imply that a strong photospheric 
component cannot be avoided practically in weakly magnetized jets.

The temperature behind the collimation shock depends on the photon production rate in the immediate downstream.
The question then arises: can dissipation at the collimation shock lead to a drastic change in $\tilde{n}$?
In the absence of photon generation the observed temperature behind the shock should equal the temperature at the origin since
$\tilde{n}$  is conserved, namely
	\begin{equation}
	kT_{obs}=\Gamma_skT_s=\Gamma_0kT_0 \approx 1.5~L_{53}^{1/4}R_7^{-1/2}\Gamma_0^{1/2} \quad {\rm MeV},
	\label{eq:T_obs}
	\end{equation}
noting that $kT_0=m_pc^2\eta_0/4\tilde{n}$  and adopting $\tilde{n}$ from Eq. (\ref{eq:ntil_GRB}).
Mixing will not alter this result, since $\tilde{n}$ will change by exactly a factor of $\eta_s/\eta_0$.
However, photon generation can lead to a gradual decline of the temperature of the advected flow behind the shock.
To estimate  $\tilde{n}$ note that the relative number of newly generated photons behind the shock 
is given by $\Delta n_{\gamma}\simeq \dot{n}_{ff} t^\prime_s$, where $t^\prime_s=r_s/\Gamma_sc$ 
is the proper flow time of the shocked plasma, $\dot{n}_{ff}\simeq \alpha_e\sigma_T c (1+x_\pm)^2n_s^2(kT_s/m_ec^2)^{-1/2}\Lambda_{ff}$ 
is the approximate free-free emission rate (see
Eq. (\ref{eq:bremss_rel})), $n_s$ and $T_s$ are the proper baryon density and temperature behind the collimation shock, respectively, $x_\pm$ 
is the pair-to-baryon ratio, and for illustration we adopt $\Lambda_{ff}\simeq10$. In terms of the pair unloaded optical depth behind the collimation shock, $\tau=\sigma_T n_s r_s/\Gamma_s$, 
the number of newly generated photons per baryon is given by: $\Delta n_\gamma/n_s\simeq 0.1(1+x_\pm)^2 \tau  (kT_s/m_ec^2)^{-1/2}\approx 0.3(1+x_\pm)^2\tau$, for the normalization adopted above.  At proper temperatures above $50$ keV roughly the pair density becomes large, $x_\pm>>1$. Thus, even a modest
$\tau$ is sufficient to increase the photon-to-baryon ratio, $n_\gamma/n_s$, well above that produced at the outflow injection point, Eq. (\ref{eq:ntil_GRB}).
 For the simulation run exhibited in Fig. \ref{fig:collim} for instance $\tau\simeq10^5$, which is quite typical, thus in
 practice it is expected that the temperature behind the collimation shock will be regulated by pair creation, and will not exceed 50 keV
 or so in the fluid rest frame (somewhat above the black body limit, $kT_{BB}= 20 (L_{53}/\Gamma_s^2 r_{s11}^2)^{1/4}$ keV ). 
The observed temperature would depend on the opening angle of jet: $kT_{obs}=\Gamma_s kT \simeq 50 \theta_0^{-1} $ keV.
For the simulation shown in Fig. \ref{fig:collim} we find $\Gamma_s\simeq4$, implying $kT_{obs}\sim 200 $ keV.  Note 
that $L_{jiso}\propto \theta_0^{-2}\propto \nu_{p}^2$ , where $h\nu_p=kT_{ obs}$ is the photon energy at the spectral peak.
Interestingly, this is consistent with the Amati relation.   A similar idea was discussed earlier by \cite{thompson2007}.

\subsection{Observational diagnostics}
The anticipated large radiative efficiency of sub-photospheric shocks, particularly the collimation shock, implies
that  they should have dominant imprints on the resulting emission.    One robust effect already mentioned above is 
photon generation behind the collimation shock, that lowers the spectral peak, and can lead to a softer spectrum below
the peak if further dissipation occurs just beneath the photosphere (e.g., by internal shocks produced through mixing or modulation
of the engine).
As noted in \cite{ito2018a}, there is 
an important difference between emission from a forward shock and reverse, as well as sub-photospheric  collimation, shocks.
While in the former case an observer detects the radiation that escapes through the upstream region and, hence, is
beamed in the forward direction (or in the forward hemisphere in the shock frame),  in the latter case the observed radiation escapes 
through the downstream region and is beamed in the backwards direction.    This gives rise to
notable differences in the observed spectra from a single shock;  in particular, the spectrum emitted from a reverse
shock extends to much higher energies than that emitted from a forward shock.   Examples are shown in 
Fig. 10 in \cite{beloborodov2017a}, and  Fig. 19 in \cite{ito2018a}, where
the integrated spectrum of photons moving with and against the flow is exhibited.  
As expected, there is a prominent hard component extending above the peak
 in the case of emission from a reverse shock, which is produced by bulk Comptonization around the RMS transition layer
 (in practice the interaction of the escaping radiation with socked gas behind the forward shock may alter the 
 transmitted spectrum, an effect not taken into account in the calculations of \cite{ito2018a}).
The spectrum emitted by a forward shock, on the other hand, lacks such 
a component (although it is broader than an exponential cutoff), since the high energy photons produced by  bulk Comptonization
move preferentially along the bulk flow.
In both cases, the portion of the spectrum below the  peak is much softer (broader) than a thermal spectrum.
 This is  due to the moderately bulk Comptonized component in which energy gain by scattering is not
 so significant, as well as due to the superposition of thermal-like spectra emitted from the upstream and downstream regions.
Broadening of the spectrum below the peak is also (independently) expected to arise from the weak internal shocks that result from the mixing of jet and cocoon material \citep{keren2014,gottlieb2019}. 

While the spectra displayed in \cite{ito2018a} are obtained by integration of the shock emission over a finite slab, the Monte-Carlo simulations 
that produce the emission assume an infinite, steady shock.   
Whether these spectra mimic the time integrated spectrum of the breakout emission is questionable. 
In reality, the structure of the shock gradually changes as the radiative losses increase, giving rise to a continuous 
adjustment of the local spectra during the breakout phase.   Making the reasonable assumption that the
shock structure evolves in a quasi-steady manner, it is possible to compute the structure and emitted spectrum at any given time 
by incorporating photon escape in the simulations.     An attempt to perform Monte-Carlo calculations of a leaking, forward RMS
is currently underway (Ito \& Levinson, in preparation).   Preliminary analysis indicates the formation and 
gradual strengthening of a collisionless subshock once a significant fraction of the shock energy starts escaping the system.   
How this affects the emitted spectrum is yet to be seen. 

\section{ Shock breakout in stellar explosions}
\label{sec:SNe}

\subsection{Breakout from a stellar surface} 
\label{sec:star_breakout}
In a typical SN an explosive release of energy at the center of the star drives a radiation mediated shock into its envelope. The shock decelerates at first at the inner parts of the envelope, but as it approaches the stellar surface, where the density descends sharply, it accelerates. The shock accelerates as long as the optical depth to the stellar edge is sufficient to support an RMS and,
if the star is not surrounded by a thick stellar wind, it breaks out of the star once the photon diffusion time to the edge becomes shorter than the shock expansion time. At this point the photons that were trapped inside the shock transition layer are released to the observer.
These photons are the first electromagnetic emission seen and they produce the so-called "shock breakout emission". The shock transition layer at the time of the breakout is called "shock breakout layer". 
After the RMS breaks out it is transformed into a collisionless shock and continues to propagate in the circum stellar medium. At the same time a rarefaction wave propagates backwards, into the shocked envelope, causing its outer parts to accelerate. As the  shocked envelope expands, photons from inner layers (behind the shock breakout layer) start diffusing out to the observer. 
This radiation, commonly termed "cooling envelope emission", readily follows the breakout episode and lasts significantly longer than the brief shock breakout signal;  in some cases it even dominates the total emitted energy (e.g., as in type IIp SNe).
 
Under the conditions prevailing in a stellar envelope the RMS is expected to be photon poor and weakly magnetized. 
Its structure  plays a dominant role in shaping the shock breakout emission, and in cases where the RMS is sufficiently fast (i.e., the radiation in the transition layer is out of thermal equilibrium, see \S \ref{sec:jump_cond}) it might also affect the early phases of the cooling envelope emission. Below we discuss the hydrodynamics and observational signature of the breakout of a spherical shock from a stellar surface and the early stages of the cooling emission,  focusing on the effects of the RMS structure on the observed signature. Due to the different nature of subrelativistic and relativistic breakouts we discuss each class separately.

\subsubsection{Subrelativistic shock breakout}
\label{sec:star_breakout_NR}
Shock breakout from a stellar surface during a SN explosion has been studied by many 
authors \cite[e.g.,][]{Colgate1974,Falk1978,Klein1978,Imshennik1981,Ensman1992,matzner1999,nakar2010,Rabinak2011,tominaga2011}. 
The evolution of the shock during the breakout phase is dictated by the density profile near a stellar edge, which can be approximated by a power-law of the distance to the edge, namely $\rho \propto x^n$ where $x=(R_*-r)/R_*$, $r$ is the distance from the center and $R_* = 10^{11} R_{*,11}$ cm is the stellar radius. For typical envelopes $n=1-3$, depending mainly on the mode of energy transfer, wherein $n \approx 1.5$ for convective envelopes (e.g., red supergiants) and $n \approx 3$ for radiative envelopes (e.g., blue-supergiants and Wolf-Rayets). The hydrodynamics of a spherical shock that propagates in such a density gradient is self-similar, with the shock velocity satisfying $v_s \propto \rho^{-\mu}$, where for RMS (downstream adiabatic index of 4/3) $\mu \approx 0.19$ with a very weak dependence on $n$ in the relevant regime \citep{sakurai1960}. The shock velocity at the outset of the acceleration phase is roughly $\sqrt{E/M_{ej}}$, where $E=10^{51} E_{51}$ ergs is the explosion energy and $M_{ej} = 5 M_{ej,5}~M_\odot$ the mass of the progenitor.
Thus, the shock velocity near the stellar edge can be approximated as $v_s \approx \sqrt{E/M_{ej}} (\rho/\rho_*)^{-0.19}$ in terms  of 
the mean stellar density $\rho_*=M_{ej}/R_*^3$.   This relation holds at densities above which the photon diffusion time, $\tau x/c$, is longer than the expansion time, $x/v_s$, here $\tau\approx \kappa \rho x $ being the optical depth near the stellar edge and $\kappa$ the opacity.
At the breakout point $\tau=c/v_s$.

The velocity profile of the ejecta (shocked gas) post breakout is dictated by the accelerating shock \citep{matzner1999}.  Approximating
the mass enclosed in a shell located at $x$ by $m(x)\approx 4\pi \rho(x) R_\star^3 x$ yields the profile $m(v) \propto v^{-\frac{n+1}{0.19n}}$, $v\le v_{bo}$, for the ejecta mass.
The breakout velocity $v_{bo}$ is obtained from the implicit equation $\tau(v_{bo})=c/v_{bo}$ with $\tau(v_{bo})=\kappa m(v_{bo})/4\pi R_\star^2$. It depends weakly on the value of $n$ (in the relevant range of $n$ values) and for $n=3$ it is \citep{nakar2010}:
\begin{equation}\label{eq:vbo_stellar}
	v_{bo} \approx 0.3c ~E_{51}^{0.58} M_{ej,5}^{-0.41} R_{*,11}^{-0.33} ~.
\end{equation}
The energy released during the shock breakout is approximately $m_{bo} v_{bo}^2$, where   $m_{bo} \approx 4 \pi R_* c/(\kappa v_{bo})$ is the mass of the breakout layer. Since for typical shock velocities $H$ and $He$ are fully ionized, the gas opacity is dominated by Thompson scattering for which $\kappa \approx 0.2-0.34{\rm~cm^2/gr}$, depending on the fraction of $H$ in the envelope. Adopting
$\kappa=0.34$ for illustration one obtains
\begin{equation}
	E_{bo} \approx m_{bo} v_{bo}^2 \approx 3\times10^{44} {\rm~erg}~ E_{51}^{0.58} M_{ej,5}^{-0.41} R_{*,11}^{1.66}.
\end{equation}
The duration of the breakout emission is roughly the light crossing time of the progenitor:
\begin{equation}
	t_{bo} \approx \frac{R_*}{c} \approx 3 ~s ~R_{*,11}~.
\end{equation}
The observed temperature is determined by the chemical potential of the radiation in the immediate downstream at the time of the breakout.
As shown in \S \ref{sec:NR_RMS}, in fast photon starved shocks the immediate  downstream temperature depends sensitively on the shock velocity and weakly on the upstream density. At typical breakout densities ($\rho_{bo} \sim 10^{-7}-10^{-9} {\rm~gr~cm^{-3}}$), the radiation is out of thermal equilibrium for $v_{bo} \gtrsim 0.04$c  (see Eq. \ref{eq:beta_crit}). From equation (\ref{eq:vbo_stellar}) it is evident that for typical SN explosions with $E_{51} \sim 1$ the radiation falls out of thermal equilibrium for $R_* \lesssim 10^{12}$ cm. Thus, in explosions of red-supergiants (RSG, $R_\star \sim 5 \times 10^{13}$ cm; e.g., type IIp SNe) the shock breakout is in thermal equilibrium, while explosions of WR stars ($R_\star \sim 10^{11}$ cm; e.g., type Ib/c SNe) it is out of thermal equilibrium and explosions of blue supergiants  (BSG, $R_\star \sim 10^{12}$ cm; e.g., 1987-like SNe) are marginal. The dependence of the shock breakout temperature on the explosion parameters is not trivial (see e.g., \citealt{nakar2010}), but in general for $E_{51} \sim 1$, the shock breakout temperature is  $\sim 1-10$keV for a WR, $\sim 0.1-1$keV for a BSG and $\sim 25$eV for a RSG.

The cooling envelope emission is divided into two phases, planar and spherical. The transition takes place roughly when the expanding breakout layer doubles its radius (namely reaches $2R_*$). During the planar phase the optical depth of the gas remains constant and photons are diffusing from just behind the breakout layer. During the spherical phase the optical depth drops quickly (as $t^{-2}$) and a diffusion wave crosses the ejecta releasing photons from increasingly deeper layers. The luminosity of the cooling phase does not depend on the RMS structure, but the temperature during the planar phase may depend on it. If the  breakout layer is out of thermal equilibrium then the radiation during the planar phase is out of thermal equilibrium as well. The emitted radiation, which emanates at this time from regions that are just behind the breakout shell, is driven slowly towards thermal equilibrium. As a result, the observed temperature during this phase drops faster than expected from adiabatic cooling alone \citep{nakar2010}. Once the spherical phase commences, deeper layers which are at thermal equilibrium are quickly exposed, changing the spectral evolution. 

\subsubsection{Relativistic shock breakout}
\label{sec:star_breakout_R}
The hydrodynamic evolution and radiation characteristics of relativistic RMS are vastly different than those of  Newtonian RMS,
partly due to a rapid creation of electron-positron pairs in the shock transition layer.
Regarding the hydrodynamics, a relativistic shock that propagates in the sharply descending density near the stellar edge, $\rho \propto x^n$, accelerates as $\gamma_{sh} \propto \rho^{-0.23}$ \citep{johnson1971,tan2001,pan2006}. Upon breakout a rarefaction wave accelerates the ejecta farther, but unlike the Newtonian case, in the relativistic regime the acceleration is highly significant.
The final Lorentz factor of a given fluid shell depends on whether the acceleration ends during the planar phase  (i.e., before the shell doubles its radius) or in the spherical phase.  Typically, shells with a terminal Lorentz factor $\lesssim 30$ end their acceleration during the planar phase (see details in \citealt{nakar2012}), in which case the final Lorentz factor of each shell is given by $\gamma_f=\gamma_{sh}^{1+\sqrt{3}} \approx  \gamma_{sh}^{2.7}$, here $\gamma_{sh}$ denotes the Lorentz factor gained by the shell upon crossing the shock \citep{johnson1971,pan2006}.  For $n=3$ the resulting mass profile of the ejecta following the acceleration phase is $m \propto \gamma_f^{-2.1}$. If the acceleration continues well into the spherical phase then the final Lorentz factor is  $\gamma_f \approx  \gamma_{sh}^{2.1}$ \citep{Yalinewich2017}.
 
The emission of a relativistic spherical breakout from a stellar edge was derived by \cite{nakar2012}.
The RMS propagates in the stellar envelope up to the point where the optical depth to the edge is too small to sustain it.  This happens when the optical depth for a photon moving from the downstream to the upstream is roughly unity. However,  under the conditions 
anticipated in stellar envelopes (photon poor plasma) the opacity of a relativistic RMS is dominated by self-generated pairs (see section \ref{sec:RRMS}),
hence the breakout does not occur at a location where the optical depth of the pre-shocked gas is $\tau_{unloaded} \approx 1$, but
rather at a much lower pair-unloaded optical depth.   
Nevertheless, the breakout emission is dominated by the $\tau_{unloaded} \approx 1$ layer, which we hereby term the "breakout layer". 
The reason is that in most cases, the radiation trapped inside this layer is released during the planar phase, despite the high pair opacity,
owing to an exponential decline of the pair density with proper temperature, that drops during the acceleration of the gas
from its value behind the shock, $\sim 200$keV, to $\sim 50$keV \citep{nakar2012}.  At this temperature the pair content becomes negligible and the radiation from the breakout layer readily escapes to the observer. 
 \cite{nakar2012} have shown that for typical parameters the radiation from the breakout layer is released after the breakout layer is accelerated to its terminal Lorentz factor. Radiation from all the layers that are external to the breakout layer (i.e., faster,  less massive, and contain less energy than the breakout layer) is also released during the planar phase.  On the other hand, layers beneath the breakout layer (i.e., slower and more massive) carry more energy but this energy is trapped during the planar phase. Thus, the breakout emission is dictated by the properties of the breakout layer.

The three major observables of the breakout pulse - total energy, characteristic observed temperature and duration can be derived based on the breakout radius, $r_{b0}=10^{13}r_{bo,13}$ cm, and the terminal Lorentz factor of the breakout layer, $\gamma_{bo,f}$  \citep{nakar2012}.  The energy released during the breakout can be estimated (after accounting for the acceleration and the rest-frame cooling to $50$keV) by:
\begin{equation}\label{eq:EboRelStar}
	E_{bo} \sim  10^{48} r_{bo,13}^{2} \gamma_{bo,f}^{1.37} {\rm~erg} .
\end{equation}
The duration is dominated by angular light travel time
\begin{equation}
	t_{bo} \sim \frac{r_{bo}}{2c\gamma_{bo,f}^2} \approx 200 \frac{r_{13,bo}}{\gamma_{bo,f}^{2}} {\rm~s},
\end{equation}
and the temperature is 
\begin{equation}\label{eq:TboRelStar}
	kT_{bo}  \sim 50 \gamma_{bo,f} {\rm~ keV}
\end{equation}
Note that the canonical breakout radius taken above ($r_{bo}=10^{13}$cm) is appropriate for low-luminosity GRBs, if those are shock breakouts (see \S\ref{sec:chokedJ}). Equations (\ref{eq:EboRelStar})-(\ref{eq:TboRelStar}) show three observables that depend on two physical parameters and therefore should satisfy a closure  relation:
\begin{equation}
	t_{bo} \sim 200 \left(\frac{E_{bo}}{10^{48}{\rm~erg}}\right)^{1/2} \left(\frac{T_{bo}}{50{\rm~keV}}\right)^{-2.7} {\rm ~ s}.
\end{equation}
Note that the relation has a strong dependence on $T_{bo}$, which is hard to constrain accurately from the observations since the observed spectrum is not a blackbody. The reason is that first, the spectrum within the shock transition layer is not a black body, and second light travel time effects mix the photons emitted at different radii between $R_{bo}$ and $2R_{bo}$. Thus, shock breakout emission is expected to agree only to within an order of magnitude with the closure relation.

The Lorentz factor of the breakout layer in a spherical explosion, after it is accelerated by the rarefaction wave is 
\begin{equation}
	\gamma_{bo,f} \approx 14~\left(\frac{E}{5\times 10^{52} {\rm~erg}}\right)^{1.7} M_{ej,5}^{-1.2} R_{*,11}^{-0.95}
	\label{eq:gamma_star}
\end{equation}  
Thus, generating a relativistic breakout in a spherical explosion requires very high energy and a compact progenitor and the resulting signal is short and relatively dim. An alternative source of a relativistic breakout that can be active at much larger breakout radii, is a shock driven by a jet. A relativistic jet that propagates through the progenitor (or other dense medium that engulfs the jet launching site, as in neutron star merger) deposits part or all of its energy into a cocoon. The breakout of the shock driven by the cocoon into the medium can be relativistic also in case that the jet energy is moderate.

\subsection{Breakout from a stellar wind} 
\label{sec:wind_breakout}

The physics of the breakout process and the characteristics of the breakout signal are different in cases
where the shock emerges from the stellar wind surrounding the progenitor, rather than the edge of the star.
This is particularly expected in compact progenitors, like Wolf-Rayet stars, that exhibit  broad emission lines, indicating fast winds
with high mass flux.  Observationally, the mass loss rate and wind terminal velocity inferred in massive
stars span a wide range, with $\dot{M}_w= 10^{-7}-10^{-4}~M_\odot$ yr$^{-1}$ and $v_w=100-3000$ km/s.   
Those winds are presumably driven  by radiative pressure. 
However, there is a growing body of evidence suggesting that many SN progenitors experience episodes of prodigious
mass loss shortly (months to years) before core collapse, with rates as high as $\dot{M}_w \sim10^{-3} - 10^{-1} ~M_\odot$ yr$^{-1}$ \citep{ofek2014b,galyam2014,svirski2014a}.
The nature of these intense eruptions is yet unclear, albeit some theoretical explanations have been offered (e.g., \citealt{shiode2014}). 

Observational evidence for shock breakout from a wind are rare and controversial. The properties of the emission depend on the breakout velocity and the duration of the signal (see below), so observations of SNe with very different properties are attributed to a breakout from a wind. Probably the most robust shock breakout candidate is the X-ray flash from SN 2008D \citep{soderberg2008,modjaz2009}. Here various properties of the X-ray flash suggest that the shock breaks out of a dense wind and not from the surface of the progenitor \citep[e.g][]{soderberg2008,balberg2011,svirski2014a}, which is presumably a WR star.  A second type of SNe where the emission is suggested to be from a shock that propagates in a wind are type IIn SNe that show a bright and blue light curve and are thought to be powered by interaction. In these SNe the rise time has been attributed to a shock breakout emission \citep[e.g.,][]{ofek2010,ofek2014}. The last type of SNe that were suggested to be a breakout through a stellar wind are bright and very long ultra-luminous SNe where the mass of the wind is so large (several solar masses or more), that the breakout signal constitues practically the entire main part of the SN light. The prototype of this class is SN2006gy  \citep[e.g.,][]{chevalier2011}.

In what follows we consider the propagation of a RMS in a dense stellar wind.
Although in reality these intense winds are likely to be clumpy and unsteady, we 
shall suppose for simplicity that the wind is stationary and spherical, with a total mass 
flux $\dot{M}_w=10^{-2} \dot{M}_{-2}~M_\odot$ yr$^{-1}$ and constant velocity
$v_w=10^3 v_{w3}$ km/s.   Under these assumptions the total optical depth of the wind can be expressed as 
\begin{equation}
\tau_{w\star}= \kappa \rho_w R_\star=10^3 \kappa_{0.2}\dot{M}_{-2}R_{\star11}^{-1}v_{w3}^{-1}
\end{equation}
in terms of the Thomson opacity per unit mass, $\kappa=0.2\kappa_{0.2}$ gr$^{-1}$ cm$^{2}$, 
and the progenitor radius $R_\star=10^{11} R_{\star11}$ cm, here assuming that the wind extends smoothly 
from the stellar edge.   The optical depth at $r>R_\star$ is given by $\tau_w(r)=\tau_{w\star}(R_\star/r)$.  
Since the optical depth must exceed $c/v_s$ in regions where the shock is mediated by radiation,
it is evident that shock breakout will occur in the wind provided 
\begin{equation}
v_{s}/c > \tau_{w\star}^{-1} = 10^{-3} (R_{\star11}v_{w3}/\kappa_{0.2}\dot{M}_{-2}).
\end{equation}
This readily implies that in relativistic explosions even a modest wind, $\dot{M}_w > 10^{-5}~M_\odot$   yr$^{-1}$, would result in a delayed shock breakout. 
This is a conservative estimate since, as explained below, the optical depth required to support a relativistic RMS in a wind is much 
smaller than unity.

The breakout radius, $r_{bo}$, and velocity, $v_{bo} \equiv  v_s(r_{bo})$, can be readily inferred from the observed energy,
$E_{bo}$, and duration, $t_{bo}$, of the breakout signal.    In relativistic shocks the downstream temperature at breakout, $T_{bo}$, 
depends on $r_{bo}$ and $v_{bo}$ alone, hence measuring the peak energy of the spectral energy distribution provides a rough 
consistency check on the model.   Furthermore, if the structure of the ejecta emerging from the star is known, or assumed, this information can 
be employed to yield a relation between the explosion energy and ejecta mass. In subrelativistic shocks the radiation must diffuse out through 
the upstream gas before reaching the observer and this can alter the observed spectrum.
In view of  the fundamental difference in the physics involved in Newtonian and relativistic RMS breakouts, we shall
discuss them separately in the following.

\subsubsection{Subrelativistic shock breakout from a wind}
\label{sec:wind_breakout_NR}
Detailed analysis of sub-relativistic breakouts from a wind is presented in \cite{svirski2012}.   Below we recapitulate the main results.
For a non-relativistic shock the optical depth at breakout satisfies $\tau_{bo}=\tau_{w\star}R_\star/r_{bo}=c/v_{bo}$
and the density $\rho_{bo}=\tau_{bo}/\kappa r_{bo}$.   The swept up mass is given by 
$m_{bo}=4\pi \rho_{bo}r_{bo}^3=4\pi cr_{bo}^2/\kappa v_{bo}$, and the swept up energy by
\begin{eqnarray}
E_{bo}=m_{bo}v_{bo}^2=\frac{4\pi c}{\kappa}v_{bo}^3 t^2_{bo}=5\times10^{44} \kappa_{0.2}^{-1}(v_{bo}/0.1c)^3 t^2_{bo,2} \quad {\rm erg},
\label{eq:E_bo_NR}
\end{eqnarray}
out of which at most $50\%$ can be radiated away,  where
\begin{equation}
t_{bo}=r_{bo}/v_{bo} = 10^2 t_{bo,2}\quad {\rm s}
\end{equation}
is the duration of the breakout signal. It is now seen that a measurement of $E_{bo}$ and $t_{bo}$ readily yields
the breakout radius, velocity and density,
\begin{eqnarray}
\rho_{bo}=\frac{3c}{\kappa}v_{bo}^{-2} t_{bo}^{-1} = 5\times10^{-10} \kappa_{0.2}^{-1}(v_{bo}/0.1 c)^{-2} t_{bo,2}^{-1}\quad {\rm g~cm^{-3}}.
\end{eqnarray}
The breakout temperature $T_{bo}$ can be computed using the RMS model.    For an infinite planar shock it is given by Eq. (\ref{eq:Theta_d}) upon substituting $\beta_u=v_{bo}/c$, $n_d=\rho_{bo}/m_p$ (see also Fig. \ref{fig:fg_Tb}).   For a thermal spectrum
one might naively expect the observed peak energy to be $\sim 3kT_{bo}$.
However, in practice this estimate suffers both from observational and theoretical uncertainties.   From an
observational perspective, the relation between the peak energy and temperature is uncertain if the spectrum substantially
deviates from thermal.  From a theoretical perspective, the shock temperature may be altered by radiative losses; in particular, it is expected to evolve during the breakout phase \citep{ioka2018}.   But the largest uncertainty comes from the interaction of the diffusing radiation with the gas
upstream of the shock \citep{svirski2012}.   Thus, using the closure relation as a consistency  test for the breakout model requires 
detailed radiative transfer calculations of the escaping radiation, using the radiation intensity just upstream of the shock as a boundary condition. 
As will be shown in the following section, this uncertainty is removed in relativistic shocks by virtue of the small opacity of the upstream gas at
the breakout radius.

%
The evolution of the shock depends on the structure of the ejecta expelled from the star.  In section \ref{sec:star_breakout}
it was shown that the energy profile of the  ejecta can be expressed in 
terms of the maximum velocity $v_0$ of the ejecta, obtained following the acceleration of the shock
in the stellar envelope, in the form:
\begin{equation}
E(v) =\frac{4\pi c v_0}{\kappa}R_\star^2(v/v_0)^{-\lambda},
\end{equation}
where $\lambda=(1+0.62 n)/0.19n$, and $1.5\le n\le3$ is the polytropic index that depends on the progenitor type.
The dependence of the shock velocity on radius can be readily obtained by equating the swept up energy, $m_{w}v_{s}^2=(4\pi \tau_{w\star} R_\star/\kappa)r v_s^2$,
with the energy injected into the shock, $E(v_{s})$.  This yields,
$v_s(r)=v_0 (c R_\star/v_0\tau_{w\star} r)^{1/(\lambda+2)}$.
At the breakout radius $r_{bo}/v_{bo}=\tau_{w\star}R_\star/c$, implying
\begin{equation}
v_{bo}=v_0(R_\star/r_{bo})^{2/(\lambda+1)}=v_0(R_\star/v_0t_{bo})^{2/(\lambda+3)}.
\end{equation}
Using Eq. \ref{eq:vbo_stellar} in section \ref{sec:star_breakout} one can express $v_0$ in terms of the 
explosion energy $E=10^{51}E_{51}$ ergs
and the ejecta mass $M_{ej}=5M_{ej,5} M_\odot$.    For a WR star with $n=3$ and $\kappa=0.2$ one finds 
\begin{equation}
v_{bo}/c\simeq 0.15~E_{51}^{0.44}M_{ej,5}^{-0.31}t_{bo,2}^{-0.25}.
\label{eq:v_bo}
\end{equation}
Consequently, a measurement of $E_{bo}$ and $t_{bo}$ imposes a constraint on the ejecta mass and the explosion energy. 
If an additional constraint can be obtained from the post breakout emission than the explosion parameters can be uniquely inferred.

The above analysis implicitly assumes that the edge of the wind exceeds the breakout radius, $r_w>r_{bo}$.  This implies
a wind age $t_w > (v_{bo}/v_w) t_{bo}$.  If this condition is not satisfied then the breakout will occur at the wind tail, where
the density gradient is much steeper than $r^{-2}$.   Predicting the properties of the breakout signal in this case would require 
a different analysis. 

\subsubsection{Relativistic shock breakout from a wind}
 \label{sec:wind_breakout_R}
As shown in detail in \cite{granot2018}, a key feature of relativistic RMS that makes them inherently distinct from non-relativistic RMS is self-generation of the shock opacity.
In section \ref{sec:RMS-with-escape} it was shown that an accelerated pair cascade ensues as long as the pair unloaded Thomson optical depth ahead of the shock 
satisfies $\tilde{\tau} \simgt m_e\gamma_{sh}/m_p$.   This can be understood by noting that in order that the shock will be mediated by radiation,
the net optical depth (including pairs and KN cross-section) $\Delta \tau$ traversed by a counter streaming photon crossing the shock (i.e., propagating from the immediate downstream to the 
upstream) should exceed unity.     The latter can be approximated as
$\Delta \tau\simeq \sigma_{KN}(n^\prime_\gamma+n^\prime_\pm) r/\gamma_{sh}$, where $\gamma_{sh}$ is the shock Lorentz factor, $r/\gamma_{sh}$ 
the width of
the shock transition layer, as measured in the shock frame, $n^\prime_\gamma$ and $n^\prime_\pm$ the photon and pair densities measured in
the shock frame, and it is assumed
that the pair-production cross section $\sigma_{\gamma\gamma}$ roughly equals the Klein-Nishina cross section $\sigma_{KN}$, which is a good approximation at relativistic energies.
Since the average energy of quanta downstream a relativistic RMS is $m_ec^2$, energy conservation implies 
$n^\prime_\gamma+n^\prime_\pm \simeq m_p n_u\gamma_{sh}^2/m_e$, where $n_u$ is the proper upstream density.    Omitting a logarithmic
factor and noting that the internal proper energy per particle of the incoming plasma is $\sim m_ec^2\gamma_{sh}$,
one finds $\sigma_{KN}\simeq \sigma_T/\gamma_{sh}^2$ (see section \ref{sec:RMS-with-escape} for further details).  The requirement $\Delta\tau\simgt1$ then
implies $\tau_{unloaded} \sim \sigma_T n_u r\simgt m_e\gamma_{sh}/m_p$, as formally obtained in section \ref{sec:RMS-with-escape} from the analytic solution.

From the above considerations it is anticipated that the breakout phase should be very gradual, with full breakout 
occurring at optical depth $\tau_{bo}\simeq m_e\gamma_{bo}/m_p$, and a corresponding 
radius $r_{bo}\simeq \tau_{w\star} R_\star/\tau_{bo} \simeq 2\times10^{13}\tau_{w\star}R_{\star11}\gamma_{bo}^{-1}$ cm,
where $\gamma_{bo}$ denotes the shock Lorentz factor at $r_{bo}$, provided the RMS remains relativistic up to this radius.    If the shock decelerates to a mildly relativistic Lorentz factor, $\gamma_{sh}\beta_{sh}\sim1$, prior to this radius, the temperature behind the shock will drop considerably below $m_ec^2/3$,  pair creation will no longer be
sufficient to maintain the required opacity and, if $\tau_w<1$ at the location where $\gamma_{sh}\beta_{sh}\sim1$,  the entire radiation 
stored inside the shock will promptly escape.   Otherwise, if the shock decelerates to a mild Lorentz factor at $\tau_w>>1$ the breakout will
be Newtonian, as described in the previous section.  We shall get back to this point later on. 

Assuming that the shock remains relativistic at breakout, simple relations can be derived between the three observables, $t_{bo}$, $E_{bo}$ and $T_{bo}$, and the breakout radius $r_{bo}=10^{13}r_{bo,13}$ cm and Lorentz factor $\gamma_{bo}=\gamma_{sh}(r_{bo})$.
Recalling that the duration is compressed by Doppler boosting one finds
\begin{equation}\label{eq:tbo1}  
	t_{bo}\approx \frac{r_{bo}}{2c\gamma_{bo}^2}  \approx 200~ r_{bo,13}  \gamma_{bo}^{-2}  {\rm ~s}.
\end{equation}  
Now, the energy is related to the swept up mass through $E_{bo}=\gamma_{bo}^2m_{bo}c^2$ (since the internal energy 
per baryon is approximately $m_pc^2\gamma_{sh}$).  The latter is given by $m_{bo}=4\pi\tau_{bo}r^2_{bo}/\kappa \simeq 4\pi m_e\gamma_{bo}r_{bo}^2/\kappa m_p$.  Thus,
\begin{equation}\label{eq:Ebo1}  
	E_{bo}=3  \times  10^{45} \kappa_{0.2}^{-1}  r_{bo,13}^2  \gamma_{bo}^3 \quad  {\rm  ~erg}.
\end{equation} 
Finally, in relativistic RMS the downstream temperature is regulated by pair creation at $T_d\simeq m_ec^2/3$ (see section \ref{sec:RRMS}).
In the observer frame this temperature is boosted by a factor $\gamma_{sh}$, yielding
\begin{equation}\label{eq:Tbo1} 
	kT_{bo} \approx 200 ~\gamma_{bo} \quad {\rm  ~keV} .
\end{equation}
 Since  these three observables  depend on two  breakout parameters  they should satisfy  a closure
relation:        
\begin{equation}\label{eq:closure}        
	E_{bo} \approx 10^{45} \kappa_{0.2}^{-1} \left(\frac{t_{bo}}{100{\rm~s}}\right)^2 \left(\frac{T_{bo}}{200{\rm~keV}}\right)^7 {\rm  ~erg}. 
\end{equation} 
%

The breakout observables can be related to the progenitor's parameters and the explosion energy by employing 
the quasi-steady shock model outlined in section \ref{sec:RMS-with-escape}.    This model assumes that prior to its complete breakout, the shock
evolves in an adiabatic manner, in the sense that it follows a sequence of steady shock solutions with increasing radiative losses. 
Under this assumption, the evolution of the shock can be quantified in terms of the escape parameter $f$, defined
as the fraction of downstream photons that escape the shock and never return (or, equivalently, the fraction of shock energy which
is radiated away).  Now, in section \ref{sec:RMS-with-escape}  it has been shown that once escape commences the shock thickness, measured in 
units of the pair unloaded Thomson length, satisfies $\Delta \tilde{\tau}\simeq (m_e/m_p)\gamma_{sh}/f$.  Since this thickness is
roughly equal to the optical depth ahead of the shock, it implies $\tau_w(r_{sh}) \simeq (m_e/m_p)\gamma_{sh}/f$ and $r_{sh}=
\tau_{w\star}R_\star/\tau_w= (m_p/m_e)\tau_{w\star}R_\star\gamma_{sh}^{-1}f$.
The evolution of $\gamma_{sh}$ can be found by equating the energy pumped into the shock by the ejecta, 
$E(\gamma_{sh})$, with the swept-up
energy $m_w(r_{sh})\gamma_{sh}^2=4\pi(m_p/\kappa m_e)\tau_{w\star}^2R_\star^2 \gamma_{sh} f$. 
The energy profile of the ejecta emerging from the star has the form (see \S \ref{sec:star_breakout_R}) $E(\gamma)=E_0(\gamma/\gamma_{0})^{-1.1}$, with $E_0=4\pi \kappa^{-1}R_\star^2\gamma_{0}$ and $\gamma_{0}$ given by Eq. (\ref{eq:gamma_star}).
This yields
\begin{equation}\label{eq:Gsh}
	\gamma_{sh}(f)
	\approx 1.2 E_{53}^{1.7} M_{ej,5}^{-1.2} R_{\star11}^{-0.95} \tau_{w\star}^{-0.95} f^{-0.48},
\end{equation} 
and
\begin{equation}\label{eq:Rsh}
	r_{sh}(f) \approx 1.7 \times 10^{14} E_{53}^{-1.7} M_{ej,5}^{1.2} R_{\star11}^{1.95} \tau_{w\star}^{1.95} f^{1.48} {\rm~cm}.
\end{equation}     
As the shock propagates $f$ increases while $\gamma_{sh}$ decreases. The breakout takes place either when $f$ approaches unity or when the shock velocity drops to a value (roughly $\beta_{sh}\simeq 0.5$) at which newly created pairs no longer dominate the shock opacity.
If the breakout occurs while shock is still relativistic then the breakout Lorentz factor and radius 
can be determined from Eqs. (\ref{eq:Gsh}) and (\ref{eq:Rsh}) with $f=1$, that is, 
$\gamma_{bo}=\gamma_{sh}(f=1)>1$, and $r_{bo}=r_{sh}(f=1)$.
The breakout observables can be related to the system parameter upon substituting $\gamma_{bo}$ and $r_{bo}$ into
equations (\ref{eq:tbo1})-(\ref{eq:Ebo1}).   Specifically,
the duration of the breakout signal is
\begin{equation}
	t_{bo}\approx 2\times10^3~ E_{53}^{-5.1} M_{ej,5}^{3.6} R_{*,11}^{3.85} \tau_{w,*}^{3.86} \quad {\rm ~s}
\end{equation}
its observed temperature at $t \sim t_{bo}$ is
\begin{equation}
	kT_{bo} \approx 250~ E_{53}^{1.7} M_{ej,5}^{-1.2} R_{*,11}^{-0.95} \tau_{w,*}^{-0.95} \quad  {\rm ~keV} 
\end{equation}
and the total emitted energy is
\begin{equation}
	E_{bo}=10^{48}~ E_{53}^{1.7} M_{ej,5}^{-1.2} R_{*,11}^{1.05} \tau_{w,*}^{1.05} \kappa_{0.2}^{-1} \quad {\rm ~ergs} .
\end{equation}

The temporal evolution of the emitted spectrum prior to a complete breakout (i.e., as $f$ evolves from $f<1$ to $f=1$) can be computed using Eqs. (\ref{eq:Gsh})-(\ref{eq:Rsh}), in terms of the time elapsed in the observer frame,
$t =r_{sh}/c\gamma_{sh}^2\propto f^{2.44}$.   In particular, the pulse bolometric luminosity and observed temperature evolve according to
\begin{equation}
L\simeq E_{sh}/t\propto t^{-0.78},
\end{equation}
and 
\begin{equation}
T_{obs}\simeq m_ec^2\gamma_{sh}/3\propto t^{-0.2}.
\end{equation}
The pulse rise time is expected to be much shorter than $t_{bo}$.

If $\gamma_{sh}(f=1)<1$ then the breakout takes place when the shock becomes mildly relativistic or Newtonian at
\begin{equation}
	r_{bo} \approx r_{sh}(\Gamma_{sh} \approx 1)\approx  3 \times 10^{14} E_{53}^{3.57} M_{ej,5}^{-2.52} R_{*,11}^{-1} \tau_{w,*}^{-1} {\rm~cm} ~~~~~(\gamma_{bo}\approx 1),
\end{equation}
assuming that $\tau_w<1$ at this location. The duration of the breakout emission is then simply
\begin{equation}
	t_{bo}\approx \frac{r_{bo}}{c} \approx 10^4 E_{53}^{3.57} M_{ej,5}^{-2.52} R_{*,11}^{-1} \tau_{w,*}^{-1}{\rm ~s}~~~~~(\gamma_{bo}\approx 1),
\end{equation}
and the temperature is
\begin{equation}
	T_{obs,bo} \approx 50-100  {\rm ~keV} ~~~~~(\gamma_{bo} \approx 1).
\end{equation}
The total emitted energy is roughly $ 4 \pi A R_{bo}c^2$, which depends only on the explosion energy and ejecta mass,
\begin{equation}
	E_{bo} \approx 2 \times 10^{48} E_{53}^{3.57} M_{ej,5}^{-2.52}  \kappa_{0.2}^{-1} {\rm ~erg} ~~~~~(\gamma_{bo}\approx1).
\end{equation}
while the luminosity depends only on the progenitor radius and wind density
\begin{equation}
   L_{bo} \approx 2 \times 10^{44} R_{*,11} \tau_{w,*}  \kappa_{0.2}^{-1} {\rm ~erg/s} ~~~~~(\gamma_{bo}\approx1).
\end{equation}

\subsection{Choked jets and low luminosity GRBs} 
\label{sec:chokedJ}
Low luminosity GRBs (henceforth llGRB) is a subclass of long GRBs (LGRBs)  that  show distinct observational properties, including substantially lower luminosities, a softer spectrum that lacks a high-energy power-law tail, and a smooth, non-variable light curve.   
Moreover, the inferred event rate of llGRBs  is much higher than that of regular LGRBs.   The distinct properties of  llGRBs 
suggest that they are likely generated by a  different mechanism  than regular LGRBs.   Yet, both llGRBs and regular LGRBs 
appear to be associated with broad-line IC supernovae - a peculiar type of CCSN - pointing towards a common origin (specifically,
a similar progenitor and explosion mechanism).  The apparent dissimilarity in emission properties despite  indications of a common origin 
have led to the suggestion of a unified picture according to which both classes share the same explosion physics but in different environments \citep{nakar2015}.

Models for llGRBs include off-axis emission from a regular LGRB jet  \citep{nakamura1998,eichler1999,yamazaki2003}, 
a long lived ($> 10^3$ s) low power central engine \citep{woosley1999,irwin2016} and shock breakout 
\citep{kulkarni1998,campana2006,li2007,waxman2007,katz2010,nakar2012,nakar2015}.    The off-axis jet model predicts strong radio emission to appear several years after the gamma-ray flash, 
which is inconsistent with the radio luminosity observed in GRB 980425 \citep[e.g.,][]{waxman2004,soderberg2004,peters2019}.   
It may also be in tension with compactness limits \citep{matsumoto2019}.   Invoking prolonged activity of 
a low power jet in llGRBs \citep{irwin2016} implicitly implies 
that the same progenitor produces vastly different central engines in regular LGRBs and llGRBs.  While this is possible, to our knowledge no specific mechanism that can account for such a disparity has been identified.  We do not attempt to provide a detailed account of llGRB 
scenarios here. Instead, in what follows we discuss the shock breakout mechanism,
which falls within the scope of this review.

According to the unified model of LGRBs the key difference between llGRBs and regular LGRBs is the outer structure of the progenitor.   In all LGRB types the progenitor consists of a compact ($\sim R_\odot$) massive ($\sim 10M_\odot$) core, however, while in llGRBs the core is ensheathed by an extended ($>100R_\odot$) low-mass ($\sim 0.01M_\odot$) envelope that chocks the jet 
as it propagates from the explosion center outwards, in regular LGRBs this envelope is absent and the jet breaks out from the 
star during the engine cycle and expands freely afterwards \footnote{In an alternative model \citep{irwin2016} the llGRB jet is not choked
but rather breaks out of the extended envelope.  This requires engine operation time of several thousands seconds.} .   As a result, while in regular LGRBs the observed gamma rays are emitted at or above
the photosphere of the highly relativistic jet, in llGRBs  the emission is released upon breakout of the mildly relativistic 
shock driven by the choked jet from the extended envelope, and is, therefore, expected to be soft and show no rapid variations, as
indeed observed.   A wide-angle, mildly relativistic shock is expected to be generated in the stellar material by the emerging cocoon also in regular LGRBs, and may contribute a weak gamma-ray signal that can be detected when observing the source at sufficiently large angles to the jet axis.
However, given the small radius of the core,  ($\sim R_\odot$), the shock breakout energy in these sources is expected to be very low and  such sources, if detected, should form a disparate class.
 
The structure suggested by the unified model for the llGRB progenitors (massive core and low-mass extended envelope) is not rare. It is seen in an increasing number of SNe, both with and without H envelope, by its hallmark signature of a double-peaked light curve, where the first peak is associated with the cooling emission of the low-mass extended envelope and the second peak is powered by radioactive decay of $^{56}$Ni \citep{hoflich1993,nakar2014,taddia2016,arcavi2017}. Thus, a unique prediction of the unified model is that llGRBs should be associated with double-peaked SNe. This model, in which the gamma-ray flash is emitted during the breakout of the shock 
from the extended envelope, also accounts for the unusual velocity profile of the ejecta inferred in this type of SNe,  particularly
the exceptionally high kinetic energy carried by the fast moving component compared with other (normal) SNe.  This
fast component is driven by the choked GRB jet and is distinct from the slower component ejected by the central SN explosion.
The association of the double-peaked SN 2006aj with llGRB 060218 lend strong support to this unified model \citep{nakar2015}.

The mass, $M_{ex}=10^{-2} M_\odot~M_{ex,-2}$, and radius, $r_{ex}=10^{13}~r_{ex,13}$ cm, of the extended envelope can 
be estimated from the time and bolometric luminosity of the first SN peak in terms of the velocity of the fast ejecta \citep{nakar2014}.  
For SN 2006aj such an analysis yields $M_{ex,-2}\simeq1$  and $r_{ex,13}>1$ \citep{nakar2015}.   
If a GRB jet having an isotropic equivalent luminosity $L_{iso}=10^{51}L_{51}$ erg s$^{-1}$ propagates through this 
medium, then the engine operation time required for a successful breakout is
$t_{e}\simgt 10^2 L_{51}^{-1/2}M_{ex,-2}^{1/2} r_{ex,13}^{1/2} $ s \citep{bromberg2011b,nakar2015}\footnote{Note that $t_e$ 
is much shorter than the light crossing time of the envelope.}, 
which for the inferred extended envelope parameters is considerably
longer than the mean duration of regular LGRBs (that reflects the engine time), unless the event  
is exceptionally powerful.    Note that the isotropic equivalent energy required for a successful breakout 
can be expressed as $E_{iso}= L_{iso}t_e > 3\times 10^{53} (t_e/30~s)^{-1}M_{ex,-2}r_{ex,13}$.  
 
In cases where the jet is choked the aspherical cocoon inflated by the jet continues to propagate inside the envelope until
breaking out.   The large optical depth of the extended envelope, $\tau\simeq 10^{3.5} M_{ex,-2} r_{ex,13}^{-2}$, implies 
that shock breakout most likely occurs near the edge of the envelope, at $r=r_{ex}$.   The energy deposited in the cocoon
by the choked jet, $E_j=10^{51}E_{j51}$ ergs, implies a bulk velocity $v\simeq 0.2 c (E_{j51}/M_{ex,-2})^{1/2}$ for a spherical
ejecta.    Any  a-sphericity of the cocoon should lead to a lateral velocity distribution, with a mildly relativistic 
matter moving at relatively small inclination angles with respect to the jet axis.  The details depend on 
the opening angle of the jet and the chocking radius.   At any rate, a mildly relativistic breakout is anticipated. 
A very rough estimate yields a breakout mass of $m_{bo}\simeq 4\pi r_{ex}^2/\kappa\sim 10^{-6} r_{ex,13}^2~M_\odot$,
 duration 
\begin{equation}
t_{bo}\sim r_{ex}/c = 300 r_{ex,13}~s,
\end{equation}
and luminosity
\begin{equation}
L_{bo}\sim m_{bo}c^2/t_{bo}\sim10^{46}~r_{ex,13}\quad {\rm erg~ s^{-1}}.
\end{equation}
The temperature of the emerging shock should lie in the range $50 - 100 $ keV.  This is consistent with observations of llGRBs.

\subsection{llGRBs as neutrino sources?} 
\label{sec:llGRB-nu}

Neutrino emission from LGRBs have been proposed in several early expositions \citep[e.g.,][]{waxman1997,dermer2003,globus2015}.  The main idea is that protons accelerated by internal
shocks to ultra-high energies (UHE) collide with either ambient matter or external radiation field to produce pions that subsequently decay
into neutrinos.   A prerequisite of this model is that a considerable fraction of the GRB jet energy is dissipated above the photosphere
in collisionless internal shocks that can accelerate protons.    Whether collisionless shocks can tap a significant fraction of the bulk energy 
to accelerate protons is yet an open issue, but at any rate, for typical bursts internal shocks must form well outside the star, at a radius $r > r_{ph}\simgt 10^{12}$ cm (depending on loading; see \S \ref{sec:GRBs} for a discussion), in order to be 
mediated by collective plasma processes rather than by radiation.  In fact, sub-photospheric RMS that form at smaller radii should be ultimately 
converted into collisionless shocks upon breakout, but the energy remaining to accelerate particles is only a fraction of the total shock energy.  
Revised estimates of neutrino production in regular LGRBs may be needed if the latter process is a dominant dissipation channel above the photosphere.
Furthermore, in models for neutrino emission from regular LGRBs the TeV neutrinos are produced via
photo-hadronic interactions of the accelerated protons with surrounding photons.  The neutrino production efficiency is then 
 limited by the photo-pion opacity at the dissipation radius, which is generally not optimal.  To date there is no evidence for neutrino emission from 
 LGRBs \citep{abbasi2011,abbasi2012}.  The strict upper limit already imposed by iceCube observations \citep{abbasi2012} 
 constrains either the photo-pion opacity or the UHECR yield in GRB internal shocks.

Alternatively, if collisionless internal shocks do indeed tap a considerable fraction of the jet energy to accelerate protons to UHE,
then efficient neutrino production can be achieved via proton-proton collisions if a sufficiently dense target is present at radii $r>r_{ph}$.
Such conditions exist in llGRBs if indeed associated with a choked jet, as proposed by \cite{nakar2015} and 
described in detail in the preceding section.   The 
optical depth for inelastic nuclear collisions is approximately $\tau_{pp}=\sigma_{pp} n r$, where $\sigma_{pp}\simeq 40$ mb at 
a (center of mass) energy of about  20 GeV, and increases logarithmically with energy.   In terms of the Thompson optical depth of the
envelope, $\tau=\sigma_T n r$, it can be expressed as $\tau_{pp}=(\sigma_{pp}/\sigma_T)\tau \sim 10^{-2}\tau$.   For an envelope 
of mass $M=0.01 M_\odot$ extending to a radius $r=10^{13} r_{13}$ cm one finds $\tau\simeq10^4 r_{13}^{-2}$, hence $\tau_{pp}>1$
if $r_{13}\simlt 10$. 

In early papers discussing neutrino production in llGRB jets \citep{murase2006,gupta2007,horiuchi2008} it has been assumed that the observed gamma ray emission
directly reflects the jet properties, specifically, that llGRB jets have low power and wide opening angle.    On the other
hand, as discussed in \S \ref{sec:chokedJ}, the unified model asserts that the jet power in llGRBs is similar to the typical power
inferred in regular LGRBs \citep{nakar2015}. Hence, the neutrino luminosity that can be potentially radiated 
while the jet is hidden inside the envelope can be much larger than previously thought.    
The accompanied, weak gamma ray signal is attributed to a shock breakout episode in this scenario
and does not reflect the power released by the central engine.   

On the other hand, the neutrino emission produced through the interaction of the narrow proton beam accelerated inside the relativistic jet 
with the envelope material is expected to have a 
beaming cone much smaller than that of the gamma ray emission.  Consequently,  in most of the observed bursts 
the neutrino signal is anticipated to be absent (i.e., beamed away from the observer).   This implies that llGRBs are not suitable 
for targeted point-sources search, similar to the search conducted for LGRBs \citep{aartsen2014b}.  However, 
they will contribute to the diffuse flux.
\cite{ahlers2014}  find that the sources of the diffuse neutrino flux produce
a total energy output of $\sim 10^{43}$ erg Mpc$^{-3}$ yr$^{-1}$ in $\sim100$ TeV
neutrinos and their volumetric rate, assuming transient sources,
must be $\simgt 10^{-8}$  Mpc$^{-3}$ yr$^{-1}$ (as inferred from the lack of
neutrino clustering). Assuming that each llGRB harbor a
relativistic jet with a typical energy of $\sim 10^{51}$ ergs, the total
energy output in such jets is roughly $3\times10^{44}$ erg Mpc$^{-3}$ yr$^{-1}$. Thus,
if 10\% of this energy is converted to high-energy protons
before the jet is choked (i.e., at radii $\simlt 10^{13}$ cm) then llGRBs
are producing the observed diffuse flux. Assuming that the
typical jet angle is $\sim10^\circ$ the rate of llGRBs for which the
neutrino beam is pointed towards Earth is
$\sim 0.5 \times 10^{-8}$ Mpc$^{-3}$  yr$^{-1}$ consistent with the limit derived by \cite{ahlers2014}. 
A high confidence association of the diffuse neutrino flux with llGRBs will provide further, strong support 
to the unified LGRB model, and will also indicate that mildly relativistic collisionless shocks accelerate protons effectively 
to UHE.


\section{Gamma-ray emission from neutron star mergers}
\label{sec:BNS}

About $1.7$ seconds after LIGO-VIRGO recorded the first-ever gravitational wave signal from a neutron star merger (\citealt{abbott2017GW} and references therein), 
a short gamma ray flash (GRB 170817A) was detected by the Fermi-GBM and Integral \citep{abbott2017grays,goldstein2017,savchenko2017}, followed by electromagnetic
emission at lower energies over much longer times.   Although GRB 170817A
was classified as a short GRB, it was fainter than the faintest sGRB previously detected by roughly three orders of magnitude,
exhibiting an isotropic equivalent energy of $E_{\gamma,iso}\simeq 4\times10^{46}$ ergs.  
Various explanations of this peculiarity of the gamma-ray signal have been offered shortly after the 
announcement of GW170817 detection.  The two that seem most plausible at present (or at least most popular)
are shock breakout emission and emission from a so-called structured jet.  In the context of this review 
we focus here on the former mechanism.   

According to the shock breakout model, the shock that produces the gamma-ray flash is driven by the relativistic jet during its 
propagation in the merger ejecta.    This breakout emission is always expected when the relativistic jet successfully 
emerges from the ejecta, and in certain circumstances also when it is choked inside.  
In the former case, gamma ray emission can also be produced through dissipation
of the relativistic jet itself;  this is most likely the source of gamma-ray emission detected in regular sGRBs, when the jet is 
observed on-axis.   However, this emission is narrowly beamed and outside the core of the ultra-relativistic jet, 
at viewing angles larger than a few degrees, appears too faint to be detected. 
The shock breakout signal on the other hand is emitted over wide angles, and for sources within the LIGO-VIRGO horizon 
can be detected at much large viewing angles.  

Since the shock-breakout dynamics and emission depend on the properties of the confining medium,
it is instructive to describe first the various components of the merger ejecta.   
We shall begin the following discussion with a brief account of the merger multi-flow structure, 
and then move on to discuss in some detail the shock dynamics and emission.   An in depth review of the origin and 
properties of the ejecta, the associated kilonova emission, and the origin and propagation of the jet  is given in \cite{nakar2019}.

\subsection{Outflow components}
The outflow which is expelled during and shortly after the collision of the two neutron stars consists of several components
that have different origins, propagate at different speeds, from $\gamma\beta \sim 0.1$ to $\gamma\beta>>1$, and interact
with each other.   These are the dynamical ejecta, that comprises a slower component ejected early on by tidal forces 
and a faster component driven by a shock that forms in the collision, the secular (post-merger) ejecta (e.g., winds from a putative
hyper-massive neutron star or a black hole disk), and the relativistic jet which is launched, presumably with some delay, by a central engine once formed. 

Mass ejection starts as the two neutron stars approach the final stages of the inspiral. Tidal forces start
ejecting mass, mostly along the equator, during the last orbit before coalescence. As the two NS collide their cores
are compressed on each other. If the total binary mass is large enough ($> 2.8 M_\odot$) the first collision leads to
a prompt collapse to a BH and dynamical mass ejection ceases.  Otherwise, tidal mass ejection from the central, fast
rotating compact object continues for several orbits while the collision of the two cores and the following bounce
drive shocks that eject more mass. Generally, tidally ejected mass is concentrated more towards the equator, it
is slower and is neutron rich. Shock driven ejecta is faster, more isotropic and its neutron-to-proton ratio is smaller.
Note that the tidal and shock driven ejecta collide, interact and affect each other so the distinction
between the two components is blurred. Yet, simulations find a clear correlation between the polar angle
and the dynamical ejecta properties, whereby material thrown out closer to the equator is slower and more neutron
rich. The net result of the tidal and shock driven ejecta is that $\sim 90\%$ of the mass moves at velocities of
$\sim 0.1c - 0.3c$,  and a small fraction of the mass, termed fast tail, moves very fast, at $> 0.6c$. The outflow
covers the entire sphere with more mass ejected near the equator and less towards the pole. The dynamical ejecta also
has a wide range of electron fraction, with roughly a uniform distribution in the range $Y_e \simeq 0.1-0.4$.
The mass and velocity of the dynamical ejecta depend strongly on the EOS and other factors in ways that will not be 
reviewed here. For a detailed account the reader is referred to \cite{nakar2019}.

As stated above,
while most of the dynamical ejecta moves at velocities $< 0.4c$, a small fraction of the mass is expected to reach
faster, possibly even relativistic, velocities. Although the mass in this fast tail is minute and may appear insignificant, it can 
nonetheless control the breakout dynamics  which, in turn, can lead to important observational consequences.  We shall get back 
to this point latter on.

In addition to the dynamical ejecta there is also the post merger ejecta that consists of disk winds and neutrino driven winds from the 
putative HMNS.  A thorough  discussion of this component lies outside the scope of this review. 

\subsection{Shock breakout and emission}
The propagation of the relativistic jet through the ejecta drives strong shocks into the ejecta, both by the jet-head and by the expanding cocoon. Strong shocks may also be driven into the ejecta by an uncollimated wind, e.g., via a magnetar spin-down.  Owing to the large optical depth of the ejecta these shocks are radiation mediated.   Once a shock, such as the forward shock driven by the cocoon, breaks out of the ejecta, the photons that are trapped 
inside the shock transition layer are released to the observer, followed by emission from shocked layers downstream of the shock. 
For a shock velocity greater than about $0.5$c the mean photon energy is in the gamma-ray regime, and the breakout episode
will appear as a gamma-ray flash to a distant observer.   The process qualitatively resembles breakout emission in supernovae, as described
in \S \ref{sec:SNe}, but differs in details.  

The breakout emission is determined by the interaction of photons released from  the shock transition layer and the downstream region with the ejecta, as they stream towards the observer. Especially in the relativistic case, this is a dynamical process, far from thermal equilibrium, that involves different species (photons, electrons, positrons and baryons) which interact on vastly different scales, and should be computed using kinetic theory.  Currently, there are no ab initio calculations of this emission, and the models provide only order of magnitude estimates of the main observables (energy, duration and typical photon frequency) of the signal emitted during the breakout  of a spherical shock that propagates in plasma with no free neutrons. \\

\subsubsection{RMS properties}
As explained in the preceding sections, the RMS structure depends on the shock velocity and the upstream conditions (particularly photon number, magnetization and composition). In the case considered here the shock velocity  is $\gtrsim 0.1$c and the upstream is most likely unmagnetized, photon poor and composed of r-process material\footnote{It is also possible that faster parts of the ejecta are composed of lighter elements, including free neutrons, although we do not consider this option here due to the lack of proper theory for RMS with free neutrons.}.  
The composition of the upstream plasma affects both the opacity and, more importantly, the photon generation rate (see derivation below Eq. (\ref{eq:bremss_lambda})).  In case of r-process material this leads to a considerable modification of the downstream temperature and pair content, as compared to
a pure hydrogen plasma, as well as stellar composition.  We therefore find it imperative to extend the calculations outlined in \S \ref{sec:NR_RMS} to the merger case, accounting for its unique composition. 

Equation (\ref{eq:ff_rate}) can be expressed in terms of the downstream density $\rho_d$ and temperature $T_d$, and the mean atomic and mass numbers, $\left<z \right>$ and  $\left<A \right>$, in the form:
\begin{equation}\label{eq:dot_nff}
	\dot{n}_{ff} \approx 4 \times 10^{36} {\rm~s^{-1}~cm^{-3}} \frac{\left<z\right>\left<z^2\right>}{\left<A\right>^2} \rho_d^2 T_d^{-1/2} \Lambda_{ff} ,
\end{equation}
where $\rho_d$ and $T_d$ are given in c.g.s units, and the averages are over the atomic fraction, e.g.,  $\left<z^2\right> = \Sigma x_j Z_j^2$ where $x_j$ is the atomic fraction of element $j$, etc.  The factor $\Lambda_{ff}(\rho_d,T_d)$ accounts for photons upscattered by inverse Compton, and it can be approximated by \citep{weaver1976,nakar2010,sapir2013}:
\begin{equation}\label{eq:Lambdaff}
	\begin{array}{l}
	\Lambda_{ff} \approx \max\left\{ 1,\frac{1}{2}\ln(y)[1.6+\ln(y)] \right\},\\
	\\
	y=500  \left(\frac{\left<z^2\right>}{\left<A\right>}\right)^{-1/2}\left(\frac{\rho_d}{10^{-9} {\rm~gr/cm^3}}\right)^{-1/2}\left(\frac{T_d}{\rm{keV}}\right)^{9/4}.
	\end{array}
\end{equation} The immediate downstream temperature $T_d$ is then found by solving the equation 
\begin{equation}\label{eq:Td}
	3k_B T_d\approx  \frac{ e_{\gamma d} }{\frac{L_{ph}}{\beta_d c} \dot{n}_{ff}(\rho_d,T_d)} = e_{\gamma d} \frac{ 3 \beta_d^2 \kappa \rho_d  c }{\dot{n}_{ff}(\rho_d,T_d)}
\end{equation}
where $L_{ph}\sim 3\beta_d\kappa \rho_d$ is the width of the layer just behind the shock within which photons can diffuse back to the 
upstream, and $\beta_d$, $\rho_d$, $e_{\gamma d}$ are the downstream velocity, density and energy density, respectively,
which are determined from the shock jump conditions in terms of the upstream density and shock velocity, as in Eq. (\ref{eq:Theta_d}), however, here we use the full solution that applies also to mildly relativistic shocks, rather than the Newtonian approximation used in \S \ref{sec:NR_RMS}. 
If the solution of equation (\ref{eq:Td}) results in temperature that is lower than the blackbody temperature, $T_{BB}=(\epsilon/a_{BB})^{1/4}$, then photon generation is rapid enough to maintain thermodynamic equilibrium in the shock transition layer and $T_d=T_{BB}$. If, however, the solution of this equation provides a temperature that is higher than $T_{BB}$, then the immediate downstream is out of thermodynamic equilibrium and its temperature is roughly
the value obtained from the solution . In that case photon generation continues to reduce the temperature as the fluid is advected away from the shock reaching thermodynamic equilibrium only at the far downstream.

\begin{figure}[ht]
	\center
	\includegraphics[width=1\textwidth]{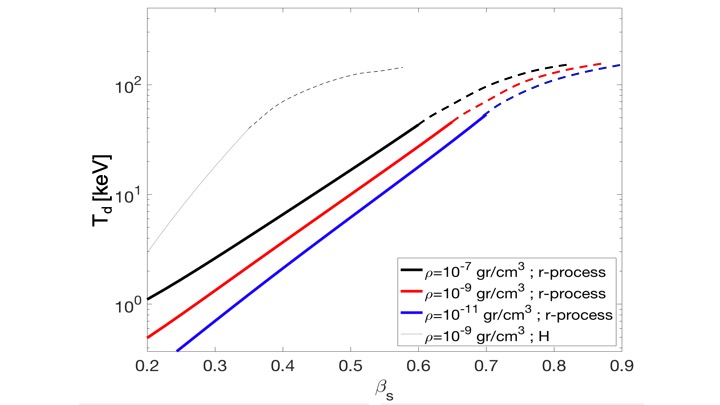}
	\caption{The temperature in the immediate downstream of a radiation dominated shock as a function of the shock velocity. The solid curve is calculated by solving numerically equations (\ref{eq:dot_nff}) - (\ref{eq:Td}). For r-process [H-rich] composition we use $\left<z\right>\left<z^2\right>/\left<A\right>^2 = 10 ~[1]$, $\left(\left<z^2\right>/\left<A\right>\right)^{1/2} = 5~ [1]$ and $\kappa= 0.16~ [0.34] {~cm^2/gr}$. This calculation is applicable for $T \lesssim 50$ keV. At higher temperatures vigorous pair production leads to increased photon generation that mitigates the rise in the temperature, setting it at $100-200$ keV, almost independent of the shock Lorentz factor. The dashed lines are illustrations of the temperature's behaviour in this regime.}%
	\label{fig:T_RMS}
\end{figure}

Equations (\ref{eq:dot_nff}) and (\ref{eq:Lambdaff}) show that the photon generation rate depends on the composition via averages on $Z$ and $A$. For fully ionized r-process elements with solar abundance and $A>85$ the values of these composition means are $\frac{\left<z\right>\left<z^2\right>}{\left<A\right>^2} \approx 10$, $\left(\frac{\left<z^2\right>}{\left<A\right>}\right)^{1/2} \approx 5$ and $\kappa=\frac{\left<z\right>}{\left<A\right>}\frac{\sigma_T}{m_p} \approx 0.16 {~\rm cm^2/gr}$. These values depend only weakly on the exact composition as long as it is dominated by r-process elements. Plugging these values into Eqs. (\ref{eq:dot_nff})-(\ref{eq:Td}) we find that the downstream radiation falls out of thermodynamic equilibrium once the shock velocity exceeds $\beta_s > 0.12 \left(\rho_d/10^{-9} {\rm~gr/cm^3}\right)^{1/30}$, where the very weak dependence on $\Lambda_{ff}$ is ignored. 
Figure \ref{fig:T_RMS} depicts the temperature in the immediate downstream as a function of the shock velocity $\beta_s$ (obtained by solving equations \ref{eq:dot_nff}-\ref{eq:Td}), for fast shocks with r-process material at several representative densities.  The results are compared with the solution obtained for H-rich plasma
at $\rho=10^{-9}$ gr/cm$^3$ in \S \ref{sec:NR_RMS} (dotted line).  
The figure shows that for r-process material  the temperature rises sharply from $< 1$ keV at $\beta_s=0.2$ to $50$ keV at $\beta_s = 0.6-0.7$.  Once the downstream temperature exceeds  $50$ keV, electron-positron pair production starts playing a role and the shock structure changes significantly. Pairs practically  affect all aspects of the shock structure, but the effect that probably has the largest impact on the observed signal is self-regulation of the photon temperature. As explained in detail in \S \ref{sec:RRMS}, exponential pair creation serves as a thermostat that controls the temperature in the immediate downstream once the number of pairs starts exceeding the number of baryons, at 50 keV roughly. 
In relativistic RMS this mechanism, that by coincidence becomes important once the shock velocity approaches the speed of light (i.e., $\Gamma_s \beta_s \gtrsim 1$), renders the proper downstream temperature (i.e., as measured in the fluid rest frame) insensitive to the shock Lorentz factor,
keeping it around  $100-200$ keV. 
The dashed lines in figure \ref{fig:T_RMS} delineate the regime where pair creation becomes important.  The flattening corresponds to the 
onset of the saturation level, at which equipartition between pairs and photons is reached.

Another important aspect of pair creation that can potentially affect the breakout signal is opacity self-generation (see \S \ref{sec:RRMS} for a detailed discussion).  
In relativistic RMS the newly created pairs dominate the optical depth within the shock transition layer, and  since the total optical depth of a relativistic RMS is $\tau_s \sim 1$, it implies a significant reduction in the physical width of the shock \citep{nakar2012,beloborodov2017a,granot2018,ito2018a}. In other words, the conversion of a relativistic RMS to a collisionless shock occurs at a radius at which the pair unloaded optical depth is much smaller than unity, owing to opacity self-generation via rapid pair production.  However, as we shall argue below, in most scenarios this does not have a significant influence on observables such as the total breakout energy or duration, since the breakout emission originates predominantly from the region were the pair unloaded optical depth is roughly unity . \\

\subsubsection{Shock dynamics}

The dynamics of a shock that crosses the ejecta depends on its driving force (the jet in the present discussion) and on the density and velocity profile of the ejecta.  It is worth noting that since shock breakout occurs when the optical depth ahead of the shock is unity or less, its properties are dictated by a minute amount of mass that is moving at the front of the ejecta. For example, if the breakout takes place at a radius of $10^{12}$ cm where the shock velocity 
is $\beta_{bo}=0.5$, then $\tau=1$ corresponds to a mass of $4 \times 10^{-8} \msun$ (adopting $\kappa= 0.16 {\rm~cm^2/gr}$) that constitutes a fraction of $\sim 10^{-6}$ of the total ejecta mass. Consequently, the breakout signal depends predominantly on the properties of the fast tail that leads the ejecta and has velocities that may significantly exceed those of the bulk of the ejecta, possibly mildly or even ultra relativistic (see, e.g., \citealt{kasliwal2017,gottlieb2018b,beloborodov2018,hotokezaka2018b,radice2018}). 
Computing the structure and velocity profile of the fast tail is a formidable task given the tiny fraction ($\sim10^{-6}$) of the total ejecta mass contained in it. 
Numerical simulations cannot reliably resolve such a minor component and the applicability of analytic methods is limited by virtue of the complex, nonlinear hydrodynamics involved in the expulsion of this mass. Nonetheless, heuristic arguments as well as state-of-the-art numerical simulations tend to indicate
that the leading parts of the ejecta are launched by energy deposition near the outer layers of the merging neutron stars during their collision, which typically yields a very steep velocity profile that extends to the relativistic regime. For example,  \cite{hotokezaka2018b} find that the fast tail reaches mildly and conceivably highly relativistic velocities (i.e., $\gamma \beta > 1$), and that the density distribution at $\beta>0.5$ is often steeper than $\rho \propto (\gamma \beta)^{-10} $.  This renders the breakout from the tail more like a breakout from a star (section \ref{sec:star_breakout}) rather than a breakout from a wind (section \ref{sec:wind_breakout}).

A key question concerning the cocoon-breakout emission model is under which conditions a shock breakout is expected, and if a breakout does occur, then at what radius and velocity. The answer to this question depends on the ejecta structure and on the source that powers the shock. 
In general, a shock breakout must always accompany the emergence of a successful jet from the merger ejecta.  However, the shock velocity 
and its lateral structure depend on the properties of the ejecta.  For ejecta consisting of a slow massive bulk with a shallow density profile 
and a low-mass, fast tail with a steep density profile, as suggestively indicated by theory, 
the jet's head propagates typically at a mildly relativistic speed within the dense bulk while the surrounding cocoon expands subrelativistically and 
is narrowly collimated.  Upon transiting to the fast tail the jet's head undergoes further acceleration and the cocoon expands sideways and 
accelerates to mildly relativistic velocities.  As a result, the breakout velocity of the shock driven by the cocoon changes laterally, ranging   
from mildly relativistic at relatively large angles to the jet axis to possibly ultra-relativistic at the head near the jet axis.

Choked jets may also lead to a shock breakout under certain conditions.  In general, a strong shock will be driven into the fast tail
if the total energy deposited in the cocoon (the net engine output in case of a choked jet) exceeds the rest energy of the mass swept up 
by the shock within the jet cone.   This is always expected when the outflow is energetic enough (isotropic equivalent energy $>10^{51}$
erg/s), or otherwise when the jet has crossed a significant part of the bulk of the ejecta before chocking.  
The dynamics of the shock once it starts crossing the tail depends on the density profile of the tail.  If it is steep  enough
($\rho \propto v^{-\alpha}$ with $\alpha>8.2$ ) a sufficiently strong shock will accelerate in the tail, ultimately breaking out, 
otherwise its fate depends on specific details (see \cite{nakar2019} for an elaborate discussion). 

The post breakout evolution is also important in shaping the observed signal. It depends mostly on whether the shock is relativistic or not and on whether pairs are produced.   In sub-relativistic shocks the pair content is negligible, particularly so for r-process composition, and post breakout acceleration is insignificant \citep{matzner1999}.  Shock breakout occurs at a radius where $\tau \approx 1/\beta_s$, at which the radiation
contained inside the shock is released. Photons from deeper layers diffuse to the observer over longer timescales.
Relativistic shocks experience significant acceleration post breakout, though likely less than in case of a breakout from a star (see \S \ref{sec:star_breakout} and references therein).  Moreover, newly created pairs dominate the opacity, hence the first signal is emitted 
once the shocked gas adiabatically cools to a temperature of $\sim 50$ keV, at which the pair density declines exponentially and 
the layer of pair-unloaded optical depth, $\tilde{\tau} \lesssim 1$, suddenly becomes transparent \citep{nakar2012}.  Photons from regions where  $\tilde{\tau} > 1$ 
cannot stream directly to the observer also after the pairs disappear and are therefore released to the observer at larger radii.

\subsubsection{Gamma-ray emission}
A simple, rough estimate of the primary breakout observables can be obtained for a spherical shock propagating in 
expanding ejecta.   The ejecta velocity and Lorentz factor, as measured in the observer frame, 
are henceforth denoted by $\beta_e$ and $\gamma_e$, respectively, and the shock velocity and Lorentz factor by $\beta_s$ and $\gamma_s$.
In the frame of the unshocked ejecta, designated by a superscript $'$, the shock velocity and Lorentz factor are given, respectively, by:
\begin{equation}
	\beta_s'=\frac{\beta_s-\beta_e}{1-\beta_s\beta_e},
\end{equation}  
and
\begin{equation}
	\gamma_s'=\gamma_s\gamma_e(1-\beta_s\beta_e).
\end{equation}  

Upon the breakout of a relativistic shock from the ejecta a short flare of gamma-rays is released to the observer from the breakout layer, where $\tilde{\tau} \sim 1/\beta_s'$. Subsequently, photons from deeper layers behind the shock start diffusing out of the expanding gas, and appear to
an observer as "cooling emission", reminiscent of SNe emission.  The cooling emission episode can be divided into two distinct phases, planar and spherical,
with the planar phase lasting roughly until the expanding material doubles its radius, and the spherical phase follows. 

The duration of the signal from the breakout layer is determined by the  {\it angular time} -  the difference in the light-travel-time of photons 
emitted from fluid elements moving at different angles to the sightline. 
The duration of the planar phase is dominated by the {\it radial time} - the difference in the arrival times of photons emitted from different radii. If the shocked material is relativistic then the radial time of the planar phase is comparable to the angular time of the breakout layer, hence, photons emitted 
at different times from the breakout layer and photons emitted during the planar phase arrive to the observer simultaneously, constituting the shock breakout $\gamma$-ray flare.  The relative contribution of each phase depends on the detailed structure at the leading edge of the ejecta and the breakout radius.  

As hinted above, a rough estimate of the energy in the shock breakout signal can be obtained by considering the emission 
from a breakout layer of dimensionless width $\tilde{\tau} \approx 1/\beta_s'$.
This is true for Newtonian as well as relativistic RMS, despite the fact that in the latter case the opacity is dominated by 
newly created pairs and the physical shock width is much smaller than that of a Newtonian shock, that is, $\tilde{\tau} << 1/\beta_s'$.
The reason is the sudden disappearance of pairs downstream of the shock 
once the temperature drops to $\sim 50$ keV \citep{nakar2012}.   Now, the pair unloaded optical depth of a layer of mass $m$ at radius $R$ 
is $\tilde{\tau} \approx \kappa m/(4\pi R^2)$, and since $\tilde{\tau} \approx 1/\beta_s'$ at the breakout radius $R_{bo}$, 
the mass of the breakout layer is estimated to be
\begin{equation}
	m_{bo} \approx \frac{4\pi R_{bo}^2}{\beta_s'\kappa} = 4 \times 10^{-8} \beta_{s,bo}'^{-1} \left(\frac{R_{bo}}{10^{12}{\rm~cm}}\right)^2 \left(\frac{\kappa}{0.16 {\rm~cm^2/gr}}\right)^{-1} \msun.
\end{equation} 
The observed energy of the breakout layer emission equals approximately the internal energy of the shocked breakout layer,
boosted to the observer frame:
\begin{equation}\label{eq:Ebo}
	E_{bo} \sim m_{bo} c^2 \gamma_{s,bo} (\gamma_{s,bo}' -1) \sim 7 \times 10^{46}  \frac{\gamma_{s,bo} (\gamma_{s,bo}'-1)}{\beta_{s,bo}'}\left(\frac{R_{bo}}{10^{12}{\rm~cm}}\right)^2 ,
\end{equation}
where  $\kappa=0.16 {\rm~cm^2/gr}$ has been adopted. As stated above, the total breakout signal might contain also 
contributions from layers deeper than the breakout layer that emit during the planar phase, in which case the breakout 
energy can be larger.

The duration of the breakout signal is determined by the angular time, which is given roughly by $R/2c\gamma^2$ for a spherical shell of radius $R$.  
Since the breakout layer accelerates following the emergence of the shock, its Lorentz factor may be substantially larger than $\gamma_s$
if the shock is relativistic.   For illustration we adopt a final Lorentz factor $\gamma_f'=\gamma_s^{\prime 2}$ in the ejecta frame, which 
translates to $\gamma_f \approx \gamma_s\gamma_s'$  in the observer frame.  This value is somewhat smaller than 
that ($\gamma_f'=\gamma_s^{\prime 2.7}$) found  for a sharp stellar edge (see \S \ref{sec:star_breakout_R}), and is closer to the one
computed by \cite{Yalinewich2017} for prolonged acceleration.
Note that since acceleration is negligible when $\gamma_s'\beta_s' \lesssim 1$, this approximation is valid also if the shock is sub-relativistic. Assuming a spherical geometry, the duration of the breakout signal is 
\begin{equation}\label{eq:tbo}
	t_{bo} \sim \frac{R_{bo}}{2c\gamma_{f,bo}^2} \approx 16 \left(\frac{R_{bo}}{10^{12}{\rm~cm}}\right) \left(\gamma_{s,bo} \gamma_{s,bo}'\right)^{-2}   {\rm~s}  ~.
\end{equation}

The breakout temperature is roughly the immediate downstream temperature of the breakout layer, as seen in the observer frame.  As evident from figure \ref{fig:T_RMS}, this temperature depends strongly on the shock velocity. If $\gamma_s'\beta_s' \lesssim 1$ then the rest-frame temperature should be calculated using equations (\ref{eq:dot_nff})-(\ref{eq:Td}) with $\beta_d \approx \beta_s'/7$, and then multiplied by $\gamma_s$ for transformation to the observer frame.  If $\gamma_s'\beta_s' \gtrsim 1$ then the rest-frame temperature at the time of emission is about $50$ keV and the Lorentz factor of the breakout layer is $\gamma_{f,bo}$, yielding and observed temperature of
\begin{equation}\label{eq:Tbo}
	T_{bo} \sim 50 \gamma_{f,bo} {\rm~keV} \sim 50 \gamma_{s,bo} \gamma_{s,bo}' {\rm~keV}~~~;~~~ \gamma_{s,bo}'\beta_{s,bo}' \gtrsim 1 ~.
\end{equation}

Equations (\ref{eq:Ebo})-(\ref{eq:Tbo}) show that when the shock is relativistic and the emission from the breakout layer is comparable to or larger than that of the planar phase, then the three main breakout observables depend on two physical parameters, $R_{bo}$ and $\gamma_{f,bo}$. In this regime the observables provide a direct measure of $\gamma_{f,bo}$ and $R_{bo}$, whereby the former is readily obtained from equation (\ref{eq:Tbo}) and the latter can be expressed as:
\begin{equation}
	R_{bo} \sim 2.5 \times 10^{11} \left(\frac{t_{bo}}{1{\rm~s}}\right)^{-1}  \left(\frac{T_{bo}}{100{\rm~keV}}\right)^2 {\rm~cm} ~~~;~~~ \gamma_{s,bo}'\beta_{s,bo}' \gtrsim 1 ~.
\end{equation}
Moreover, three observables that depend on two physical parameters must satisfy a closure relation:
\begin{equation}\label{eq:closure_BNS}
	t_{bo} \sim 1 \left(\frac{E_{bo}}{10^{46}{\rm~erg}}\right)^{1/2} \left(\frac{T_{bo}}{100{\rm~keV}}\right)^{-2.5} {\rm~s} ~~~;~~~ \gamma_{s,bo}'\beta_{s,bo}' \gtrsim 1 ~.
\end{equation}
These relations provide only gross approximations for the observables, even in cases where the breakout signal is dominated by emission 
from the breakout layer.   First, any contribution from the planar phase will render the estimate of the breakout energy uncertain. Second, and more importantly, the duration and radius depend sensitively on the temperature, which is difficult to estimate from the observations since 
the spectrum is expected to differ considerably from  a black body (see below), in which case the peak energy of the spectral energy distribution
may be determined by additional physics (e.g., bulk Comptonization at the shock).    These insights motivate detailed calculations of the spectral 
evolution inside and downstream of the shock at the breakout radius.

Predicting the spectrum of the breakout signal is a far more involved task, since unlike the total energy and duration, the spectrum depends on the dynamical evolution of the shock transition layer during the breakout phase.  At present, we are unaware of any detailed calculations of the emitted spectrum.  Nevertheless, some spectral features can be inferred from recent analyses.
First, the angle averaged spectrum in the shock transition layer differs from a blackbody or Wein (see Fig. \ref{fig:spect_starved} and \citealt{budnik2010})
 and, therefore, the shock breakout spectrum is expected to deviate significantly from thermal.   However, in the frame of the shock the spectrum has
 a strong angular dependence and detailed calculations are needed to predict its observed shape, particularly in cases where the upstream moves 
 relativistically with respect to the observer. 
Second, RMS are not expected to accelerate particles and, consequently, the spectrum should not have a high-energy, power-law extension over 
many energy decades above $\nu F_\nu$ peak, although broadening over a limited spectral range is certainly possible. Third, light-travel-time effects 
can give rise to simultaneous detection of photons originating from different radii and directions, and this might considerably affect the shape of the observed
spectrum.  Detailed calculations of the observed spectrum of the shock breakout emission 
need to take into account the structure, geometry of the breakout layer and the lateral variation of  the shock parameters.


\section{Summary and Outlook}
\label{sec:summary}
The early emission observed in a plethora of extreme cosmic transients, including GRBs, SNe, and neutron star
mergers, is dictated by the structure and dynamics of a radiation mediated shock upon breakout from the dense matter 
surrounding the blast center.   
The duration of the breakout signal ranges from seconds to days
and the observed temperature (or peak energy in case 
of non-thermal emission) from extreme UV to gamma rays, depending on the
environmental conditions and shock velocity at the breakout radius.    Although the physics of RMS is rather universal,
the details depend on the specific characteristics of the system in which they form, and therefore deserve to be considered
separately.  
A summary of applications to different classes of objects follows:\\

{\bf Supernovae}:  In most regular SNe the explosion is  quasi-spherical.  The characteristics of the breakout emission 
depend on the type of progenitor, the explosion energy, and on whether shock breakout occurs at the stellar 
surface or in a stellar wind if opaque enough to sustain the RMS after its emergence from the star, as anticipated in
compact progenitors like WR stars.  
The short duration of the breakout episode renders the detection of the breakout signal difficult.   To date
there are only a handful of candidates.  The wide field of view, high cadence transient surveys that started running recently and 
those that are planned for the near future (e.g., ZTF, BlackGem, LSST and others) are likely to detect many more.\\

{\bf Low luminosity GRBs}:  The distinct features of this GRB class, their apparent association with double-peaked
SNe, and the lack of a bright radio afterglow lend strong support to the hypothesis that, while their central
engines are similar to those of regular LGRBs, the observed gamma ray
emission originates from a mildly relativistic shock breakout driven by a choked jet.  This mechanism requires
the progenitors of llGRBs to be surrounded by an extended, low mass envelope, which is absent in progenitors 
of regular LGRBs.   The a-sphericity of the cocoon driven by the choked jet likely leads to a mildly relativistic
breakout, with duration and luminosity that depend on the extension of the envelope.  For envelope radius of
 $\sim 10^{13}$ cm and mass of $\sim 0.01 M_\odot$ the predicted signal is consistent with observations.   In the case of 
 SN2006aj, the envelope parameters required to account for the double-peaked light curve are remarkably consistent with those 
 needed to produce the observed gamma ray flash.  More associations of llGRBs with SNe of this type are needed to firmly
 establish this model.   In addition, such envelopes also provide a thick target for inelastic nuclear collisions and are, therefore,
 optimal environments for neutrino production via the interaction of ultra-high energy protons accelerated in the relativistic jet 
during its propagation inside the envelope with the ambient gas.    The vastly different beaming cones of the neutrino and gamma ray emissions
anticipated in this model render llGRBs potential sources of the diffuse neutrino flux, provided a significant fraction of the jet bulk energy can be tapped for the  acceleration of protons, but not suitable sources for targeted observations. \\

{\bf Regular long GRBs}:  The origin of the prompt GRB emission is yet unresolved. Recent analysis, confirmed by 
ab-initio 3D hydrodynamic simulations, indicates that
if the outflow becomes sufficiently weakly magnetized inside the star, such that a strong collimation shock forms, 
then the observed signal should be robustly dominated by photospheric emission under conditions commonly envisaged.   
Moreover, rapid pair creation inside and just
downstream of the shock and consequent photon generation, acts as a thermostat that fixes the temperature, offering a 
natural explanation for the observed peaks of the prompt emission spectrum.  
A pure adiabatic expansion of the flow above the collimation shock should yield a quasi-thermal emission spectrum, 
and it is evident that some additional, mild dissipation below the photosphere is required to modify the spectrum.  
The details as to how the observed spectrum is established are not entirely clear at present, and there are recent attempts to address this issue. 
Post processing, radiative transfer simulations indicate that multiple, sub-photospheric RMS, as well as sheared flow regions, 
produced by modulations of the central engine and/or mixing at the collimation throat are expected to modify the spectrum as they emerge from the photosphere.   These methods are currently limited by resolution and require further improvements and convergence tests for a better performance. 
Monte-Carlo shock simulations demonstrate 
that broad spectra are produced even in a single RMS during the prompt GRB phase, however, further analysis is needed to assess whether
these models can reproduce the observed Band spectrum, and under which conditions, and such efforts are currently underway.   
Moreover,  sub-photospheric shocks 
are expected to undergo a transition from radiation mediated to collisionless shocks upon breakout, leading, subsequently,  to synchrotron 
emission by nonthermal electrons accelerated at the collisionless shock front (if not suppressed by too high magnetization).  
The contribution of this spectral component would depend on the fraction of the shock energy remaining above the photosphere, and on 
whether mildly relativistic shocks ($\Gamma \beta \sim 1$) with significant magnetization ($\sigma > 10^{-3}$) are capable of accelerating
particles.   Since this emission can contribute significantly to the high-energy tail above the peak its analysis is highly desirable, albeit challenging. 

Avoidance of a strong photospheric component requires the jet to remain highly magnetized above the stellar envelope.    On what scales 
dissipation occurs, and how the observed spectrum is produced in these class of models is also unclear.    Since astrophysical relativistic jets
are likely launched by magnetic fields and, therefore, Poynting flux dominated at their origin, their stability and dissipation is the most 
pressing issue.   In the context of aforementioned discussion, if it will turn out that magnetic field conversion occurs well below the edge of the 
star then it seems that dominant photospheric emission is unavoidable. \\

{\bf Neutron star mergers}:
While the concomitant detection of a gravitational wave burst (GW 170818) and a gamma-ray flash (GRB 170817A) has confirmed an old
prediction \citep{eichler1989}, that sGRBs are produced in BNS (or NS-BH) mergers, the unusual faintness of GRB 170817A  indicated 
that the origin of the gamma-ray emission in this source is different than in regular sGRBs.  Off-axis emission from a relativistic jet, high inclination emission from a stratified jet, and cocoon breakout emission have been considered as plausible mechanisms and discussed extensively in the recent literature.    In this review (section  \S \ref{sec:BNS}) we focused on the latter mechanism, as it fits to the main review topic.

Given the observational constraints on the ejecta mass and speed, and the strong evidence for the presence of a relativistic jet in GW 170817, 
cocoon breakout emission seems quite likely to account for the observed gamma-ray flash in this source.  
The emission, in this scenario, originates from shocked plasma downstream of a RMS driven by the jet-cocoon system, as it breaks out of the merger ejecta.  It subtends a wide angle relative to the relativistic jet core, and its energy constitutes, quite generally, a tiny fraction of the total explosion energy. The low energy and wide emission cone predicted for the breakout signal naturally account for the  exceptionally low brightness of GRB 170817A, given the inferred viewing angle ($\sim 20^\circ$ with respect to the jet axis).

The dynamics of the shock, its geometry and the breakout radius can be computed using 3D HD simulations of jet propagation, 
for given ejecta properties and
delay time between the expulsion of the ejecta and the launching of the relativistic jet.  The breakout signal along different sight lines can then be computed by exploiting RMS theory with the shock parameters (i.e., upstream velocity and density, etc.) taken from the HD simulations. Such an
approach has been adopted in, e.g., \cite{kasliwal2017} and \cite{gottlieb2018b}.  There is very little freedom in this analysis (the only free parameters are essentially the delay time and velocity profile of the ejecta fast tail), particularly given that the structure and dynamics of the emerging jet-cocoon system also determine the afterglow lightcurve. 
The breakout radius required to explain the characteristics of GRB 170817A was found to be $\sim 2\times10^{11}$ cm , 
and the shock Lorentz factor with respect to the observer was found to be a few.  To achieve such a large breakout radius
the existence of a fast tail with a steep density profile, as suggestively indicated by merger simulations, 
has been posited.  The breakout radius can somewhat vary in general, depending on the fast tail structure.

An important point is that the temperature behind the RMS is vastly larger than the black-body temperature.  In fact, if the shock is mildly 
relativistic the proper temperature should be $\sim 50$ keV, below which the shocked plasma in the breakout layer suddenly becomes transparent 
by virtue of the exponential suppression of the pair content with decreasing temperature.   Consequently, shock breakouts in BNS mergers 
naturally produce a short, dim gamma-ray flash.  To account for the observed 
SED ($\nu F_\nu$) peak in GW 170817, the Lorentz factor of the emitting plasma (along the inferred sightline) should be around $5$, consistent
with the results of 3D HD simulations.

\section*{Acknowledgement} 
This work was supported by Israel Science Foundation Grant 1114/17.  EN is partially supported by an ERC grant (JetNS).
We thank Ore Gottlieb and  Hirotaka Ito for valuable discussions, help and permission to reproduce some figures.


\end{document}